\documentclass[10pt,
               floatfix,
               aps,
               pra,
               showpacs,
               amssymb,
               nofootinbib,
               superscriptaddress,
               twocolumn,
               longbibliography]{revtex4-2}
\usepackage{amsbsy}
\usepackage{amsthm}
\usepackage{amssymb}
\usepackage{amsfonts}
\usepackage{amsmath, mathtools}
\usepackage{bbold}
\usepackage{bbm}
\usepackage{bm}
\usepackage{bigints}
\usepackage{braket}
\usepackage{dsfont}
\usepackage{capt-of}
\usepackage{comment}
\usepackage{graphicx}
\usepackage{dsfont}
\usepackage{enumerate}
\usepackage{float}
\usepackage[colorlinks]{hyperref}
\usepackage[figure,table]{hypcap}
\usepackage[ansinew]{inputenc}
\usepackage{mathrsfs} 
\usepackage{multibib}
\usepackage{MnSymbol}
\usepackage{mathtools}
\usepackage{multirow} 
\usepackage{orcidlink}
\usepackage{subfigure}
\usepackage{siunitx}
\usepackage{soul}
\usepackage[most]{tcolorbox}
\usepackage{xparse}

\makeatother
\hypersetup{
	bookmarksnumbered,
	pdfstartview={FitH},
	citecolor={darkgreen},
	linkcolor={darkred},
	urlcolor={darkblue},
	pdfpagemode={UseOutlines}}

\makeatletter 
\newcommand{\raisem@th}[3]{\raisebox{#1}{$#2#3$}}
\newcommand{\raisemath}[1]{\mathpalette{\raisem@th{#1}}}
\makeatother
\NewDocumentCommand{\newhbar}{O{0pt} O{0pt}}{
  \ensuremath{\mathrlap{\raisemath{#2}{\hspace*{#1}{\mathchar'26\mkern-9mu}}}h}
}

\newcommand{\scpr}[2]{\ensuremath{\left\langle\right.\hspace*{-1pt} #1 \hspace*{-1pt}\left|\right.\hspace*{-1pt} #2 \hspace*{-1pt}\left.\right\rangle}}

\newcommand{\subtiny}[3]{\ensuremath{_{\hspace{#1 pt}\protect\raisebox{#2 pt}{\tiny{$ #3$}}}}}
\newcommand{\suptiny}[3]{\ensuremath{^{\hspace{#1 pt}\protect\raisebox{#2 pt}{\tiny{$ #3$}}}}}

\newcommand{\cP}{\mathcal{P}}

\mathchardef\mhyphen="2D 
\def\hbar{\newhbar[0.4pt][-0.35pt]}
\DeclareSIUnit \hb  {\textit{\hbar}}
\newcommand{\parag}[1]{\vspace{-6pt}\noindent\textit{#1{\textemdash}}}
\newcommand{\authG}[3]{\author{\href{https://scholar.google.com/citations?user=#3}{\textcolor{black}{#1}}\,\orcidlink{#2}}}
\newcommand{\auth}[2]{\author{#1\,\orcidlink{#2}}}

\renewcommand{\thesection}{\Roman{section}}
\renewcommand{\thesubsection}{\Roman{section}.\Alph{subsection}}
\renewcommand{\thesubsubsection}{\Roman{section}.\Alph{subsection}.\arabic{subsubsection}}
\makeatletter
\renewcommand{\p@subsection}{}
\renewcommand{\p@subsubsection}{}

\makeatother

\definecolor{lightbeige}{rgb}{0.996, 0.973, 0.914}
\definecolor{darkbeige}{rgb}{1, 0.933, 0.788}
\definecolor{darkgreen}{RGB}{50,190,50}
\definecolor{darkblue}{RGB}{0,0,190}
\definecolor{darkred}{RGB}{238,0,0}

\begin{document}
\title{State-Agnostic Approach to Certifying Electron-Photon Entanglement in Electron Microscopy}

\authG{Phila Rembold}{0000-0003-1405-730X}{htguzkAAAAAJ}
\email{phila.rembold@tuwien.ac.at}
\thanks{These authors contributed equally}
\affiliation{Atominstitut, TU Wien, Stadionallee 2, 1020 Vienna, Austria}
\authG{Santiago Beltr{\'a}n-Romero}{0009-0000-0310-5551}{W65xBVgAAAAJ}
\thanks{These authors contributed equally}
\affiliation{Atominstitut, TU Wien, Stadionallee 2, 1020 Vienna, Austria}
\affiliation{University Service Centre for Transmission Electron Microscopy, TU Wien, Wiedner Hauptstra\ss e 8-10/E057-02, 1040 Vienna, Austria}
\auth{Alexander Preimesberger}{0009-0000-7404-6165}
\thanks{These authors contributed equally}
\affiliation{Atominstitut, TU Wien, Stadionallee 2, 1020 Vienna, Austria}
\affiliation{University Service Centre for Transmission Electron Microscopy, TU Wien, Wiedner Hauptstra\ss e 8-10/E057-02, 1040 Vienna, Austria}
\authG{Sergei Bogdanov}{0009-0003-4393-6685}{h5LnuzAAAAAJ}
\affiliation{Atominstitut, TU Wien, Stadionallee 2, 1020 Vienna, Austria}
\affiliation{University Service Centre for Transmission Electron Microscopy, TU Wien, Wiedner Hauptstra\ss e 8-10/E057-02, 1040 Vienna, Austria}
\authG{Isobel C. Bicket}{0000-0002-5255-4481}{PvVECsIAAAAJ}
\affiliation{Atominstitut, TU Wien, Stadionallee 2, 1020 Vienna, Austria}
\affiliation{University Service Centre for Transmission Electron Microscopy, TU Wien, Wiedner Hauptstra\ss e 8-10/E057-02, 1040 Vienna, Austria}
\authG{Nicolai Friis}{0000-0003-1950-8640}{bW2vxuMAAAAJ}
\affiliation{Atominstitut, TU Wien, Stadionallee 2, 1020 Vienna, Austria}
\authG{Elizabeth Agudelo}{0000-0002-5604-9407}{MeKj42gAAAAJ}
\affiliation{Atominstitut, TU Wien, Stadionallee 2, 1020 Vienna, Austria}
\authG{Dennis R{\"a}tzel}{0000-0003-3452-6222}{ucB2mZcAAAAJ}
\affiliation{Atominstitut, TU Wien, Stadionallee 2, 1020 Vienna, Austria}
\affiliation{University Service Centre for Transmission Electron Microscopy, TU Wien, Wiedner Hauptstra\ss e 8-10/E057-02, 1040 Vienna, Austria}
\affiliation{ZARM, University of Bremen, 28359 Bremen, Germany}
\authG{Philipp Haslinger}{0000-0002-2911-4787}{17FQIcgAAAAJ}
\affiliation{Atominstitut, TU Wien, Stadionallee 2, 1020 Vienna, Austria}
\affiliation{University Service Centre for Transmission Electron Microscopy, TU Wien, Wiedner Hauptstra\ss e 8-10/E057-02, 1040 Vienna, Austria}

\begin{abstract}
{Transmission electron microscopes (TEMs) enable atomic-scale imaging and characterisation, driving advances across fields from materials science to biology. 
Quantum correlations, specifically entanglement, may provide a basis for novel hybrid sensing techniques to make TEMs compatible with sensitive samples prone to radiation damage.
We present a protocol to certify entanglement between electrons and photons naturally arising from certain coherent cathodoluminescence (CL) processes. 
Using mutually unbiased bases in position and momentum, our method allows robust, state-agnostic entanglement verification and provides a lower bound on the entanglement of formation, enabling quantitative comparisons across platforms. 
Simulations under experiment-inspired conditions and preliminary experimental data highlight the feasibility of implementing this approach in modern TEM systems with optical specimen access. 
Our work integrates photonic quantum information techniques with electron microscopy. 
It establishes a foundation for entanglement-based imaging at the atomic scale, offering a potential pathway to reduce radiation exposure.}
\end{abstract}

\maketitle

\section{Introduction}

The concept of entanglement is at the very heart of foundational questions in quantum mechanics~\cite{EinsteinPodolskyRosen1935, Bell1964, HorodeckiEntanglementReview2009}. 
It plays a central role in many modern quantum technologies, from quantum communication~\cite{Ursin-Zeilinger2007}, through quantum computing~\cite{Preskill2018}, to quantum metrology~\cite{GiovannettiLloydMaccone2011}. 
Setups for generating entangled photon pairs have become extremely compact and robust~\cite{BittermannBullaEckerNeumannFinkBohmannFriisHuberUrsin2024}, and the certification of their entanglement a routine problem~\cite{FriisVitaglianoMalikHuber2019}. 
At the same time, entangled photons have long been explored as a resource for enhanced imaging techniques~\cite{PittmanShihStrekalovSergienko1995,gilaberte_basset_perspectives_2019}, an area in which significant progress has been made recently~\cite{NdaganoDefienneLyonsStarshynovVillaTisaFaccio2020, DefienneNdaganoLyonsFaccio2021, CameronCourmeVernierePandyaFaccioDefienne2024}. 
Here, we focus on entanglement in a very different type of imaging device: the transmission electron microscope (TEM)~\cite{ReimerKohl2008, Schattschneider2018Entanglement}.

Despite their remarkable ability to produce atomic-resolution images, TEMs face limitations due to the radiation damage they inflict on samples~\cite{egerton_radiation_2004, ilett_analysis_2020}. 
This issue is particularly acute in the study of fragile radiation-sensitive biological specimens, which are quickly damaged by high-energy electrons. 
One way to address this issue is to reduce the electron dose. However, at low electron doses, shot noise becomes a significant limiting factor~\cite{egerton_spatial_2022, baxter_determination_2009}. 
In photon-based systems, quantum correlations have been successfully used to mitigate shot noise effects~\cite{bridaSystematicAnalysisSignaltonoise2011,bridaExperimentalRealizationSubshotnoise2010,onoEntanglementenhancedMicroscope2013} and improve phase-contrast imaging~\cite{ortolanoQuantumEnhancedNoninterferometric2023}. 
{In TEMs, quantum correlation studies have been proposed\protect{~\cite{Konecna2022entanglement_electron-photon, kazakevich2024spatial}}, including quantum eraser experiments with electron-photon pairs\protect{~\cite{HenkeJengRopers2024}} and the transfer of optical coherence to emitted photons using ultrashort electron-pulse trains via photon-induced near-fields~\protect{\cite{Kfir2021}}. Within the last few months the first experimental evidence for electron-photon entanglement has been presented given certain assumptions on the state structure~\mbox{\cite{preimesbergerExperimentalVerificationElectronPhoton2025,henke_observation_2025}}. Here, we provide a state-agnostic protocol bounding the amount of entanglement and closing assumption-associated loopholes.
The ongoing advancements demonstrate the growing interest in quantum correlations in electron-photon pairs. Fundamentally, entanglement grants access to higher-resolution measurements in multiple parameters and to post-processing strategies that reveal structures beyond classically available information~\mbox{\cite{rotunno_one-dimensional_2023}}.} All together these properties could provide a significant breakthrough in high-resolution imaging of sensitive samples.

A promising approach to generate the pairs is provided by cathodoluminescence~(CL), a process during which electrons traversing a material sample lead to the emission of photons. 
Such processes are typically classified by the coherence of their optical fields with the field of the exciting electron~\mbox{\cite{2010deAbajo}}. Incoherent CL processes, such as inter- and intra-band transitions, colour centre de-excitations, exciton decay, etc., typically arise via intermediary excitations, such as bulk plasmons~\mbox{\cite{varkentinaCathodoluminescenceExcitationSpectroscopy2022, yanagimoto_unveiling_2025}}. 
In contrast, in coherent CL processes energy and momentum are conserved within the electron-photon pair, thus offering a compelling avenue for exploring quantum correlations. 

Motivated by the recent technological advances of TEMs in terms of time-correlated detection of electron-CL-photon pairs~\cite{yanagimotoTimecorrelatedElectronPhoton2023,feistCavitymediatedElectronphotonPairs2022,preimesbergerExploringSinglePhotonRecoil2025}, we propose a certification protocol for the entanglement of these pairs, drawing from techniques in quantum optics and quantum information~\cite{TascaRudnickiAspdenPadgettSoutoRibeiroWalborn2018, BavarescoEtAl2018, MorelliHuberTavakoli2023}. 
As an example, we present a detailed model of Cherenkov radiation, focusing on the interaction between an incident electron and a dielectric material. 
By taking into account energy and momentum conservation, we simulate the resulting electron-photon state and its entanglement.
State-of-the-art TEM devices can image electrons and CL-photons in parallel when modified accordingly. 
Hence, the verification of electron-photon entanglement becomes a matter of performing suitable measurements.

\begin{figure}[t]
    \centering
    \includegraphics[width=0.95\columnwidth]{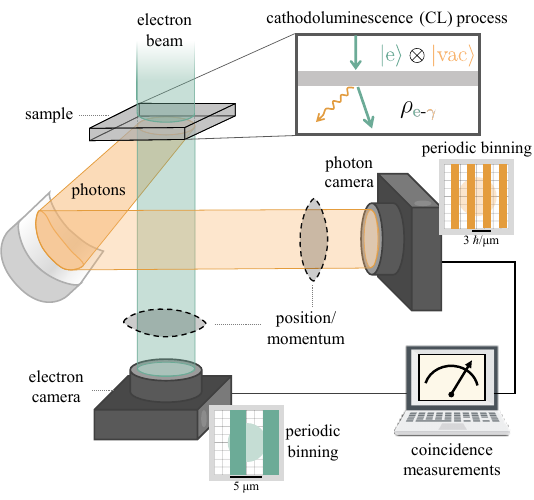}

    \caption{{Schematic overview of the measurement apparatus required for the proposed certification technique:} 
    A collimated electron beam shown in green hits a sample before progressing to a camera. 
    The inlay illustrates a generic coherent CL process, where electrons travelling through the sample produce photons, such that energy and momentum are conserved. 
    The emitted photons and transmitted electrons form electron-photon pairs. 
    The photons, shown in orange, are collected using a mirror and directed through a lens towards a photon detector. 
    Optional lens transformations convert between position and momentum measurements. 
    Using periodic binning, the particle pairs can then be measured in mutually unbiased bases. 
    Both electrons and photons are measured using a pixelated camera, where spatially periodic binning is applied in post processing. 
    Electron detections are then post-selected according to their temporal correlation with the photons. 
    Finally, the coincidences between electrons and photons are counted within the different bases and used to bound the entanglement of the pairs.}
    \label{fig:prop-exp}
\end{figure}

To provide an overview, we review a range of entanglement-detection methods~\cite{BertaChristandlColbeckRenesRenner2010,SpenglerHuberBrierleyAdaktylosHiesmayr2012,BavarescoEtAl2018,Herrera-ValenciaSrivastavPivoluskaHuberFriisMcCutcheonMalik2020} that rely on correlation measurements for pairs of complementary local observables corresponding to positions and momenta of electrons and photons. 
Taking into account subtleties with the selection of coarse-grained detection regions, we identify suitable bases.
To illustrate such a measurement protocol, we sketch an idealized, hypothetical  experimental implementation (see Fig.~\ref{fig:prop-exp}). 
We introduce prerequisites in terms of position and momentum resolution that any successful implementation of this entanglement witness requires and show that they are readily attainable when comparing to different experiments in previous literature. 
We include preliminary data from an experimental implementation in order to show that the required conditions are also attainable in a single experiment specific to this purpose. 

The strategy outlined in this proposal is aimed to be as general as possible{\textemdash}producing results that are independent of the exact form of the produced states, even in the data analysis{\textemdash}while providing quantitative information on the entanglement, rather than ``just'' confirming its presence. 
We distinguish between the tasks of entanglement detection, certification, and quantification, with detection referring purely to a dichotomic statement about the presence or absence of entanglement, and quantification referring to precise statements about the amount of entanglement as captured by a suitable entanglement measure carrying operational meaning. 
When only lower bounds on such measures can be provided, we speak of entanglement certification~\cite{FriisVitaglianoMalikHuber2019}. 
In addition, these three tasks may depend on assumptions about the underlying states and measurements. 
If no assumptions on either are necessary, one typically speaks of a device-independent setting.
Some entanglement-detection strategies rely on the assumption that the state is pure (see, e.g., Refs.~\cite{adiv_observation_2023, HenkeJengRopers2024} in the context of electron-photon entanglement) to identify entanglement, but extending such conclusions to arbitrary states is challenging~\cite{DAngeloKimKulikShih2004, HenkeJengRopers2024, ChaikovskaiaKarlovetsSerbo2024}. 
Here, we focus on entanglement certification without any underlying assumptions about the measured state, extending our methods to bound the amount of entanglement in the system. 
This allows us to benchmark the system. 
{The applied discretisation introduces a dimensionality which ensures a meaningful comparison with platforms of both continuous and discrete degrees of freedom.}
Our model provides concrete numbers that reproduce the effects and noise that influence the state during the measurement process. 

\vspace{-4mm}
\section{What is Entanglement?}\label{sec:what_is_entanglement} 

Entanglement is a form of correlation between two or more quantum systems that{\textemdash}in contrast to classical correlations{\textemdash}manifests in correlated measurement outcomes for, in principle, arbitrarily many different local measurements. 

\begin{tcolorbox}[breakable, enhanced,boxrule=0pt,title=Bell State]
    A paradigmatic example of an entangled state is the Bell state 
    \[
    \ket{\Phi^{+}}=\frac{\ket{0}\otimes\ket{0}+\ket{1}\otimes\ket{1}}{\sqrt{2}} = \tfrac{1}{\sqrt{2}}\left(\begin{smallmatrix}
            1\\0\\0\\1\end{smallmatrix}\right)
    \]
    of two parties. The parties are two-level quantum systems{\textemdash}qubits{\textemdash}describing, e.g., spin-$\tfrac{1}{2}$ particles or photon polarization, with orthogonal basis states $\ket{0}$ and $\ket{1}$\@. 
    Note that the ``$\otimes$" is often omitted.
    The Bell state is a pure state and its full density matrix is given by
    \[
    \hat{\rho} = \ket{\Phi^{+}}\bra{\Phi^+}
    = \tfrac{1}{2}\left(\begin{smallmatrix}
            1&0&0&1\\
            0&0&0&0\\
            0&0&0&0\\
            1&0&0&1
            \end{smallmatrix} \right).
    \]
    Since $\ket{\Phi^{+}}$ has the symmetry $\hat{U}\otimes \hat{U}^*\ket{\Phi^{+}}=\ket{\Phi^{+}}$ for all (local) unitaries $\hat{U}$, it is perfectly correlated in infinitely many pairs of bases. 
\end{tcolorbox}

The effect of decoherence due to an external environment can be described via mixing, i.e., the state becomes a statistical mixture of pure states. 
Any (pure or mixed) quantum state is described by a density operator $\hat{\rho}$. 
For pure states, the density operator is a (rank-one) projector, $\hat{\rho}^{2}=\hat{\rho}=\ket{\rho}\bra{\rho}$.  
In general, the diagonal entries of the density matrix with respect to any basis form a probability distribution. 
In particular, the eigenvalues of $\hat{\rho}$ are required to be non-negative ($\hat{\rho}$ is positive semi-definite) and sum to one ($\hat{\rho}$ has unit trace).

A bipartite system is described as a state in a Hilbert space $\mathcal{H}\subtiny{0}{0}{A}\otimes\mathcal{H}\subtiny{0}{0}{B}$, where $A$ and $B$ may represent a photon and an electron.
A state $\hat{\rho}$ is called \emph{separable} with respect to the bipartition $A|B$ if it is a mixture of product states, i.e., if there are density operators $\rho\suptiny{0}{0}{A}_{i}$ and $\rho\suptiny{0}{0}{B}_{i}$, and probabilities $p_i$ such that $\hat{\rho}=\sum_{i}p_{i}\,\rho\suptiny{0}{0}{A}_{i}\otimes\rho\suptiny{0}{0}{B}_{i}$\@. 
States that are not separable are called \emph{entangled}~\cite{Werner1989} and play a crucial role in enabling quantum information-processing protocols such as teleportation~\cite{BennettEtAl1993,Bouwmeester-etal1997} or entanglement-based imaging techniques~\cite{NdaganoDefienneLyonsStarshynovVillaTisaFaccio2020,DefienneNdaganoLyonsFaccio2021,CameronCourmeVernierePandyaFaccioDefienne2024}.

Despite the seemingly simple definition, conclusively determining whether or not a given mixed state $\hat{\rho}$ is separable is in general a difficult task, both mathematically and in practice (see, e.g., Refs.~\cite{Bruss2002,HorodeckiEntanglementReview2009,GuehneToth2009,FriisVitaglianoMalikHuber2019} for reviews). 
If the full description of the system is known, one can apply several criteria to check for entanglement, such as the so-called positive partial-transpose (PPT) criterion~\cite{Peres1996, HorodeckiMPR1996}. 
It states that, if the partial transpose on one of the subsystems leads to negative eigenvalues of the resulting matrix, the system is entangled. 
However, often we do not have the full state description at our disposal, as the number of measurements required for a complete state tomography{\textemdash}a full reconstruction of the state{\textemdash}of a multi-particle system scales exponentially with the number of particles. 
This exponential increase makes the procedure unfeasible even for systems of modest size.

In particular, the experimental verification of entanglement is often limited by the type and number of measurements that can be performed. 
A common restriction is that only local measurements with respect to a small set of bases of the individual Hilbert spaces can be carried out. 
Formally, this means that one has access to matrix elements of the form $\rho\suptiny{1}{0}{(i)}_{mn}:=\bra{m\suptiny{0}{0}{(i)}n\suptiny{0}{0}{(i)}}\hat{\rho}\ket{m\suptiny{0}{0}{(i)}n\suptiny{0}{0}{(i)}}$ where $i=1,2,\ldots,M$ labels the~$M$ different available local bases, while $m,n=0,1,\ldots,d-1$ label the elements of each basis in the $d$-dimensional Hilbert spaces $\mathcal{H}\subtiny{0}{0}{A}$ and~$\mathcal{H}\subtiny{0}{0}{B}$\@. 
\emph{Perfect correlation} with respect to any pair of bases labelled by~$i$ means that, for some labelling of the respective vectors $\ket{m\suptiny{0}{0}{(i)}}\subtiny{0}{0}{A}$ and $\ket{n\suptiny{0}{0}{(i)}}\subtiny{0}{0}{B}$, we have $\sum_{n}\rho\suptiny{1}{0}{(i)}_{nn}=1$, and these correlations are considered to be strongest if $\rho\suptiny{1}{0}{(i)}_{nn}=\tfrac{1}{d}$ for all~$n$, that is, when all~$d$ measurement outcomes for either party are equally likely, yet perfectly correlated.  
{If correlations in two or more different basis pairs are sufficiently strong, they can imply entanglement}~{\cite{LiHuberFriis2024}}.
\\

\parag{MUBs}For entangled states, these (perfect) correlations can persist for many bases, even when the respective pairs of bases labelled by $i$ and $j\neq i$ are \emph{mutually unbiased bases} (MUBs), i.e., when 
\begin{align}
    |\scpr{m\suptiny{0}{0}{(i)}}{n\suptiny{0}{0}{(j)}}|^{2}=\tfrac{1}{d}\ \ \ \text{for all $m$ and~$n$}.
    \label{eq:MUB}
\end{align} 
For separable states, in contrast, correlations cannot be arbitrarily perfect in such complementary bases. For~$M$ pairs of MUBs, separable states satisfy~\cite{SpenglerHuberBrierleyAdaktylosHiesmayr2012}
\begin{align}
    \sum\limits_{i=1}^{M}\sum\limits_{n=0}^{d-1}\rho\suptiny{1}{0}{(i)}_{nn}&\leq\,1\,+\,\tfrac{M-1}{d}.
    \label{eq:mubW}
\end{align}

The violation of this bound represents an entanglement-certification criterion which is independent of the purity of the state.
It can also be extended to certify high-dimensional entanglement via the so-called Schmidt number~\cite{MorelliHuberTavakoli2023}. Crucially, this method works even with measurements in only two product bases $\{\ket{m\suptiny{0}{0}{(i)}}\otimes\ket{n\suptiny{0}{0}{(i)}}\}_{m,n}$ and the knowledge that they are locally mutually unbiased, i.e., that Eq.~(\ref{eq:MUB}) holds. 
In some cases this may be guaranteed without having further information on, or control over the phases of~$\scpr{m\suptiny{0}{0}{(i)}}{n\suptiny{0}{0}{(j)}}$. Moreover, this technique can also be extended to measurements in arbitrary local bases~\cite{LiHuberFriis2024}.

\vspace*{1.2mm}
\begin{tcolorbox}[breakable, enhanced,boxrule=0pt,title=Correlations]
    Consider the following local bases for subsystems $A$ and $B$:
    \[
    \begin{split}
        \ket{0\suptiny{0}{0}{(1)}}_A&=\ket{0\suptiny{0}{0}{(1)}}_B=\ket{0},\\
        \ket{1\suptiny{0}{0}{(1)}}_A&=\ket{1\suptiny{0}{0}{(1)}}_B=\ket{1},\\
        \ket{0\suptiny{0}{0}{(2)}}_A&=\ket{0\suptiny{0}{0}{(2)}}_B=\ket{+}=\tfrac{1}{\sqrt{2}}(\ket{0}+\ket{1}),\\
        \ket{1\suptiny{0}{0}{(2)}}_A&=\ket{1\suptiny{0}{0}{(2)}}_B=\ket{-}=\tfrac{1}{\sqrt{2}}(\ket{0}-\ket{1}).
    \end{split}
    \]
    The Bell state $\ket{\Phi^+}$ is perfectly correlated in both bases, as all elements $\rho\suptiny{1}{0}{(i)}_{nn}=\bra{00}\hat{\rho}\ket{00}=\bra{11}\hat{\rho}\ket{11}=\bra{++}\hat{\rho}\ket{++}=\bra{--}\hat{\rho}\ket{--}=\tfrac{1}{2}$ are the same and equal to $\tfrac{1}{d}$, where $d=2$\@. 
    Note that in principle, the basis states of $A$ and $B$ carrying the same label (e.g., $\ket{0\suptiny{0}{0}{(1)}}_A$ and $\ket{0\suptiny{0}{0}{(1)}}_B$) do not have to have any relation~\cite{BertlmannFriis2023}. 
    Indeed, $A$ and $B$ could be completely different physical systems with different degrees of freedom. 
    As the bases above are mutually unbiased, i.e., $|\langle m\suptiny{0}{0}{(1)}|n\suptiny{0}{0}{(2)}\rangle|^2=\tfrac{1}{2}$, the correlations in these bases imply entanglement, because $\sum_n \rho\suptiny{1}{0}{(i)}_{nn}=2$, in violation of the condition $\sum_n \rho\suptiny{1}{0}{(i)}_{nn}\leq\tfrac{3}{2}$ satisfied by all separable states. 
\end{tcolorbox}
\vspace*{3.5mm}

\parag{Fidelity Criterion}When one has sufficient control and knowledge of the measurements to pick specific MUBs, the corresponding measurement outcomes for $M$ such basis pairs can be used to evaluate a tighter entanglement-detection bound based on a lower bound $\tilde{\mathcal{F}}$, where $\tilde{\mathcal{F}}(\hat{\rho},\Phi^{+})\leq \mathcal{F}(\hat{\rho},\Phi^{+})$ for the fidelity $\mathcal{F}(\hat{\rho},\Phi^{+}):=\bra{\Phi^{+}}\hat{\rho}\ket{\Phi^{+}}$ with the maximally entangled state $\ket{\Phi^{+}}=\frac{1}{\sqrt{d}}\sum_{n}\ket{n,n}$. 
For all separable states $\hat{\rho}$, this fidelity satisfies $\mathcal{F}(\hat{\rho},\Phi^{+})\leq\tfrac{1}{d}$, such that finding a value $\tilde{\mathcal{F}}(\hat{\rho},\Phi^{+})>\tfrac{1}{d}$, certifies entanglement~\cite{BavarescoEtAl2018, Herrera-ValenciaSrivastavPivoluskaHuberFriisMcCutcheonMalik2020}. 
For general local dimensions~$d$ the lower bound $\tilde{\mathcal{F}}(\hat{\rho},\Phi^{+})$ has an unwieldy expression (see, e.g.,~\cite[pp.~580{\textendash}592]{BertlmannFriis2023}), and requires the construction of MUBs with phases according to the prescription in Ref.~\cite{WoottersFields1989}. 

For two qubits ($d=2$) and only two pairs of local bases, where we write $\ket{n\suptiny{0}{0}{(1)}}\equiv\ket{n}$ for $n=0,1$ and $\ket{n\suptiny{0}{0}{(2)}}=\tfrac{1}{\sqrt{2}}\bigl(\ket{0}+(-1)^{n}e^{i\varphi}\ket{1}\bigr)$, the relative phase $\varphi$ can be absorbed into the definition of $\ket{1}$ and can thus be set to zero, $\varphi=0$, without loss of generality. The bound then takes the simple form
\begin{align} 
    &\tilde{\mathcal{F}}(\hat{\rho},\Phi^{+})\,:=\,
    \tfrac{1}{2}\bigl[\bra{00}\hat{\rho}\ket{00}+\bra{11}\hat{\rho}\ket{11}-1\bigr]  
    \label{eq:tilde F k0 d2}\\[1mm]
	&\ +
    \sum\limits_{n=0,1}\bra{n\suptiny{0}{0}{(2)}n\suptiny{0}{0}{(2)}}\hat{\rho}\ket{n\suptiny{0}{0}{(2)}n\suptiny{0}{0}{(2)}}
	- \sqrt{\bra{01}\hat{\rho}\ket{01}\bra{10}\hat{\rho}\ket{10}}\,, \nonumber 
\end{align}
with values of the right-hand side $\tilde{\mathcal{F}}(\hat{\rho},\Phi^{+})$ larger than~$\tfrac{1}{2}$ detecting entanglement. 

Moreover, using this bound one is not only able to certify but also to quantify entanglement~\cite{BavarescoEtAl2018}{\textemdash}providing a comparative benchmark across implementations and platforms: 
a value $\tilde{\mathcal{F}}(\hat{\rho},\Phi^{+})>\tfrac{k}{d}$ implies that $\hat{\rho}$ has a Schmidt number of (at least) $k+1$, while the entanglement of formation $E_{\rm{F}}(\hat{\rho})$ can be bounded by $E_{\rm{F}}(\hat{\rho}) \,\geq\,-\,\log\bigl(1\,-\,I^{2}(\hat{\rho})\bigr)$, where (for two qubits, $d=2$), the quantity $I(\hat{\rho})$ is given by
\begin{align}
    I(\hat{\rho}) &\geq\tfrac{1}{\sqrt{2}}
    \bigl(2
    \tilde{\mathcal{F}}(\hat{\rho},\Phi^{+})
    -\bra{00}\hat{\rho}\ket{00}-\bra{11}\hat{\rho}\ket{11}\nonumber\\
     &\ \ \ \ \ \ \ \ \  -2\sqrt{\bra{01}\hat{\rho}\ket{01}\bra{10}\hat{\rho}\ket{10}}
     \bigr).
     \label{eq:lower bound on I of rho 2}
\end{align}

However, not all setups allow for the measurements to be designed to be mutually unbiased, or for the specific phase relations to be controlled sufficiently well. 
But even in such a case, entanglement detection is possible in principle.  
For instance, knowing just the largest value $\max_{m,n}|\scpr{m\suptiny{0}{0}{(i)}}{n\suptiny{0}{0}{(j)}}|$ provides a way to construct an (albeit less sensitive) entanglement test based on entropic uncertainty relations~\cite{BertaChristandlColbeckRenesRenner2010,ColesBertaTomamichelWehner2017}, which provides a lower bound on an entanglement measure called the distillable entanglement~\cite{DevetakWinter2005}. 
Alternatively, a generalization of the bounds in Ineq.~\eqref{eq:mubW} and Ref.~\cite{MorelliHuberTavakoli2023} to arbitrary bases has recently been derived~\cite{LiHuberFriis2024}.

In the following we will focus on the (generally optimal) scenario where measurements can be performed in MUBs. 
In particular, we will describe how the two-qubit entanglement criterion in Ineq.~\eqref{eq:mubW} can be evaluated for continuous-variable (CV) systems, which we discuss in the next section.

\section{Infinite-Dimensional Systems} 
\label{sec:infinite_dimensional_systems}

So far, our description has primarily focused on correlations in finite-dimensional systems, i.e., systems with discrete spectra capturing degrees of freedom like, e.g., polarisation. 
However, the quantities to be measured in this proposal, namely position~$x$ and momentum~$p$, represent infinite-dimensional degrees of freedom with continuous spectra. 
Traditionally, there are a number of approaches to characterise entanglement in CV systems. 
Some of these make use of the fact that a two-party quantum state is entangled if its partial transpose is negative, or variations of this condition~\cite{Peres1996,HorodeckiMPR1996, Simon2000,ShchukinVogel2005}.
Another common approach is to employ uncertainty relations to detect CV entanglement~\cite{ReidDrummond1989, ReidDrummondBowen2009, CourmeVerniereSvihraGiganNomerotskiDefienne2023, preimesbergerExperimentalVerificationElectronPhoton2025}, {which tend to rely on an underlying Gaussian model to analyse the data~\mbox{\cite{SchneelochHowell2016,SchneelochHowland2018}}.
The resulting entanglement bounds are difficult to compare to their discrete counterparts.}

One can also exploit the fact that the operators whose spectra are continuous come in complementary pairs like $\hat{x}$ and $\hat{p}$ with commutation relations of the form $[\hat{x},\hat{p}]=i\hbar$, which allows us to formulate a set of CV MUBs~\cite{WeigertWilkinson2008} to certify the entanglement of electron-photon pairs in the TEM\@. 
In principle, these conjugate variables can be used to design entanglement-witnessing conditions in the form of Ineq.~\eqref{eq:mubW}. 
Considering that measurement devices register outcomes with finite precision and within a limited range of values, experimental assessments of CV observables may diverge from theoretical predictions. 
To accurately describe the system's correlations, proper coarse-graining of these measurements is necessary, effectively discretising the CVs. 
Although challenging, this can be achieved through the careful design of measurement bases~\cite{TascaSanchezPieroWalbornRudnicki2018, Herrera-ValenciaSrivastavPivoluskaHuberFriisMcCutcheonMalik2020}, in particular MUBs.
Once we have properly discretised the measured CV basis, we can proceed beyond entanglement detection and move towards certification of high-dimensional entanglement. 
Here, one requires more robust protocols, like those developed using single-outcome measurements~\cite{BavarescoEtAl2018} for discrete scenarios.

\begin{figure}[t]
    \centering
    \includegraphics[width=0.9\columnwidth]{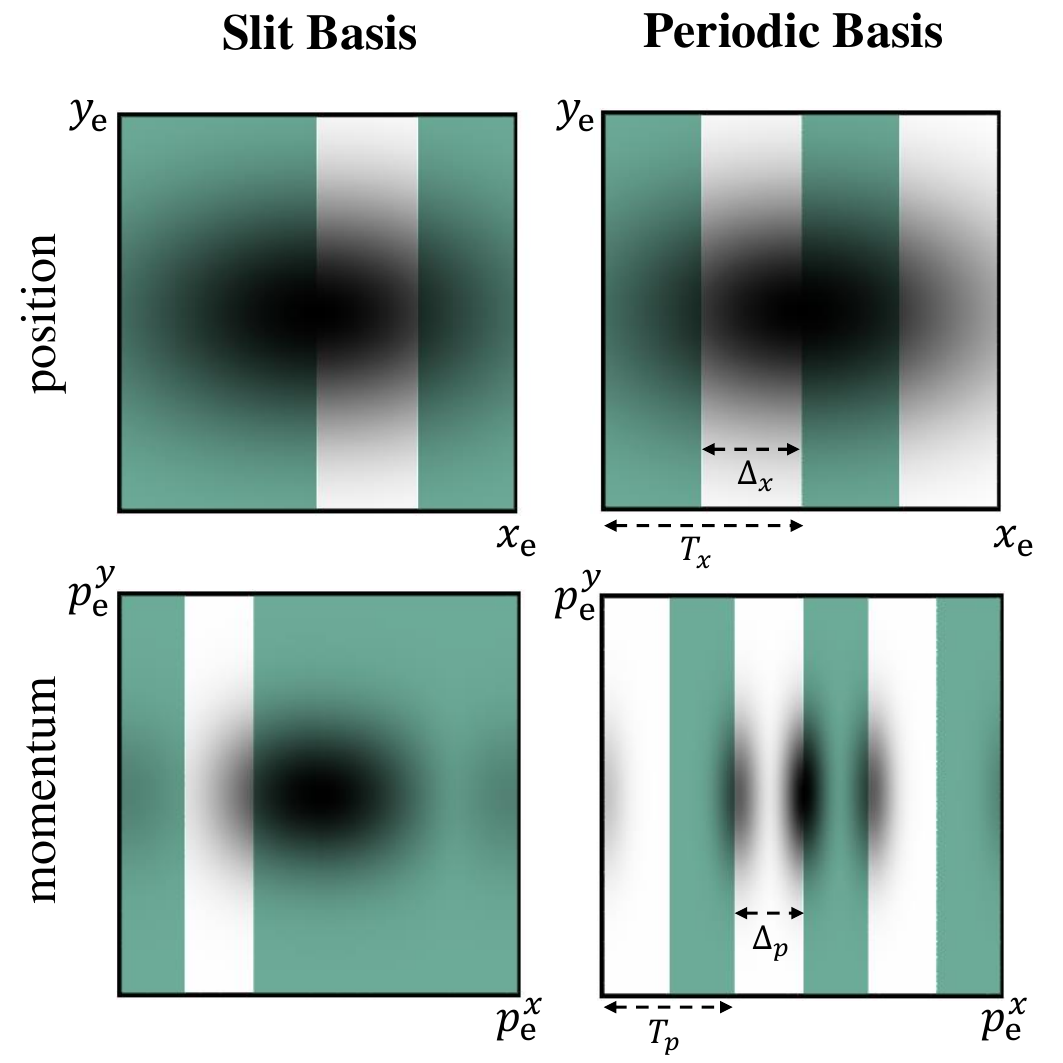}
    \caption{
    Illustration of the unbiasedness of periodic basis elements in contrast to slits, considering position and momentum of an electron. 
    Note that the elements effectively project the wave function onto the transparent areas.
    The top row shows the intensity of an exemplary wave function in position plotted over the basis element corresponding to a slit basis (left) and periodic basis (right). 
    Subsequently, the wave function is projected onto that element. 
    The bottom row shows the intensity of the momentum representation of the post-projection wave function plotted over a given momentum-basis element.
    The periodic basis can be seen as giving the wave function a ``kick'' in momentum space, which is exactly $\Delta_p$ of the complementary basis, thereby just large enough to switch it from one basis element to the other.
    An unbiased set leads to an equal distribution between all elements in momentum.
    However, the slit-basis example clearly shows that certain elements have a higher probability to be hit than others, which corresponds to entanglement-independent correlations between the different bases.
    The periodic basis, on the other hand, produces unbiased elements.
    }
    \label{fig:slit_masks}
\end{figure}

In the following analysis, we consider a coarse-graining model that involves periodic binning~\cite{TascaSanchezPieroWalbornRudnicki2018}. 
If the projections have a finite width, information is preserved in the conjugate counterpart, making arbitrary binning approaches biased and, hence, inadequate for our purposes. 
This is exemplified on the left-hand side of Fig.~\ref{fig:slit_masks}. 
However, by using periodic binning with a carefully chosen relationship between the position and momentum periodicities, one can construct an unbiased set of bases.
The piecewise subsets $\mathcal{R}_n$ which describe each of the~$d$ basis elements, are defined over the whole set of real numbers~$\mathbb{R}$. 
For position measurements, they are defined as $\mathcal{R}_n:=\left\{ x \in \mathbb{R} \, | \,  x_{\rm cen}+ n\Delta_x \leqslant x ({\rm mod} \, T_x) <  \ x_{\rm cen} + (n+1)\Delta_x \right\}$, with $n=0, \cdots, d-1$ and $x_\text{cen}$ representing the center of the periodic pattern. 
The parameter $\Delta_x$ acts as a bin width and the index $n$ is adapted in the following for $d=2$ and to carry an identifier for the variable under consideration. 
As such, $n=\{0\suptiny{0}{0}{(x)},1\suptiny{0}{0}{(x)}\}$ labels the detection outcomes, akin to the typical method used for discrete-variable quantum systems.
The parameter $T_x$ represents the periodicity of the discretisation as illustrated in Fig.~{\ref{fig:slit_masks}}, and $d=T_x/\Delta_x=2$ is the number of detection outcomes. 
A notable advantage of this method is that the number of detection outcomes can be fully adjusted by selecting appropriate parameters \(T_x\) and \(\Delta_x\), independent of the detection range, i.e., the area which is accessible for the detector. 
{This method relies on the projections to form a complete set, so that every considered state has a non-zero overlap with at least one projector and is hence detectable.}
The projection operators for the discretised position are then written as
\begin{equation}\label{eq:projection_op_discretized_position}
\hat{\Pi}\suptiny{0}{0}{(x)}_n 
= \int_{\mathcal{R}_n}  |x\rangle \langle x| dx 
= \sum_{u \in \mathbb{Z}}  \int_{x_{\rm cen} + n\Delta_x + uT_x}^{x_{\rm cen} + (n+1)\Delta_x + uT_x}  |x\rangle \langle x| dx.
\end{equation}
Analogously, the operators over the conjugate variable $p$ are given by
\begin{equation}\label{eq:projection_op_discretized_momentum}
\hat{\Pi}\suptiny{0}{0}{(p)}_m 
= \int_{\mathcal{R}_m}  |p\rangle \langle p| dp 
= \sum_{u \in \mathbb{Z}}  \int_{p_{\rm cen} + m\Delta_p + uT_p}^{p_{\rm cen} + (m+1)\Delta_p + uT_p}  |p\rangle \langle p| dp,
\end{equation}
where the bin width and periodicity of $p$ are $\Delta_p$ and $T_p$, respectively, and the basis-element index is given by $m=\{0\suptiny{0}{0}{(p)},1\suptiny{0}{0}{(p)}\}$.

The requirement for mutual unbiasedness is that the outcome probabilities for one basis are uniformly distributed when the quantum state is localised with respect to the other basis (and vice versa).
In our case, we have sets of projectors $\{ \hat{\Pi}\suptiny{0}{0}{(x)}_n \}$ and $\{ \hat{\Pi}\suptiny{0}{0}{(p)}_m \}$ that define coarse-graining measurements in the complementary variables~$x$ and~$p$ of a CV quantum system described by $\hat{\rho}$.
The measurement designs yield~$d$ outcomes in each domain, with probabilities given by $\textrm{P}\suptiny{0}{0}{(x)}_n (\hat{\rho}) = {\rm Tr}  (\hat{\rho}\hat{\Pi}\suptiny{0}{0}{(x)}_n)$ and $\textrm{P}\suptiny{0}{0}{(p)}_m (\hat{\rho}) ={\rm Tr}  (\hat{\rho}\hat{\Pi}\suptiny{0}{0}{(p)}_m)$.
Finally, the unbiasedness condition reads~\cite{TascaSanchezPieroWalbornRudnicki2018}:
\begin{align}
\text{if }\textrm{P}\suptiny{0}{0}{(x)}_n (\hat{\rho}) =\delta_{n_0n} \hspace{0.1in} &\Rightarrow   \hspace{0.1in}  \textrm{P}\suptiny{0}{0}{(p)}_m (\hat{\rho}) = d^{-1},\\
\text{if }\textrm{P}\suptiny{0}{0}{(p)}_m (\hat{\rho}) =\delta_{m_0m}   \hspace{0.1in}  &\Rightarrow   \hspace{0.1in}   \textrm{P}\suptiny{0}{0}{(x)}_n (\hat{\rho}) = d^{-1}.
\end{align}
We impose the following relation between the periodicities $T_x$ and $T_p$:     $\frac{T_xT_p}{2\pi \hbar} = \frac{d}{u}$, where $u \in \mathbb{N}$ such that $\forall\; v\in\{1, \cdots, d-1\}$ $\frac{uv}{d} \notin \mathbb{N}$.
Please note that this condition is independent of $x_\text{cen}$ and $p_\text{cen}$, making it insensitive to transversal alignment errors.
The simplest and most significant case is the condition with $u = 1$, as it is valid for all~$d$. 
This condition offers the optimal balance between experimentally accessible periodicities, that is, 
\begin{equation}
    T_x T_p = d\cdot 2\pi \hbar.
    \label{eq:periodicities}
\end{equation}

Using these relationships, we can now define MUBs for position and momentum that are measurable via masks~\cite{TascaSanchezPieroWalbornRudnicki2018}.
Coincidence measurements between the electron and photon provide access to the joint probabilities in position, $\mathcal{P}(n_\text{e},n_\gamma)$, and momentum, $\mathcal{P}(m_\text{e},m_\gamma)$, where e and $\gamma$ stand for the electron and the photon system (explicit expressions in \mbox{Sec.~\ref{sec:experimental_setup}} and~\mbox{\ref{sec:epr_electron_photon_certification_entanglement}}). Their joint measurements allow the certification of entanglement between the electron and photon. 

However, until recently, entanglement between such pairs, or between a free electron and sample excitations was considered to be an inaccessible resource due to the fast decoherence in electron microscopy~\cite{Schattschneider2018Entanglement}. 
Consequently, making use of the pair's quantum features seemed to be beyond current technology. 
Recent proposals have opened up new ways to explore the quantumness of free electrons: entangling individual electrons through sequential interactions with guided Cherenkov photons~\cite{Kfir2019entanglement_electron-electron}, creating free-electron qubits~\cite{Reinhardt2021entanglement_electron-photon}, and verifying electron-photon entanglement via cavity excitations~\cite{Konecna2022entanglement_electron-photon}. 
Recently, both an imaging-based scheme~\mbox{\cite{preimesbergerExperimentalVerificationElectronPhoton2025}} and a protocol based on the quantum eraser~\mbox{\cite{henke_observation_2025}} have been implemented, showing the feasibility of the described approach. 
While each previous study illuminates a different angle of electron-photon quantum correlations, they make assumptions on the underlying quantum systems.
Next, we will describe how state-agnostic entanglement certification can be implemented in a way that does not rely on specific properties of the involved wave functions.

\section{Experimental Setup for MUB Measurements on Electron-Photon Pairs}\label{sec:experimental_setup}

In this section, we will first illustrate the proposed technique by discussing a basic implementation in a standard TEM adapted to allow optical access to the specimen {and equipped with a time-resolved electron detector}. 
The experimental viability of any such implementation is mostly determined by the position and momentum accuracy for measurements of the two particles.
Drawing from experimental data, we will provide realistic resolution limits. In Section~\ref{sec:electron_cherenkov_photon_pair}
we determine whether these values are sufficient for entanglement certification for the exemplary case of electron-Cherenkov-photon pairs.

\begin{figure}[b]
    \includegraphics[width=\linewidth]{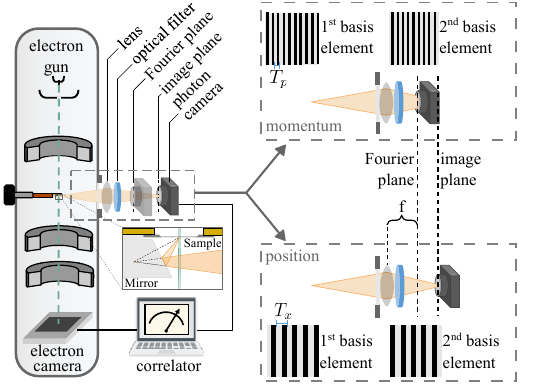}
    \caption{Simplified experimental implementation of the proposed entanglement witness. Using a mirror and a lens, CL photons  are collected. An optical filter restricts the accepted photon energy. The desired measurement basis is selected by placing the pixelated detector, either at the image plane (position) or the Fourier plane (momentum) of the lens. The appropriate periodic coarse-graining is applied via binning on the recorded data.}
    \label{fig:setup}
\end{figure}
We consider the setup depicted in Fig.~\ref{fig:setup}: The sample is illuminated with a collimated beam over a wide area in order to closely approximate an incident plane wave. 
Upon interaction with the sample, an electron may produce a CL photon which is captured by the photon-collection system. 
In the basic setup, this system comprises a mirror, a single lens, an optical bandpass filter and a time-resolved, single-photon camera. 
Equivalently, the camera and binning could be replaced with a tailored photon mask, along with a single-photon bucket detector.

We are interested in the electron-photon wave function at the sample plane, which is projected to the image plane of the lens.
Spatially resolving the photon in this plane corresponds to a measurement of its initial state in position space.
Photons are measured in momentum space by sending them through a narrow-bandpass wavelength filter and detecting them in the Fourier plane.

The same two options are available for the transmitted electron: 
In imaging mode the magnetic lenses inside the TEM are set to produce an image of the transmitted wave function at the detector plane, resulting in a position measurement of the electron. 
When recording a diffraction pattern, the back focal plane of the objective lens is projected onto the detector, thus implementing a momentum measurement of the electron. 
The recoil due to the emission of a single photon is on the micro- to nanoradian scale~\mbox{\cite{preimesbergerExploringSinglePhotonRecoil2025}}, requiring very high momentum resolution compared to typical electron-diffraction applications. 
The resulting distribution is closely correlated to the angular distribution of the photon.

Practical experimental implementations will typically measure deflection angles that need to be interpreted in momentum space. 
In Appendix~\ref{ap:deflection_vs_momenta}, the relation between these angles and momentum conservation is clarified. 
One consequence is the need to consider as narrow an energy window as possible to establish a one-to-one mapping between deflection angle and momentum, thereby maintaining closer adherence to a perfect momentum measurement. 
The energy range of the coincident electrons is determined by the wavelength filtering of the photons due to energy conservation in the coherent CL processes. 
By detecting both electrons and photons in a time-resolved manner, we can identify electron-photon pairs~\cite{preimesbergerExploringSinglePhotonRecoil2025, feistCavitymediatedElectronphotonPairs2022, varkentinaCathodoluminescenceExcitationSpectroscopy2022} and test their entanglement.  

As described in Sec.~\ref{sec:infinite_dimensional_systems}, any measurement of continuous variables, such as position and momentum, requires coarse-graining in a practical setup. 
In one spatial dimension (here the $x$-direction), this is achieved by assigning one of two measurement outcomes depending on where the particle hits the detector.
In the following, we will consider a coarse-grained measurement in the position-position basis as an example. 
To start, one can equivalently consider either the photon or the electron.
The entire position range is divided into two complementary regions $\mathcal{R}_{n_\zeta}$, where $n_\zeta\in\{0\suptiny{0}{0}{(x)},1\suptiny{0}{0}{(x)}\}$ and $\zeta\in\{\mathrm{e},\gamma\}$ indicates the subsystem. 
Taking the magnification factor $A\suptiny{0}{0}{(x)}_\zeta$ into account, the one-to-one mapping of an idealised lens between the original sample position $x,\,y$ and the image position $x',\,y'$ on the detector is given by 
$x=x'/A\suptiny{0}{0}{(x)}_\zeta$, and equivalently $y=y'/A\suptiny{0}{0}{(x)}_\zeta$. 

Translating the regions which define the basis elements into areas on the detector, we thus get a position-dependent measurement outcome $n_\zeta(x')$:
\begin{align}
    n_\zeta(x') &= \begin{cases} 
      0 & x'/A\suptiny{0}{0}{(x)}_\zeta\in \mathcal{R}_{0}\,,\\[2pt]
      1 & x'/A\suptiny{0}{0}{(x)}_\zeta\in \mathcal{R}_{1}\,.\\
    \end{cases}
   \label{eq:pos_mask}
\end{align}
On a pixelated detector, this coarse-graining can be represented as a binning which assigns one of the basis elements to each pixel. 
Compatible coarse-grainings indexed by $n_\gamma \in \{0\suptiny{0}{0}{(x)},1\suptiny{0}{0}{(x)}\}$ and $n_\mathrm{e} \in \{0\suptiny{0}{0}{(x)},1\suptiny{0}{0}{(x)}\}$ are defined for both the electron and the photon. 
By measuring the coincidence count rates $C_{ n_\mathrm{e}n_\gamma}$ for equal and opposite indices, one can now calculate the joint probabilities $\mathcal{P}({n_\mathrm{e}, n_\gamma})$ for the experimental wave function and projectors.
\begin{equation}
    \mathcal{P}({n_\mathrm{e}, n_\gamma})=\frac{C_{n_\mathrm{e}n_\gamma}}{\sum_{n_\mathrm{e},n_\gamma}C_{ n_\mathrm{e}n_\gamma}}.
\end{equation}
Entangled CL pairs ideally produce correlated outcomes with $\mathcal{P}(0\suptiny{0}{0}{(x)},0\suptiny{0}{0}{(x)})$ and $\mathcal{P}(1\suptiny{0}{0}{(x)},1\suptiny{0}{0}{(x)})$ close to $0.5$. More details on the expected outcomes and how they might be affected by detector resolutions will be given in the following sections.

In momentum space, the appropriate coarse-graining has to be mutually unbiased with respect to the position basis. 
This condition is met by enforcing the relationship from Eq.~\eqref{eq:periodicities}. 
We refer to the resulting complementary regions in momentum space as $\mathcal{R}_{m_\zeta}$ where $m_\zeta\in\{0\suptiny{0}{0}{(p)},1\suptiny{0}{0}{(p)}\}$ and $\zeta\in\{\mathrm{e},\gamma\}$ again identifies the subsystem. 
Placing the detector in the Fourier plane maps the momentum of either particle, $p=p_x$, for a given photon energy onto the position on the detector, $x',\,y'$, such that 
\begin{equation}
    p(x',y')=|\vec{p}|\frac{x'/A\suptiny{0}{0}{(p)}_\zeta}{\sqrt{(x'/A\suptiny{0}{0}{(p)}_\zeta)^2+(y'/A\suptiny{0}{0}{(p)}_\zeta)^2+1}},
\end{equation}
where $A\suptiny{0}{0}{(p)}_\zeta$ is the momentum-magnification factor. 
{Similar to its position equivalent, it represents the effective distance from the sample to the plane of the observed diffraction pattern (known as the camera length in electron microscopy).  }
This relationship can be used to assign the appropriate measurement outcome to each pixel, similar to Eq.~\eqref{eq:pos_mask}.
Joint measurements in position and momentum provide coincidence count rates, and therefore joint probabilities, for all 16 combinations of settings $\mathcal{P}({i\suptiny{0}{0}{(\kappa_\mathrm{e})}, j\suptiny{0}{0}{(\kappa_\gamma)}})$. 
This allows us to both evaluate the entanglement measures described in Sec.~\ref{sec:what_is_entanglement} and verify that the measured probabilities are compatible with the mutual unbiasedness of the measurement basis.  

\begin{table}[h]
    \centering
    \begin{tabular}{lcccc}
         \multicolumn{1}{p{1.8cm}}{\centering Measurement} & Symbol & \multicolumn{1}{p{2cm}}{\centering Experimental \\ Value} & \multicolumn{1}{p{1.8cm}}{\centering Literature \\ Value} & Ref. \\
        \hline
         $\mathrm{e}$ position & $\delta\suptiny{0}{0}{(x)}_\mathrm{e}$ & $<\;0.1\,$\textmu m & $<\;0.1\,$\textmu m &~\cite{KisielowskietAl2008}\\
         $\mathrm{e}$ momentum & $\delta\suptiny{0}{0}{(p)}_\mathrm{e}$ & $<\;0.2\frac{\hbar}{\text{\textmu} \text{m}}$ & 
         &\\
         $\gamma$ position   & $\delta\suptiny{0}{0}{(x)}_\gamma$ & $<\;1.2\,$\textmu m & $<\;1\,$\textmu m &~\cite{MatsukataOguraGarciadeAbajo2022}\\
         $\gamma$ momentum   & $\delta\suptiny{0}{0}{(p)}_\gamma$ & $<\;0.2\frac{\hbar}{\text{\textmu} \text{m}}$ & $<\;0.1\frac{\hbar}{\text{\textmu} \text{m}}$
         &~\cite{Yamamoto2016_Highres_CL_STEM}\\
    \end{tabular}
    \caption{Full width at half maximum (FWHM) of the point spread function for position and momentum measurements. We assume these realistically achievable values to evaluate the feasibility of the proposed technique for entanglement certification.}
         \label{tab:FWHMs}
\end{table}

Once a sufficient electron-photon coincidence count rate has been achieved, the biggest challenge to successfully implement the presented protocol is the limited resolution of the measurement for each particle in each basis. 
{The effect of finite resolution on the measurements can be modelled by convolving a point spread function~(PSF)  with the probability density.
Each measurement's PSF can be modelled as a Gaussian distribution $\text{PSF}(\kappa_\zeta;\delta\suptiny{0}{0}{(\kappa)}_\zeta)$ over $\kappa_\zeta$ characterised by its full width at half maximum (FWHM), $\delta\suptiny{0}{0}{(\kappa)}_\zeta$, and mean value at zero. 
As before, $\kappa\in\{x,p\}$ and $\zeta\in\{\mathrm{e},\gamma\}$. }
The PSF allows us to jointly represent signal reduction due to low resolution (mislabelling of measurement results) as well as decoherence (broadening of the correlations). However, we do not include material-specific decoherence effects in our analysis. Further details can be found in Appendix~\mbox{\ref{ap:assumptions}}.

Table~\mbox{\ref{tab:FWHMs}} summarises FWHM values from the literature, which have previously been achieved in TEMs or CL measurement systems installed in TEMs. 
Additionally, it gives preliminary values from an experimental setup currently being developed by some of the authors at USTEM, TU Wien (see Appendix ~\mbox{\ref{ap:experimenta_resolution_values}} for more details). 
This apparatus has recently been used to show electron-photon entanglement using a different  criterion~\mbox{\cite{preimesbergerExperimentalVerificationElectronPhoton2025}} based on uncertainty relations. Those measurements already fulfil most of the experimental requirements needed for the presented protocol. The setup implements free-space, coincidence-matched measurements of electron-photon pairs for each of the required joint bases, with a coincidence count rate of approximately 10~counts per second. In order to go beyond the results of Ref.~\mbox{\cite{preimesbergerExperimentalVerificationElectronPhoton2025}} and realise this fully state-agnostic approach, the remaining challenge consists of implementing the binning for the MUBs and further optimizing the resolutions.

\section{Detecting entanglement in CL Electron-Photon Pairs with MUBs}\label{sec:epr_electron_photon_certification_entanglement}

Einstein-Podolsky-Rosen (EPR) type entanglement is one of the first quantum correlations described in the literature~\cite{EinsteinPodolskyRosen1935}. The corresponding states exhibit correlations in one CV and anticorrelations in its conjugate. 
In the context of electron--photon pairs, such entanglement manifests naturally in position and momentum.

A prominent physical setting for this phenomenon is coherent CL, where a photon is emitted as an electron traverses or passes near a dielectric material. 
The electron's velocity $\vec{v}$ defines the longitudinal direction (typically aligned with the $z$-axis), while the interaction with the electromagnetic field often exhibits translational invariance in the transverse $(x, y)$ plane. 
This symmetry imposes conservation of transverse momentum, leading to strong correlations and, thus, entanglement between the electron and photon in these degrees of freedom.

In momentum space, the electron state is represented as $\ket{\vec{p}, s}_{\mathrm{e}}$, where $\vec{p} = (p_x, p_y, p_z)$ denotes the momentum and $s$ the spin. 
Similarly, the photon state is given by $\ket{\vec{k}, \epsilon}_\gamma$, with $\vec{k} = (k_x, k_y, k_z)$ the wave vector and $\epsilon$ the polarization. 
The entangled electron-photon pair state---under transversal momentum conservation---is given by the expression
\begin{align}\label{eq:final_state_electron-photon}
    \ket{\psi_{\mathrm{e}\mhyphen\gamma}} = 
    & \sum_{\epsilon, s}\int d^2k_\perp \int dk_z\, dp_z\, \Psi(\vec{k}_\perp, k_z, p_z, \epsilon, s) \\
    &\quad \cdot \ket{-\hbar k_x, -\hbar k_y, p_z, s}_\text{e} \otimes \ket{k_x, k_y, k_z, \epsilon}_\gamma, \nonumber
\end{align}
where $\hbar \vec{k}_\perp=\hbar(k_x, k_y)=-(p_x, p_y)$, and $\Psi(\vec{k}_\perp, k_z, p_z, \epsilon, s)$ are the respective probability amplitudes.

When coherence is preserved during the emission process, initially defined classically via a stable phase relationship between the electron and the emitted photon~\cite{2010deAbajo}, quantum coherence can also emerge. 
In this regime, pure states of the electron and the electromagnetic field evolve into a pure entangled state without involving the sample or environment. 
Conservation of energy and momentum then provides a sufficient condition for generating entanglement detectable in the position and momentum representations.

As exemplified later by the discussion of the Cherenkov effect in Sec.~\ref{sec:electron_cherenkov_photon_pair}, energy and \textit{partial} momentum conservation is also sufficient under the right experimental conditions.
Suitable CL phenomena for certifying CV entanglement in position and momentum include transition radiation~\cite{StogerPollach2016}, the Smith-Purcell effect~\cite{SmithPurcell1953}, radiative plasmon decay~\cite{TengStern1967}, and Cherenkov radiation~\cite{cherenkov_1937}.
Contributions from incoherent CL can be minimised by choosing a sample which naturally has low incoherent emission rates.
As incoherent emission often results from the excitation of more energetic intermediary particles (e.g. bulk plasmons~\mbox{\cite{varkentinaCathodoluminescenceExcitationSpectroscopy2022, yanagimoto_unveiling_2025}}), electron-energy filtering allows us to remove many of the unwanted incoherent contributions.
Additionally, incoherent processes with long lifetimes can be temporally excluded by coincidence matching~\mbox{\cite{Scheucher2022}}.

While this work focuses on low-energy photon emission in dielectric media, high-energy scattering processes are also expected to produce electron-photon entanglement, as shown by recent theoretical investigations of photoelectrons and X-ray photons~\mbox{\cite{Ahrens2017, Tanaka2024, NagChowdhury2021}}. 
These developments highlight the broader applicability of our certification approach across energy regimes.

In the following, we discuss entanglement certification using the MUB approach outlined in Sec.~\ref{sec:infinite_dimensional_systems} for the EPR-type electron-photon pair state in Eq.~\eqref{eq:final_state_electron-photon}. 
Our considerations are restricted to entanglement in position and momentum along an arbitrary transverse direction denoted as $x$.
To construct the reduced state $\hat{\rho}_{\mathrm{e}\mhyphen\gamma, x}$ depending only on $x$, we trace $\hat{\rho}_{\mathrm{e}\mhyphen\gamma}=\ket{{\psi}_{\mathrm{e}\mhyphen\gamma}}\bra{{\psi}_{\mathrm{e}\mhyphen\gamma}}$ over the variables that are not measured, namely the $y$- and $z$-momenta of both particles, the electron spin and photon polarization. 
The resultant reduced state is then given by
\begin{equation}\label{eq:red_Density_x}\allowdisplaybreaks
    \begin{aligned}
    \hat{\rho}_{\mathrm{e}\mhyphen\gamma, x}=\int dk_{1, x}&\, dk_{2, x} \frac{\hbar}{L_\perp}\tilde{f}(k_{1, x}, k_{2, x}) \\
&\ket{-\hbar k_{1, x}}_\text{e}\ket{k_{1, x}}_\gamma\bra{-\hbar k_{2, x}}_\text{e}\bra{k_{2,x}}_\gamma,
    \end{aligned}
\end{equation}
where $\tilde{f}(k_{1, x}, k_{2, x})$ are the matrix elements associated with plane-wave functions confined within a transverse length $L_\perp$ (see Appendix~\ref{ap:correlated_measurements_EPR} for details).
These elements depend only on the photon wave numbers in the $x$-direction ($k_{j,x}$), as the transverse electron momentum is fixed by momentum conservation. Here, $j = 1$ refers to the ket states, while $j = 2$ refers to the corresponding bra states.

As discussed in Sec.~\ref{sec:infinite_dimensional_systems}, entanglement certification based on CV MUBs requires coarse-grained correlation measurements. 
Hence, the available position and momentum ranges are split into two regions, i.e., basis elements, each. 
They are defined by local projection operators $\hat{\Pi}_{i\suptiny{0}{0}{(\kappa)}}\suptiny{0}{0}{(\kappa)}$, where $i\in \{0, 1\}$ represents the element index and $\kappa,\kappa'\in\{x,p\}$ position or momentum for a given subsystem. 
In the following paragraph we use superscripts on the indices, otherwise denoted as $n_\zeta$ and $m_\zeta$, to clearly distinguish between the bases.
Please note that $p=p_x$ and $p_\gamma=\hbar k_\gamma$.
Therefore, the joint probability $\mathcal{P}(i\suptiny{0}{0}{(\kappa_\mathrm{e})},j\suptiny{0}{0}{(\kappa_\gamma)})$ for the electron to be detected in element $i\suptiny{0}{0}{(\kappa_\mathrm{e})}\in\{n_\mathrm{e}\suptiny{0}{0}{(x_\mathrm{e})},m_\mathrm{e}\suptiny{0}{0}{(p_\mathrm{e})}\}$ and for the photon to be detected in element $j\suptiny{0}{0}{(\kappa_\gamma)}\in\{n_\gamma\suptiny{0}{0}{(x_\gamma)},m_\gamma\suptiny{0}{0}{(p_\gamma)}\}$ is given by 
\begin{equation}
    \begin{split}
        \mathcal{P}({i\suptiny{0}{0}{(\kappa_\mathrm{e})}, j\suptiny{0}{0}{(\kappa'_\gamma)}})=&{\rm Tr} \left(\hat{\rho}_{\mathrm{e}\mhyphen\gamma, x}\cdot\hat{\Pi}_{i\suptiny{0}{0}{(\kappa_\mathrm{e})}}\suptiny{0}{0}{(\kappa_\mathrm{e})}\otimes \hat{\Pi}_{j\suptiny{0}{0}{(\kappa'_\gamma)}}\suptiny{0}{0}{(\kappa'_\gamma)}\right)\\
        =&\int_{\mathcal{R}_{i\suptiny{0}{0}{(\kappa_\mathrm{e})}}}\int_{\mathcal{R}_{j\suptiny{0}{0}{(\kappa'_\gamma)}}} \tilde{\Pi}\left(\kappa_\mathrm{e}, \kappa'_\gamma\right) d\kappa_\mathrm{e} d\kappa'_\gamma,  
    \end{split}
    \label{eq:probs_basis_els}
\end{equation}
where $\tilde{\Pi}\left(\kappa_\mathrm{e}, \kappa'_\gamma\right)$ are the probability densities corresponding to the projective measurements, as illustrated in  
Fig.~\ref{fig:graphs_certification}. 
Here, the projections associate every part of the probability density with a joint basis element, resulting in a joint probability.
The correlated measurement outcomes correspond to the green ($i\suptiny{0}{0}{(\kappa_\mathrm{e})}=j\suptiny{0}{0}{(\kappa_\gamma)}=0\suptiny{0}{0}{(\kappa)}$) and yellow ($i\suptiny{0}{0}{(\kappa_\mathrm{e})}=j\suptiny{0}{0}{(\kappa_\gamma)}=1\suptiny{0}{0}{(\kappa)}$) areas in Fig.~\ref{fig:graphs_certification}, while the anti-correlated outcomes are indicated in beige ($i\suptiny{0}{0}{(\kappa_\mathrm{e})}=0\suptiny{0}{0}{(\kappa)}$, $j\suptiny{0}{0}{(\kappa_\gamma)}=1\suptiny{0}{0}{(\kappa)}$) and white ($i\suptiny{0}{0}{(\kappa_\mathrm{e})}=1\suptiny{0}{0}{(\kappa)}$, $j\suptiny{0}{0}{(\kappa_\gamma)}=0\suptiny{0}{0}{(\kappa)}$). 

The projective measurements corresponding to the experimental settings we consider result in probability densities
\begin{subequations}\label{eq:measurements}\allowdisplaybreaks
    \begin{align}
        \tilde{\Pi}(p_\mathrm{e}, \hbar k_\gamma)&=\,\frac{1}{N_{p-p}}\tilde{f}(k_\gamma, k_\gamma) \delta(\hbar k_\gamma+p_{\mathrm{e}}),\label{eq:mom_mom_measurement}\\[1mm]
        \tilde{\Pi}(x_\mathrm{e}, \hbar k_\gamma)&=\,\frac{1}{2x_{\text{max}}\hbar N_{p-p}}\tilde{f}(k_\gamma, k_\gamma),\label{eq:pos_mom_measurement}\\[1mm]
        \tilde{\Pi}(p_\mathrm{e},\; x_\gamma\;)&=\,\frac{1}{2x_{\text{max}}\hbar N_{p-p}}\tilde{f}(p_{\mathrm{e}}/\hbar, p_{\mathrm{e}}/\hbar),\label{eq:mom_pos_measurement}\\
            \tilde{\Pi}(x_\mathrm{e}, \; x_\gamma \;) &\propto\!\!\! \int\limits_{-k_{x, \text{max}}}^{k_{x, \text{max}}} \!\!\!\!dk'_{x} 
        \!\!\int\limits_{-k_{x, \text{max}}}^{k_{x, \text{max}}}\!\!\!\! dk_{x}\,  \tilde{f}(k_{x}, k'_{x}) e^{i(k_{x}-k'_{x})(x_\mathrm{e}-x_\gamma)}\,,\label{eq:pos_pos_measurement}
    \end{align}
\end{subequations}
where $p_\mathrm{e}$ and $\hbar k_\gamma$ ($x_{\mathbf{e}}$ and $x_\gamma$) represent the momentum (position) along the $x$ axis for the electron and the photon, respectively. 

\begin{tcolorbox}[breakable, enhanced,boxrule=0pt,title=Modeling Entanglement Certification]
\centering
            \includegraphics[width = 0.97\textwidth]{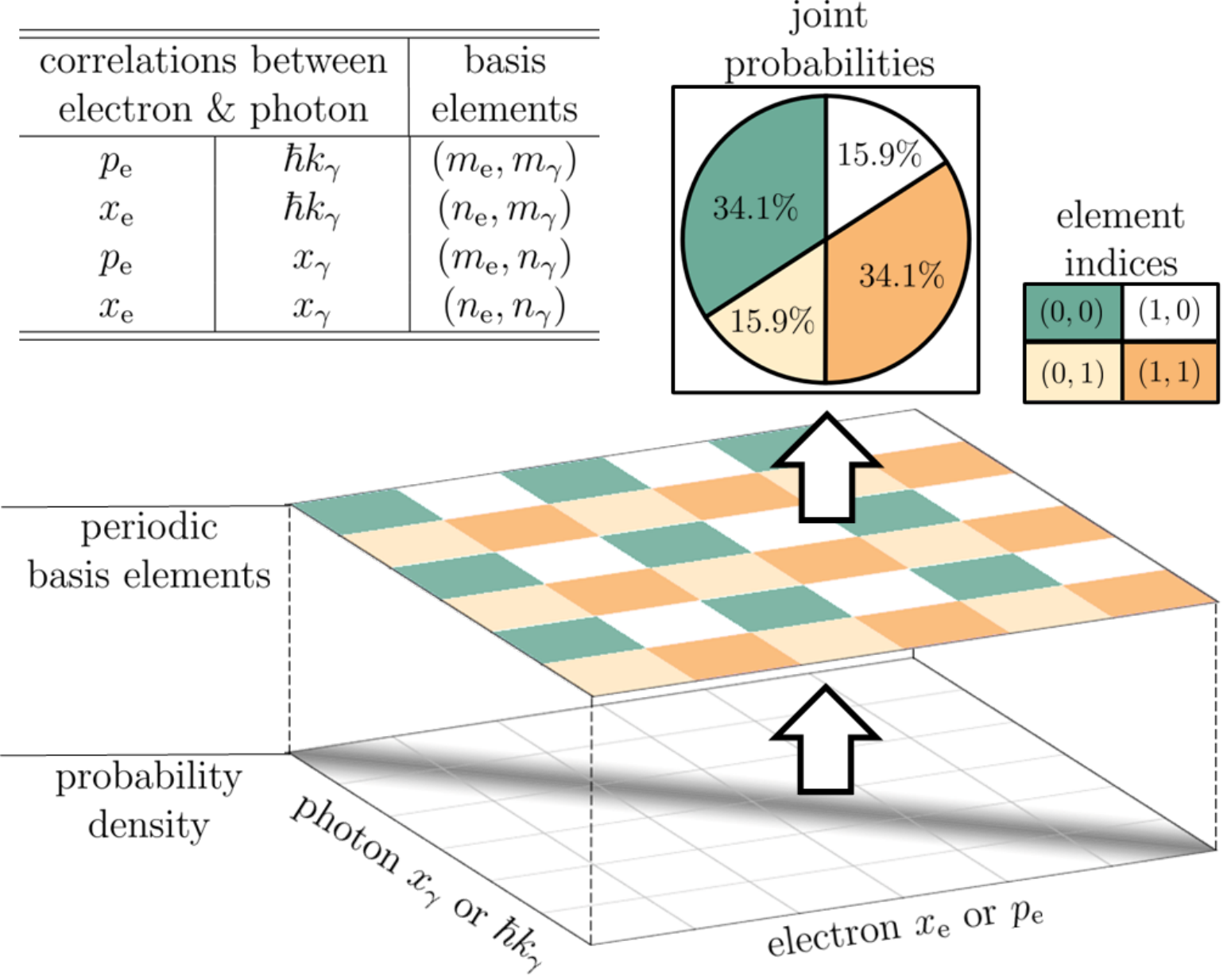}
    \captionof{figure}{
    {
    Illustrating correlation measurements shown in Figs.~{\ref{fig:witness_measurements}} and~{\ref{fig:witness_blurred}}. The certification of entanglement relies on the joint probability density (in greyscale, bottom of stack plot) of two observables (position or momentum) of the electron and the photon. 
    Measurements are carried out with respect to one basis element each. 
    The elements are interposed forming a chequerboard pattern (top of the stack plot) to illustrate the regions corresponding to each joint probability. 
    The combinations of elements are represented by coloured areas: green $(0,0)$, white $(1,0)$, beige $(0,1)$, and orange $(1,1)$.
    The measurements effectively sort every part of the probability density into a joint basis element. 
    Finally, the probabilities of finding both particles in a joint basis element are added, as described in Eq.~{\eqref{eq:probs_basis_els}}, with the results shown in the inset pie chart.
    These joint probabilities sum up the electron-photon correlation corresponding to each combination of basis elements.}} \label{fig:graphs_certification}
\end{tcolorbox}

The derivation of this result is detailed in Appendix~\ref{ap:position_momentum_measurements_electron_photon}, {where both positions and momenta are bounded within $x_{\mathbf{e}}, x_\gamma \in [-x_{\text{max}}, x_{\text{max}}]$ $p_{\mathbf{e}}, \hbar k_\gamma \in [-k_{x,\text{max}}, k_{x,\text{max}}]$ defining sufficiently large position ranges to have a convergence in the joint probabilities, resulting in the normalization constant $N_{p-p}$ given explicitly in Eq.~\eqref{eq:Normalization_Np}.}

The presented projectors are infinitely wide, that is, the complete sets of projectors $\ket{x}\bra{x}$ and $\ket{p}\bra{p}$ are obtained for all $x,p\in\mathbb{R}$.
However, in the experiment the area of detection is finite. 
The application of the projectors in this area is valid when the considered wave functions are constrained to that same area. 
A typical electron wave function produced in a TEM, for example, is localised in position and high-momentum components produced during the CL process are filtered by internal reflection, leaving us with wave functions that are strongly localised in both position and momentum.
This way the bases do not need to cover the entire Hilbert space, but rather the occupiable part.

Finally, we consider the detection efficiency to be independent of the position on the detector to form the discussed bases.
An example where this might not be the case is a camera with blind spots in between pixels or non-uniform pixel response functions.
An alternative setting with a potentially higher detection efficiency could include filters in the shape of the basis elements and bucket detectors, similar to Ref.~\mbox{\cite{TascaSanchezPieroWalbornRudnicki2018}}.

As noise from different parts of the setup blurs out the correlations, entanglement becomes more difficult to detect.
However, experimental measurement imperfections can be taken into account in post processing or via adjusted entanglement-detection bounds~\mbox{\cite{LiHuberFriis2024,morelli_entanglement_2022}} dependent on their characterisation (see Appendix~\mbox{\ref{app:robustness}}).

In the following, we consider a potential physical realization to further exemplify some practicalities of this proposal.

\section{Example: Electron-Cherenkov-Photon Pairs}\label{sec:electron_cherenkov_photon_pair}
\begin{figure}
    \centering
    \includegraphics[width=\linewidth]{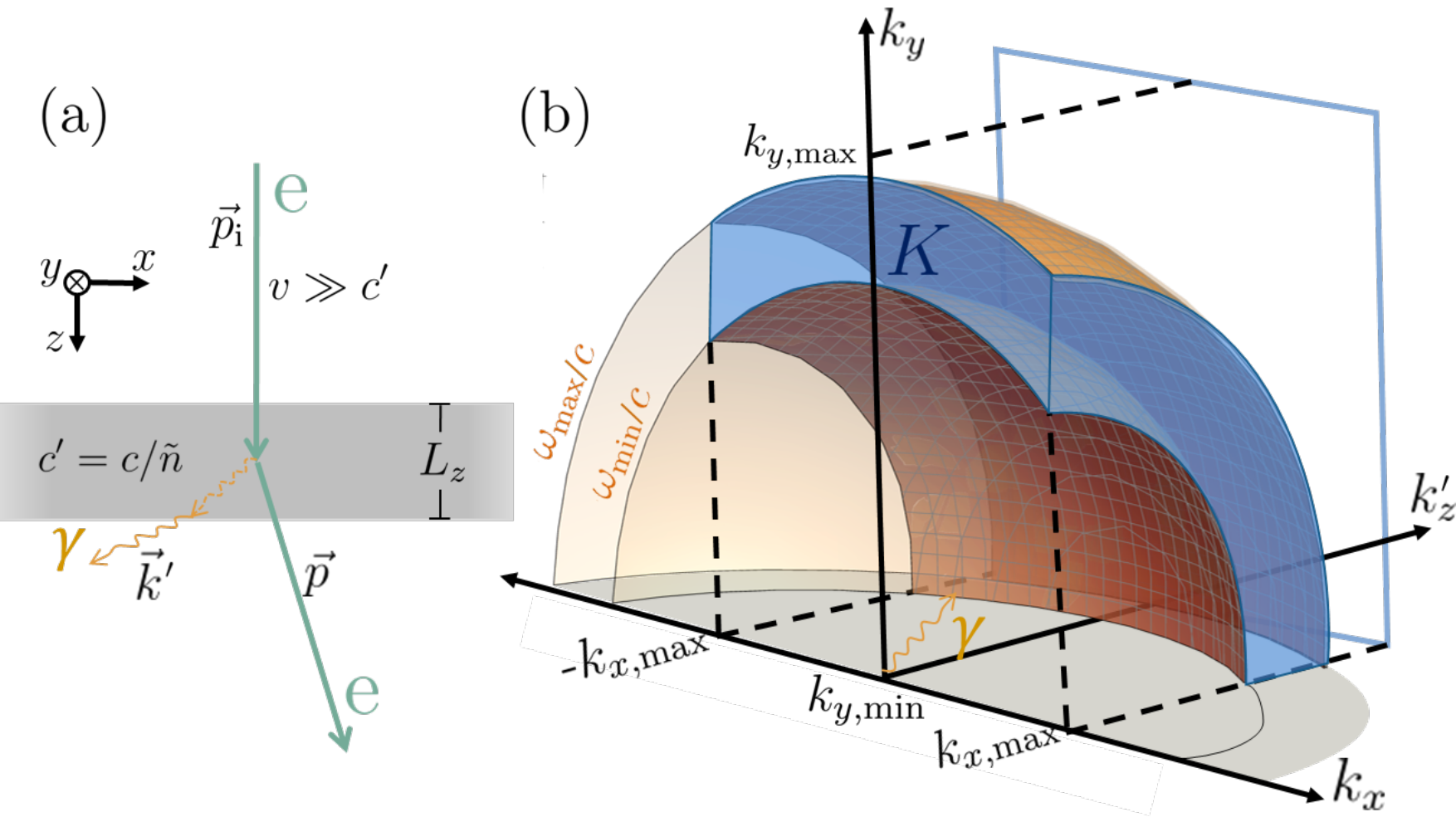}
    \caption{(a) Photon emission due to the Cherenkov effect: The incident electron beam (green) interacts with a dielectric material (gray rectangle), producing a photon (yellow). This interaction deflects the incident electron.  
    Initially, the electron is described by a longitudinal Gaussian wave packet with momentum $\vec{p}_\text{i} = p_{\text{i}, z} \vec{e}_z$, where $\vec{e}_z$ is the unit vector along~$z$.
    During the interaction, the Cherenkov effect deflects the momentum to $\vec{p}$, generating a photon with corresponding wave vector~$\vec{k}$. 
    (b) Representation of the photon wave vector space outside the dielectric slab: the emitted photons are filtered by energy, defining the minimum and maximum wave vectors, as indicated by the dark orange concentric spherical sections with radii $\omega_{\text{min}}/c$ and $\omega_{\text{max}}/c$, respectively. Additionally, only photons with incident angles that are not subject to total internal reflection inside the sample ($\theta_\gamma \leq \theta_{\text{crit}}$) are considered. 
    {{The range of momenta in the $x$-$y$ plane is further restricted under typical experimental detection conditions, as illustrated by the blue rectangle in the figure. Specific cut-off values, $k_{x, \text{max}}$ and $k_{y, \text{min(max)}}$, are applied, with $k_{y, \text{min}} = 0$ in the diagram.}} As a result, the momentum eigenvalues under consideration are confined to the blue-highlighted region $K$.
    }
    \label{fig:simulations_model}
\end{figure}

In this section, we discuss entangled pairs generated via Cherenkov radiation in a thin dielectric membrane. We illustrate how this specific mechanism can give rise to the desired quantum correlations and provide an estimate of the entanglement that could be observed under experimental conditions.
The Cherenkov effect is a specific instance of a coherent CL process that results in electron-photon pairs. 
The process comprises the emission of a photon when a charged particle moves into or past a dielectric material at a velocity $v$ exceeding the speed of light $c$ within the material, that is, the velocity ratio $\beta = v/c > 1/\tilde{n}$~\cite{2004VavilovCherenkov, 2010deAbajo}, where $\tilde{n}$ is the refractive index of the material. 
As a result of the photon emission, the electron is scattered, experiencing recoil, as shown in Fig.~\ref{fig:simulations_model}~a. 
The Cherenkov condition is easily fulfilled for typical samples in a TEM, where electrons are accelerated to energies on the order of $30-300\,$keV. 
For sufficiently thick dielectrics with respect to the photon wavelengths, momentum conservation is approximately fulfilled.
The resulting momentum change can be assumed to be equivalent in absolute value but opposite to the momentum of the emitted photon. 
As a result, the electron and the photon are entangled, effectively representing an EPR pair.\\

\parag{Inside the Sample}Based on the framework of quantum electrodynamics~(QED), one can describe the Cherenkov process and analyse the state of electron-Cherenkov-photon pairs (referred to as ``Cherenkov pairs" for brevity) in detail. 
This has been demonstrated by numerous studies~\cite{2016Kaminer, 2021Karnieli, 2018Roques-Carmes, ChaikovskaiaKarlovetsSerbo2024} exploring, for example, the feasibility of producing light with various polarizations and electrons with distinct spin states. 
In the following, we will present an idealised model that captures the essential features of the Cherenkov process in a sample represented by a dielectric slab of finite thickness, leading to a finite photon-emission probability (Appendix~\ref{ap:emmsion_angle}). 
To obtain an expression for the pair state, we use previously derived transition amplitudes~\cite{2016Kaminer, ChaikovskaiaKarlovetsSerbo2024}. 
At the end of this section, the resulting state is used to assess the feasibility of certifying entanglement between the electron and the photon produced. 

The electron initially propagates along the $z$-axis. 
In this direction, it is described by a narrow Gaussian wave packet $\psi_G(p_{\text{i}, z})$ with a momentum distribution centred on the average initial momentum $\bar{p}_{\text{i}, z}$, corresponding to energy $E_0=\sqrt{m_\mathrm{e}^2c^4+\bar{p}_{\text{i}, z}^2c^2}$, and standard deviation $\Delta p_z = 10^{-6}\, \bar{p}_{\text{i}, z}$. 
Its momentum in the transverse directions is close to zero. 
Such a wave function would be expected from the parallel illumination used in TEMs, where the electron beam is collimated to minimise transverse spread. 
We consider the dielectric to be infinitely wide in the transverse direction, $L_\perp\rightarrow \infty$, and of finite longitudinal width, $L_z$, which corresponds to the sample thickness. 
It should be noted that the finite width of the dielectric along $z$ breaks translation invariance, leading to imperfect longitudinal momentum conservation, a factor that is considered in the analysis of the resulting Cherenkov pairs.

The state of the produced Cherenkov pair is a linear combination of states $|\frac{\vec{p}}{\hbar}, s\rangle_\mathrm{e}\otimes |\vec{k}, \epsilon \rangle_\gamma$ with amplitudes described in Ref.~\cite{2016Kaminer} (see Appendix~\ref{ap:reduce_matrix} for details). 
We specifically focus on Cherenkov photons with much lower energy than that of the incident electron, i.e., in the optical and near-UV spectrum, where quantum emission rates tend toward classical rates~\cite{2016Kaminer, 2018Roques-Carmes}. 
In the low-energy photon regime, the Cherenkov-pair state is independent of electron spin and photon polarization because the dominant contribution to the transition amplitude does not correspond to a spin flip and always results in radially polarised light (cf. Appendix~\ref{ap:reduce_matrix}).
In this regime, the Cherenkov-pair state follows the same mathematical form presented in Eq.~\eqref{eq:final_state_electron-photon} and it is given by the expression
\begin{align}\label{eq:final_state_finiteLz_main}
\ket{\psi_{\mathrm{e}\mhyphen\gamma}}\propto 
& \int \!\!\!d^3k \, dp_{\text{i},z} \ \mathcal{A}(\vec{k})\, \psi_G(p_{\text{i}, z}) \\
&\qquad\qquad\qquad\cdot\big|-\hbar k_x, -\hbar k_y, p_{z}^{\text{con}}\big\rangle_\text{e}\otimes \big| k_x,k_y,k_z\big\rangle_\gamma ,\nonumber
\end{align}
where $p_{z}^{\text{con}} \approx p_{\text{i}, z} - \frac{\hbar k E_\text{i}}{\tilde{n}c p_{\text{i}, z}}$ from the energy-conservation condition, with energy {$E_\text{i}=\sqrt{m_\mathrm{e}^2c^4+{p}_{\text{i}, z}^2c^2}$}, and the Cherenkov amplitudes
\begin{equation}
    \begin{aligned}
\mathcal{A}(\vec{k}) \approx &  \sqrt{\frac{\alpha }{2\pi^2 \tilde{n}}} \frac{k_\perp}{k^{3/2}} L_z \, \text{sinc}\left(\frac{L_z}{2}\left(k_z - \frac{k}{\beta \tilde{n}}\right)\right),
    \end{aligned}
\end{equation}
with fine-structure constant $\alpha$, $k_\perp = \sqrt{k_x^2 + k_y^2}$ and $k = \sqrt{k_\perp^2 + k_z^2}$. 
The sinc term results from the finite integration length $L_z$ during the scattering process.
For an infinite integration length, the sinc term approaches a Dirac delta function, ensuring perfect momentum conservation along $z$. 
We can identify the state in Eq.~\eqref{eq:final_state_finiteLz_main} as a superposition of EPR pairs in the transverse momenta.\\

\parag{Outside the Sample}So far, we have described the state inside the sample.
However, measurements on electrons and photons are typically performed outside of the sample. 
Due to the dielectric's thickness, multi-scattering processes are negligible. The lower-layer interface is assumed to have no effect on the electron, thus leaving its wave function unchanged.

Still, the refraction of the photons at the sample boundary must be taken into account. 
This transformation is represented by a map $\hat{\rho}_{\mathrm{e}\mhyphen\gamma} \rightarrow \hat{\rho}_{\mathrm{e}\mhyphen\gamma, \mathrm{out}} = \mathcal{B}[\hat{\rho}_{\mathrm{e}\mhyphen\gamma}]$ applied to the density matrix $\hat{\rho}_{\mathrm{e}\mhyphen\gamma} = \ket{\psi_{\mathrm{e}\mhyphen\gamma}}\bra{\psi_{\mathrm{e}\mhyphen\gamma}}$. 
The explicit form of this map depends on the specific experimental setup and requires detailed modeling.

In our idealised model, we assume that the boundary is a planar surface orthogonal to the $z$-direction, separating the dielectric sample from the vacuum. 
This assumption is supported by the fact that certain samples generally appear optically flat to the emitted light~\mbox{\cite{lilienfeld_optical_1986}}, as atomic-scale imperfections are negligible at the relevant photon wavelengths.
At this boundary, photon refraction and total internal reflection are treated as the dominant effects, resulting in a state $\ket{{\psi}_{\mathrm{e}\mhyphen\gamma, \text{out}}} = \mathcal{\hat{M}} \ket{{\psi}_{\mathrm{e}\mhyphen\gamma}}$, where $\mathcal{\hat{M}}$ is defined as
\begin{equation}\label{eq:mapping_outside}
    \mathcal{\hat{M}} = \int_{\theta_\gamma \leq \theta_{\text{crit}}} d^3k \, {\tau_\parallel(\theta_\gamma)}\ket{k_x, k_y, k'_z}\bra{\vec{k}}.
\end{equation}
Here, $\tau_\parallel(\theta_{\gamma})$ represents the Fresnel transmission amplitude for p-polarised light~\cite{hecht2002optics}, as Cherenkov radiation is classically emitted with this polarization~\cite{yamamoto1996} (see Appendix~\ref{ap:red_density_x}).
As a result, only wave vectors satisfying $\theta_\gamma = \arccos(k_z/k) \leq \theta_{\text{crit}} = \arcsin(1/\tilde{n})$ are transmitted through the boundary. 
According to the law of refraction, the photon's momentum components in the $x$- and $y$-directions remain constant across the boundary, while the $z$-component decreases, resulting in the modified wave number outside the dielectric, i.e., 
$k'_z = \frac{1}{\tilde{n}} \sqrt{(1 - \tilde{n}^2)(k_x^2 + k_y^2) + k_z^2}$.\\

\parag{Certifying Entanglement}We now apply the entanglement certification strategy described in Sec.~\ref{sec:epr_electron_photon_certification_entanglement} to the case of Cherenkov pairs. 
The reduced state $\hat{\rho}_{\mathrm{e}\mhyphen\gamma, x}$ is constructed analogously to Eq.~\eqref{eq:red_Density_x} by tracing over the unmeasured degrees of freedom. However, in the Cherenkov case, this trace is applied to $\hat{\rho}_{\mathrm{e}\mhyphen\gamma, \mathrm{out}}$ and includes an additional restriction: the photon momentum is limited to a finite region $K$, defined by the cutoffs $[-k_{x,\text{max}}, k_{x,\text{max}}] \times [k_{y,\text{min}}, k_{y,\text{max}}]$ and a detectable photon energy range $E_\gamma \in [\hbar \omega_{\text{min}}, \hbar \omega_{\text{max}}]$ (see Fig.~\ref{fig:simulations_model}b and Eq.~\eqref{eq:momentum_limits}). These constraints reflect practical detection limitations and the effects of total internal reflection.
The reduced density operator retains the same structural form as in Eq.~\eqref{eq:red_Density_x}, but the azimuthal symmetry of its probability amplitudes allows it to be written as

\begin{equation}\label{eq:red_Density_x_cherenkov}
    \begin{aligned}
        \hat{\rho}_{\mathrm{e}\mhyphen\gamma, x}= &\int_0^{k_{x, \text{max}}} dk_{1, x} \int_0^{k_{x, \text{max}}} dk_{2, x} \frac{\hbar}{L_\perp} \tilde{f}(k_{1, x}, k_{2, x}) \\
        &\cdot \left(\ket{-\hbar k_{1, x}}_\mathrm{e} \ket{k_{1, x}}_\gamma + \ket{\hbar k_{1, x}}_\mathrm{e} \ket{-k_{1, x}}_\gamma \right) \\
        &\ \left(\bra{-\hbar k_{2, x}}_\mathrm{e} \bra{k_{2, x}}_\gamma + \bra{\hbar k_{2, x}}_\mathrm{e} \bra{-k_{2, x}}_\gamma \right),
    \end{aligned}
\end{equation}
where its matrix elements $\tilde{f}(k_{1, x}, k_{2, x})$ are now given by
\begin{equation} \label{eq:felements_finitetime}
    \begin{aligned}
        \tilde{f}(k_{1, x}, k_{2, x})
    \approx \frac{1}{\tilde{N}}\int_{K_{y\mhyphen z}} &{dk_{y} \, dk'_{z}}\, \mathcal{A}(\vec{k}_1)\mathcal{A}(\vec{k}_2)\\
    &\cdot \tau_\parallel(\theta_{\gamma, 1})\tau_\parallel(\theta_{\gamma, 2})\mathcal{G}(\vec{k}_1, \vec{k}_2),
    \end{aligned}
\end{equation}
{where $\vec{k}_j=(k_{j, x}, k_y, k_{j,z})$ and $\cos\theta_{\gamma, j}=\frac{k_{j, z}}{|\vec{k}_j|}$ with $k_{j,z}=\sqrt{(\tilde{n}^2-1)(k_{j,x}^2+k_y^2)+\tilde{n}^2k^{'2}_{z}}$} (see the derivation in Appendix~\ref{ap:red_density_x}).
The matrix elements and the normalization constant $\tilde{N}$, defined in Eq.~\eqref{eq:norm_Ntilde}, are obtained through numerical integration, which depends on the integration limits within the region $K_{y\mhyphen z}$, an area defined by fixed values of $k_{1, x}$ and $k_{2, x}$. 
The Gaussian function $\mathcal{G}(\vec{k}_1, \vec{k}_2)$ in the second line results from the initial Gaussian electron wave packet and is explicitly given in Eq.~\eqref{eq:gaussian_spread_term}. 

The dominant terms of the density operator in Eq.~\eqref{eq:red_Density_x_cherenkov} are centred on the diagonal ($\ket{-\hbar k_x}_\mathrm{e}\ket{k_x}_\gamma\bra{-\hbar k_x}_\mathrm{e}\bra{k_x}_\gamma$) and anti-diagonal ($\ket{-\hbar k_x}_\mathrm{e}\ket{k_x}_\gamma\bra{\hbar k_x}_\mathrm{e}\bra{-k_x}_\gamma$) elements, similar to the Bell state discussed in Sec.~\ref{sec:what_is_entanglement}. 
An explicit example and visualization are provided in Appendix~\ref{ap:red_density_x}.
Generally, the finite width of the incident momentum wave packet and finite sample width lead to increased momentum uncertainty, introducing additional contributions to the density operator. 
In the limiting case where $\Delta p_z \rightarrow 0$ and $L_z \rightarrow \infty$, the density operator closely approximates a statistical mixture of maximally entangled states within disjoint subspaces, such that their respective partial transposes do not overlap.\\

\parag{Modelling Measurements}As the model provides a complete state description via the reduced density operator [cf.~Eq.~\eqref{eq:red_Density_x_cherenkov}], we can evaluate entanglement using formal separability criteria such as the PPT criterion~\cite{1998Horodecki}. Taking the partial transpose of the reduced density matrix (see Appendix~\ref{app:criterion}) reveals negative eigenvalues, thereby confirming the presence of entanglement. Beyond this formal certification, we now assess whether entanglement can be verified using only accessible measurements in position and momentum space.

\begin{figure}[t]
    \centering
    \includegraphics[width=\linewidth]{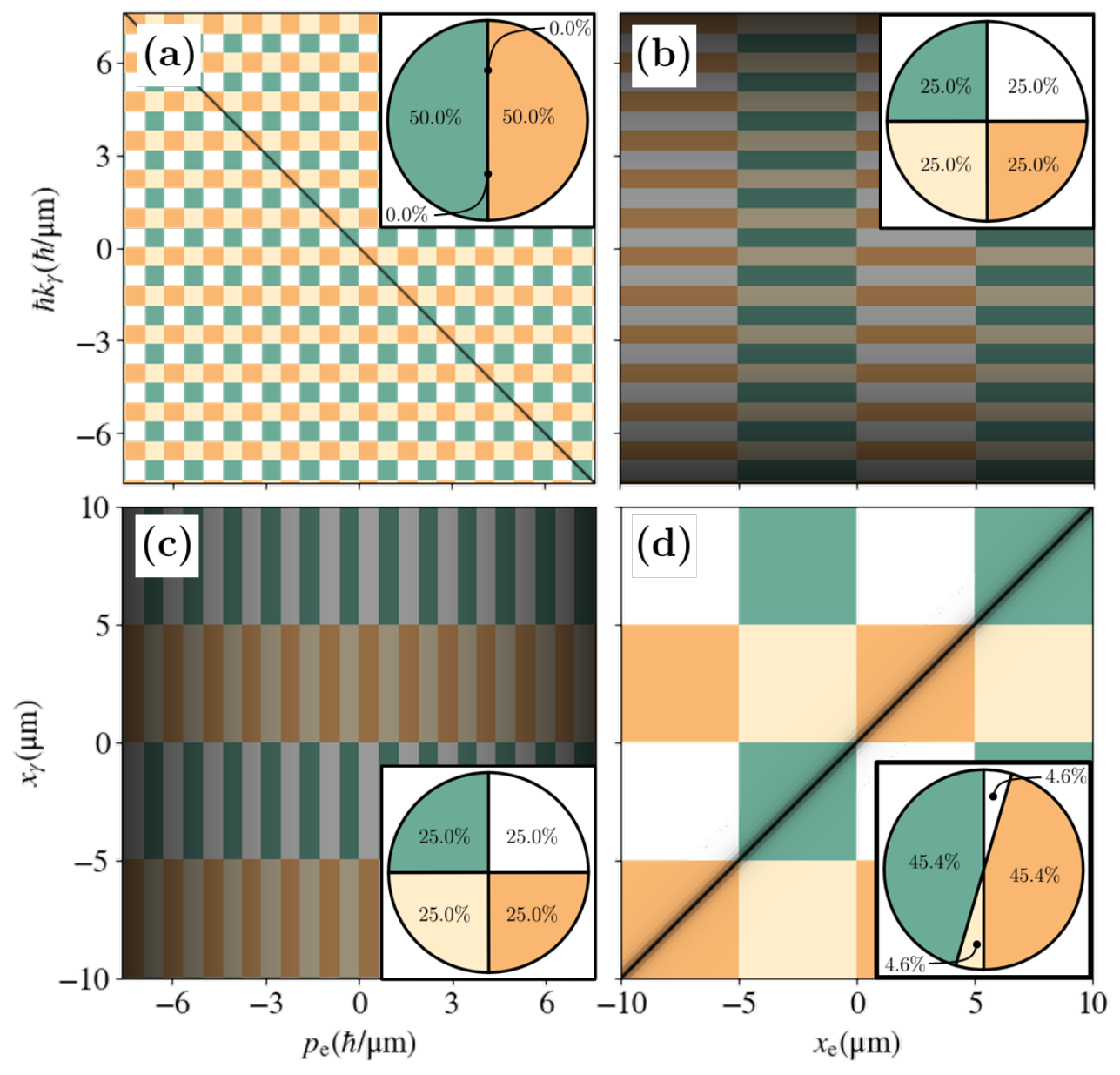}
    \caption{Correlation measurements assuming perfect detectors. 
    Measurements are taken in the combined position and momentum bases for each particle, with the measured variables being: 
    (a) $p_\mathrm{e}, \hbar k_\gamma$, (b) $x_\mathrm{e}, \hbar k_\gamma$, (c) $p_\mathrm{e}, x_\gamma$, and (d) $x_\mathrm{e}, x_\gamma$. 
    The probability density is represented by the dark shading intensity (a.u.) following Eq.~\eqref{eq:measurements}. Each colour indicates one combination of basis elements with periods $T_p=1.26\;\frac{\hbar}{\text{\textmu} \text{m}}$ and $T_x=\;10\,$\textmu m. 
    The composition of the plots, insets, and colouring is explained in Fig.~\ref{fig:graphs_certification}. 
    {Here, we show a smaller, representative range in position than that used to calculate the joint probabilities.}
    The underlying model is based on the following parameters: 
    sample thickness $L_z =\;\text{200\,nm}$ and refractive index $\tilde{n} = 1.6$; incident electron with kinetic energy $E_{\text{kin}}=E_0-m_\mathrm{e}c^2 =\;200\,$keV ($\beta = 0.7$); and photon energy range $[3.5,\,4.0]\,$eV. 
    These parameters and their experimental feasibility are discussed at the end of this section.
}
\label{fig:witness_measurements}
\end{figure}

To this end, Fig.~\ref{fig:witness_measurements} illustrates the probability densities obtained from the modelled state according to Eq.~\eqref{eq:measurements} and the matrix elements in Eq.~\eqref{eq:felements_finitetime}, along with the periodic basis elements that define the MUB measurements. The momentum-momentum distribution [Eq.~\eqref{eq:mom_mom_measurement}, Fig.~\ref{fig:witness_measurements}a] is sharply peaked around zero total momentum, corresponding to a Dirac delta function and indicating perfect anti-correlation, consistent with an incident electron wave packet closely approximating a plane wave. In contrast, the position-momentum distributions [Eqs.~\eqref{eq:pos_mom_measurement} and~\eqref{eq:mom_pos_measurement}, Figs.~\ref{fig:witness_measurements}~b and~c] exhibit no visible correlations, as they depend solely on the momentum component. Finally, the position-position distribution [Eq.~\eqref{eq:pos_pos_measurement}, Fig.~\ref{fig:witness_measurements}d] shows a pronounced peak at $|x_\mathrm{e} - x_\gamma| = 0$, indicating strong spatial correlations. These contrasting features reflect the structure of the underlying entangled state and form the basis for entanglement certification using MUB-based observables.

Hence, an idealised Cherenkov-pair state exhibits perfect anti-correlation in momenta and almost perfect correlation in positions. 
The corresponding joint probabilities for momentum-momentum measurements are $\mathcal{P}(0\suptiny{0}{0}{(p)},0\suptiny{0}{0}{(p)}) = \mathcal{P}(1\suptiny{0}{0}{(p)},1\suptiny{0}{0}{(p)}) = 1/2$. Similarly, for position-position measurements, we have $\mathcal{P}(0\suptiny{0}{0}{(x)},0\suptiny{0}{0}{(x)}) = \mathcal{P}(1\suptiny{0}{0}{(x)},1\suptiny{0}{0}{(x)}) \approx 1/2$.
The position-correlation does not reach that ideal value as our model is fundamentally constrained by the sample's thickness, which defocuses emission points along the $z$-axis, effectively blurring the transverse position projection.
Similarly, entanglement in larger Hilbert spaces would lead to an effective mixing and consequent blurring of the joint probability distribution.
To increase the measured position-correlation, $T_x$ could be increased (decreasing $T_p$), but we opted for realistic values. 
For mixed measurements, the joint probabilities are $\mathcal{P}(0\suptiny{0}{0}{(x)},0\suptiny{0}{0}{(p)}) = \mathcal{P}(1\suptiny{0}{0}{(x)},1\suptiny{0}{0}{(p)}) = \mathcal{P}(0\suptiny{0}{0}{(x)},1\suptiny{0}{0}{(p)}) = \mathcal{P}(1\suptiny{0}{0}{(x)},0\suptiny{0}{0}{(p)}) = 1/4$, and vice versa, due to the mutual unbiasedness of the bases. 
Generally, these values would change because of experimental factors such as noise. 
Limited detector accuracy, for example, leads to blurring of the distributions.\\

\begin{figure}[h]
    \centering
    \includegraphics[width=1\linewidth]{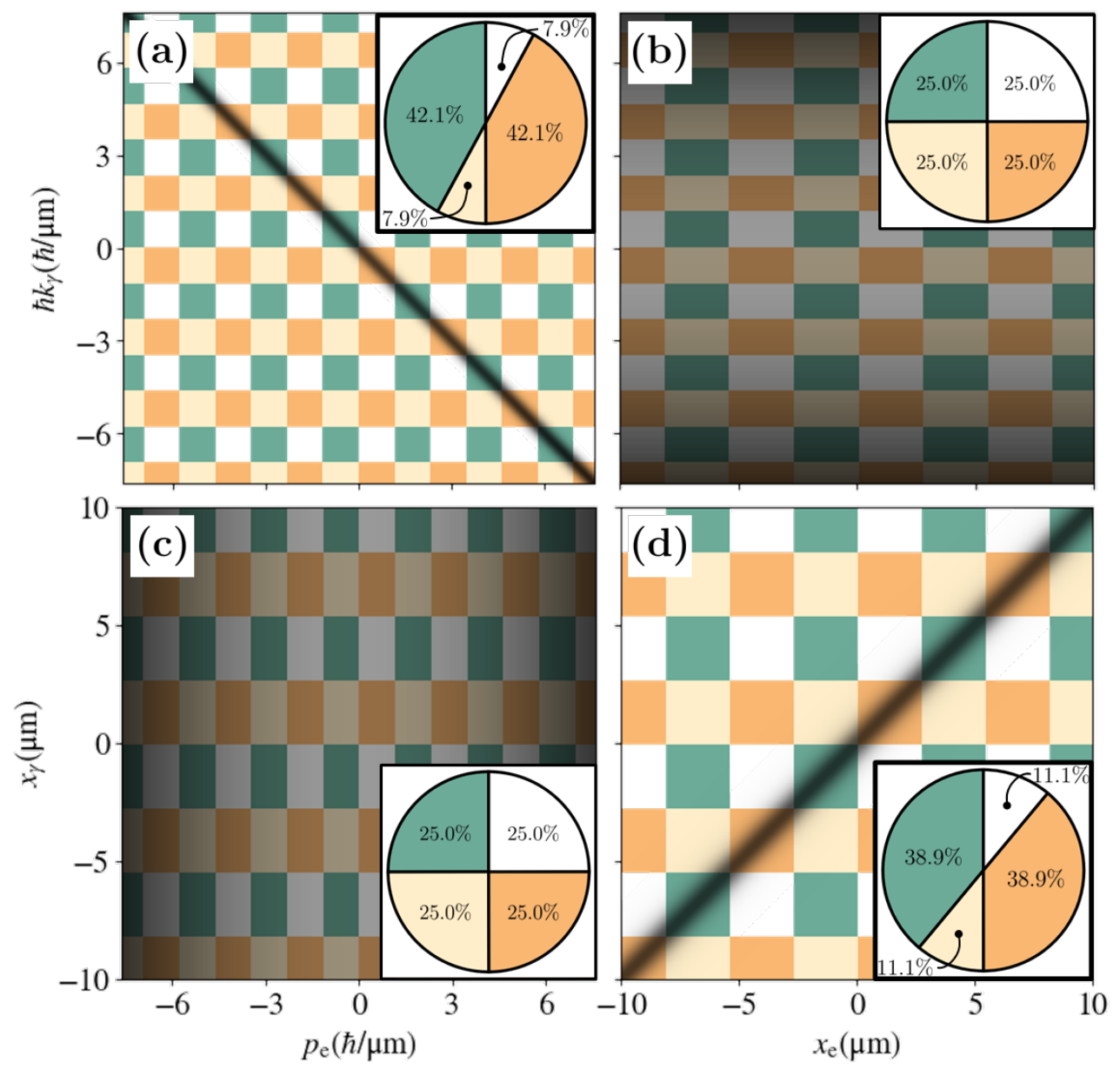}
    \caption{
    Correlation measurements for realistic detector resolutions (using the experimental values in Table~\ref{tab:FWHMs}) with optimised periodicities. 
    The finite resolution is considered by convolving the probability densities from Eq.~\eqref{eq:probs_basis_els} with distributions corresponding to Table~\ref{tab:FWHMs}. 
    As specified in Fig.~\ref{fig:graphs_certification}, probability densities are represented by the shaded intensity of the overlaid distribution, while the joint periodic basis elements are indicated by the coloured background. 
    The insets give the joint probabilities. 
    {As in} Fig.~\ref{fig:witness_measurements}{, the range of position values used for the numerical evaluation is larger than the ones shown in the plots.}
    The applied optimal basis periods are $T_p=\;2.31\frac{\hbar}{\text{\textmu} \text{m}}$ and $T_x=5.44\,$\textmu m.}
    \label{fig:witness_blurred}
\end{figure}

\parag{Measuring Electron-Cherenkov-Photon Pairs}The effect of the limited resolution can be understood as convolving the PSF characterising the measurement apparatus with the probability density of the measured state. For the example of electron-Cherenkov-photon pairs, we use the model state shown in Fig.~\ref{fig:witness_measurements}. Hence, the probability densities accounting for detector resolution are
\begin{equation}
    \Pi_{\text{exp}}(\kappa_\mathrm{e}, \kappa'_\gamma)=\tilde{\Pi}(\kappa_\mathrm{e}, \kappa'_\gamma)*\text{PSF}(\kappa_\mathrm{e};\delta^\kappa_\mathrm{e})\text{PSF}(\kappa'_\gamma;\delta^{\kappa'}_\gamma),
\end{equation}
{resulting in the joint probabilities presented in Fig.~{\ref{fig:witness_blurred}}.}

In contrast to Fig.~\ref{fig:witness_measurements}, the probability densities in Fig.~\ref{fig:witness_blurred} are more blurred.
This is especially clear in Fig.~\ref{fig:witness_blurred}~a, where the initial Dirac delta changed into a distribution with finite width, and consequently worse joint probabilities.
However, the blurring is strongest in photon position (Fig.~\ref{fig:witness_blurred}~c and~d).
To keep the effect on the joint probability to a minimum, the position periodicity of the basis, $T_x$, must be high enough to cover a large area of the distribution. 
Keeping in mind that $T_x$ is directly related to the momentum periodicity, $T_p$, via Eq.~\eqref{eq:periodicities}, we optimise the two of them together to maximise the joint probabilities in momentum-momentum and position-position spaces without compromising the unbiasedness. 

The blurring leads to reduced joint probabilities of momentum-momentum ($0.421<0.5$) and position-position ($0.389<0.5$) measurements partially mitigated by optimising the basis.

Meanwhile, the equal joint probabilities in position-momentum measurements indicate no change in unbiasedness. 
While unbiasedness is a property of the basis and valid for all states, the constancy implies that, as expected, the blurring only reduces the correlation but does not introduce new ones.
Using the presented probabilities, the criterion from Ineq.~\eqref{eq:mubW} is violated for $M = d = 2$, namely
\vspace{-2mm}
\begin{align}
    &\mathcal{P}(0\suptiny{0}{0}{(x)},0\suptiny{0}{0}{(x)})
    \,+\,\mathcal{P}(1\suptiny{0}{0}{(x)},1\suptiny{0}{0}{(x)})\,+\,\mathcal{P}(0\suptiny{0}{0}{(p)},0\suptiny{0}{0}{(p)})
    \,+\,\mathcal{P}(1\suptiny{0}{0}{(p)},1\suptiny{0}{0}{(p)})
    \nonumber\\[1mm]
    &\ =\,2\cdot 0.421+2\cdot 0.389 \,=\, 1.620
    >1.5\;,
\end{align}
\vspace{-6mm}

\noindent
successfully certifying the state's entanglement. 
Similarly, the fidelity-based criterion from Eq.~\eqref{eq:tilde F k0 d2} would give $\tilde{\mathcal{F}}=0.620>0.5$ also confirming the pair's entanglement. 
As a result, the entanglement of formation can be estimated to be $E_F\geq 0.03$.
\begin{tcolorbox}[breakable, enhanced,boxrule=0pt,title=Model-Complementary Experimental Parameter Ranges]

{So far, we have considered an idealised system; this box provides some practical considerations. 
In order to design an experiment that uses Cherenkov radiation as a source of entangled electron-photon pairs, we need to consider a wide range of parameters, most notably: the sample's thickness and refractive index connected with the required energy range and electron acceleration voltage. 
While many other parameters can be chosen independently, these four need to be balanced in order for the experiment to be viable.} \\

\parag{\textbf{Sample Thickness \& Acceleration Voltage}}
Photon production increases linearly with sample thickness, $L_z$, but a thicker sample increases the likelihood of an electron to experience an undesirable secondary inelastic interaction. 
This effect can be mitigated by increasing the acceleration voltage, which reduces the inelastic mean free path.
An electron accelerated with 200$\,$keV, for example, has an inelastic mean free path of approximately 150$\,$nm in Si and 180$\,$nm in SiO$_2$~{\cite{lee2002}}. 
Typically, TEM samples stay under 300$\,$nm, depending on the acceleration voltage.\\

\parag{\textbf{Refractive Index \& Energy Range}}
The Cherenkov angle, given by $\cos(\theta_{\text{CR}}) = (\beta \tilde{n})^{-1}$, depends on the sample{'}s refractive index $\tilde{n}$ and the electron's velocity ratio $\beta$. 
{
The ratio between sample thickness and photon wavelength determines the dominant emission angle. 
The higher this ratio is, the closer the angle becomes to the Cherenkov angle $\theta_{\text{CR}}$}~{\cite{yamamoto1996}} {exhibiting a narrower angular distribution (see Appendix~{\ref{ap:emmsion_angle}}).
For a photon to be detected, its emission angle should be below the critical angle for total internal reflection. }
Hence, the optimal refractive index range is bounded by $\beta^{-1} \leq \tilde{n} \leq \sqrt{1 + 1/\beta^2}$. 
For 200$\,$keV electrons ($\beta = 0.7$), this range is approximately $1.43 \leq \tilde{n} \leq 1.74$. 
{In the optical regime, the sample thickness and wavelength inside the material are of comparable size. 
While this leads to an undesirably broad distribution of the emission angle, the optical wavelength range aligns with readily available optics and detectors, facilitating efficient photon detection.
At optical wavelengths many materials, such as quartz, calcite, and various glasses, also fall within the desired range of refractive index and have low dispersion.
Estimates based on photon-emission probability and error propagation for 350$\,$nm suggest that acquisition times on the order of hours should be sufficient to collect the necessary amount of signal.}
\end{tcolorbox}
\vspace{2mm}
\parag{Model vs. Experiment}Any practical implementation of the discussed entanglement-certification scheme will be subject to experimental conditions beyond the presented idealised model (see Appendix~\mbox{\ref{ap:assumptions}}). 
As mentioned earlier, one such factor is the refraction at the sample boundaries, which induces interference effects that alter the mode structure of the electromagnetic field. 
A corresponding QED model of the Cherenkov process is provided in Ref.~\cite{2018Roques-Carmes}. 
Additional factors that may change the presented distributions are: 
the characteristics of the incident electron wave function such as beam diameter and energy spread differing from a plane wave (the electron's refraction is expected to be negligible), sample defects or geometric imperfections, charging effects, multi-scattering events including multi-photon emission, elastic scattering within the sample, and other excitation processes. This may include both coherent mechanisms (e.g., transition radiation, plasmons) and incoherent mechanisms (e.g., electron-hole pair generation)~\cite{2010deAbajo}.

Cherenkov pairs serve as a representative example of the general considerations discussed at the end of Sec.~\ref{sec:epr_electron_photon_certification_entanglement}, where measuring transverse momentum corresponds to detecting deflection angles in the far field. In typical Cherenkov scenarios, the relationship between deflection angle and transverse momentum transfer is straightforward when photons are emitted near the Cherenkov angle, corresponding to the direction of peak intensity in an infinite dielectric slab. However, when the photon wavelength becomes comparable to or larger than the sample thickness, emission occurs over a broader angular range (see Appendix~\ref{ap:emmsion_angle}), which effectively broadens the distribution of longitudinal momentum components ($p_z$). Therefore, the necessity of narrow energy filtering or joint angle-energy detection still holds for the momentum-space MUB projectors over Cherenkov pairs (see Appendix~\ref{ap:deflection_vs_momenta}).

As we concluded in Section~\ref{sec:epr_electron_photon_certification_entanglement}, the set of projectors must cover the occupiable part of the electron-photon pair wave function.
In the presented model, only the momentum contribution is strongly localized for both particles. 
Still, reasonable estimates can be made, because the position-position probability density is uniform in $x_\text{e}+x_\gamma$, that is, the correlations only depend on $|x_\text{e}-x_\gamma|$. This allows us to obtain approximate results for the joint probabilities for position-position measurements while restricting both $x_{\mathbf{e}}$ and  $x_\gamma$ to $[-x_{\text{max}}, x_{\text{max}}]$ as shown in detail in Appendix~{\ref{ap:pos_mom_measurement}}.

\section{Conclusion}
Hybrid systems, such as entangled electron-photon pairs, provide a promising direction for advancing quantum technologies by combining complementary features of distinct particles. 
In particular, the electron's picometer-scale de~Broglie wavelength allows one to achieve atomic resolution, while photons can be easily guided and detected in a phase-sensitive manner.
However, to exploit the hybrid system's quantum features the subsystems need to be verifiably entangled.
In this work, we have described a protocol for bounding electron-photon entanglement, specifically the entanglement of formation, in a state-of-the-art TEM adapted to provide optical access to the sample. 
Previous theoretical protocols for similar systems detect entanglement~\cite{HenkeJengRopers2024, kazakevich2024spatial, Konecna2022entanglement_electron-photon} relying on certain underlying features of the system, rather than describing state-agnostic measurements.
Similarly, recent experimental studies have made assumptions on the state to infer entanglement~\mbox{\cite{adiv_observation_2023, henke_observation_2025,preimesbergerExperimentalVerificationElectronPhoton2025}}.
Here we review different ways to verify entanglement and show how it can be certified without assumptions on the state.

While we have used the Cherenkov effect as a example, any coherent and energy-conserving CL process can, in principle, produce entangled electron-photon pairs. 
Indeed, the central concepts and techniques presented in this work are similarly applicable to correlated electron-electron pairs, such as the ones investigated in Ref.~\cite{haindl2023}. 
By using the appropriate periodic coarse-graining and a time-resolved direct-detection camera to identify coincident electron-electron pairs, their entanglement{\textemdash}if present{\textemdash}could be certified and benchmarked. In relativistic regimes or when spin-orbit interactions become significant, one could also explore Bohm-type spin-momentum entanglement~\mbox{\cite{NagChowdhury2021}}, potentially extending the MUB-based certification approach. It remains to be seen which processes will be the most easily exploitable to produce useful entangled pairs. 

Our model takes into account a number of experimental constraints, including noise and blurring effects, and proposes a way to optimise MUBs for the detection and certification of entanglement. 
To increase the entanglement bounds described in this manuscript, it is crucial to improve the resolutions as much as possible. 
Indeed, resolution is the most important factor to consider when testing equivalent or alternative setups.
Furthermore, rather than taking into account the full Hilbert space, i.e. any possible wavefunction, bases could be designed for unbiasedness within the occupiable spatial and momentum subspace to reduce the measurement complexity.

Generally, the created states could provide the basis for various entanglement-based protocols which have so far only been accessible to purely photonic setups.
One such application is entanglement-based sensing which could reduce sample damage by adapting quantum imaging techniques from photonics~\cite{gilaberte_basset_perspectives_2019,
bridaExperimentalRealizationSubshotnoise2010,onoEntanglementenhancedMicroscope2013} to conventional TEM imaging, diffraction, or spectroscopy techniques.
Especially in bio-technology~\cite{casacio_quantum-enhanced_2021} such techniques already enable beyond-classical resolution and non-invasive sensing. 
Furthermore, the correlations could be used to shape the electron wave function~\mbox{\cite{rotunno_one-dimensional_2023, preimesbergerExperimentalVerificationElectronPhoton2025}} or to further characterise the CL produced by a given sample. 
The high resolving power and flexibility of electron microscopy may be complemented by quantum imaging, offering new directions for quantum technology at the atomic level.

\vspace{5pt}
\begin{acknowledgements}
We acknowledge financial support from the Austrian Science Fund (FWF) [10.55776/P36478] funded by the European Union{\textemdash}NextGenerationEU, as well as by the Austrian Federal Ministry of Education, Science and Research via the Austrian Research Promotion Agency (FFG) through the flagship project FO999897481 (HPQC), the project FO999914030 (MUSIQ), and the project FO999921407 (HDcode) funded by the European Union{\textemdash}NextGenerationEU.
E.A. acknowledges funding from the FWF [10.55776/V1037].
P.H. thanks the FWF: [10.55776/Y1121; 10.55776/P36041; 10.55776/P35953]. This project was supported by the FFG project AQUTEM.
\end{acknowledgements}


\bibliographystyle{apsrev4-1fixed_with_article_titles_full_names_new}
\bibliography{Master_Bib_File}

\hypertarget{sec:appendix}
\appendix

\renewcommand{\thesubsubsection}{\Alph{section}.\Roman{subsection}.\arabic{subsubsection}}
\renewcommand{\thesubsection}{\Alph{section}.\Roman{subsection}}
\renewcommand{\thesection}{\Alph{section}}
\setcounter{equation}{0}
\numberwithin{equation}{section}
\setcounter{figure}{0}
\numberwithin{figure}{section}
\renewcommand{\theequation}{\Alph{section}.\arabic{equation}}
\renewcommand{\thefigure}{\Alph{section}.\arabic{figure}}


\onecolumngrid

\newpage

\section{Correlated Measurements over EPR Type Entangled Electron-Photon Pair State}\label{ap:correlated_measurements_EPR}

In this Appendix, we implement the MUB certification protocol for a general EPR-type electron-photon pair state, subject to transverse momentum conservation. Specifically, we describe the probability densities in both position and momentum representations, which are essential for evaluating CV entanglement within the protocol.

We begin with the general EPR-type electron-photon pair state given in Eq.\eqref{eq:final_state_electron-photon}. As discussed in Sec.~\ref{sec:epr_electron_photon_certification_entanglement}, the protocol is constrained to assess entanglement in the position and momentum degrees of freedom along a single spatial axis. In this case, we focus on the transverse $x$-axis, where momentum correlations naturally arise due to conservation laws. 
Therefore, the reduced density matrix over which the measurements are applied is obtained by tracing over the $y$ and $z$ momenta of both particles and the electron spin and photon polarization. 
Then, we obtain the following density matrix:
\begin{equation}
    \begin{split}
\hat{\rho}_{\mathrm{e}\mhyphen\gamma}=&\text{Tr}_{p_y, p_z, s, k_y, k_z, \epsilon}\left(\ket{\psi_{\mathrm{e}\mhyphen \gamma}}\bra{\psi_{\mathrm{e}\mhyphen \gamma}}\right)\\
=& \sum_{\epsilon', s'} \int dk'_y dk'_z \int dp'_y dp'_z \bra{p'_y, p'_z, s'}_\text{e}\otimes   \bra{k'_y, k'_z, \epsilon'}_\gamma \ket{\psi_{\mathrm{e}\mhyphen \gamma}}\bra{\psi_{\mathrm{e}\mhyphen \gamma}} \ket{p'_y, p'_z, s'}_\text{e}\otimes   \ket{k'_y, k'_z, \epsilon'}_\gamma, 
    \end{split}
\end{equation}
where the new amplitudes depend on the overlap
\begin{equation}
    \begin{split}
\bra{k'_y, k'_z, \epsilon'}_\gamma \ket{\psi_{\mathrm{e}\mhyphen \gamma}} =& \sum_{\epsilon, s}\int d^2k_\perp \int dk_z\, dp_z \Psi(\vec{k}_\perp, k_z, p_z, \epsilon, s)\delta(p'_y+\hbar k_y)\delta(p'_z-p_z)\delta_{s, s'}\delta(k'_y-k_y)\delta(k'_z-k_z) \delta_{\epsilon, \epsilon'}\\
=&\int dk_x \Psi(k_x, k'_y, k'_z, p'_z, \epsilon', s') \delta(p'_y+\hbar k'_y)\ket{-\hbar k_x}_\text{e}\ket{k_x}_\gamma.
    \end{split}
\end{equation}
Inserting both overlaps into the reduced density matrix we end up with the reduced density matrix in Eq.~\eqref{eq:red_Density_x}:
\begin{equation*}
    \begin{split}
\hat{\rho}_{\mathrm{e}\mhyphen\gamma, x}=\int dk_{1, x}\, dk_{2, x} \frac{\hbar}{L_\perp}\tilde{f}(k_{1, x}, k_{2, x}) \ket{-\hbar k_{1, x}}_\text{e}\ket{k_{1, x}}_\gamma\bra{-\hbar k_{2, x}}_\text{e}\bra{k_{2,x}}_\gamma,
    \end{split}
\end{equation*}
where $j = 1$ and $j = 2$ labels the ket and bra states respectively. Additionally, we use the fact that:
\begin{equation}\label{eq:doubledeltaremoval}
   \lim_{L_\perp\rightarrow \infty} \frac{L_\perp}{\hbar} \delta_{L_\perp/\hbar} (p'_y+\hbar k'_{y}) = \lim_{L_\perp\rightarrow \infty} \delta^2_{L_\perp/\hbar} (p'_y+\hbar k'_{y}) = \delta^2(p'_y+\hbar k'_{y})\,.
\end{equation}
The matrix elements depend on the EPR pair amplitudes as follows
\begin{equation}
   \tilde{f}(k_{1, x}, k_{2, x})=\frac{L_\perp^2}{\hbar^2}\sum_{\epsilon', s'}\int dk'_y dk'_z dp'_z \Psi(k_{1, x}, k'_y, k'_z, p'_z, \epsilon', s')\Psi^*(k_{2, x}, k'_y, k'_z, p'_z, \epsilon', s').
\end{equation}
Due to these amplitudes, the dependence on the transversal length is canceled out, as this length appears in the normalization of the confined plane-wave functions.

\subsection{Position and Momentum Measurements in EPR type Electron-Photon Pairs}\label{ap:position_momentum_measurements_electron_photon}

Now if we use Eq.~\eqref{eq:red_Density_x} for the reduced density matrix, we find that the simultaneous measurement of momentum variables, restricting the detected $x$ momenta of both particles to an interval $K_x=[-k_{x, \text{max}}, k_{x, \text{max}}]$, leads to the probability density
\begin{equation}\label{eq:Pi_pp_general}
    \begin{aligned}
\tilde{\Pi}(p_{\mathrm{e}}, k_\gamma)=&\frac{1}{N_{p-p}}\bra{p_{\mathrm{e}}, k_\gamma} \int_{K_x\times K_x} dk_{1, x}dk_{2, x} \frac{\hbar}{L_\perp} \tilde{f}(k_{1, x}, k_{2, x}) \ket{-\hbar k_{1, x}, k_{1, x}} \bra{-\hbar k_{2, x}, k_{2, x}} p_{\mathrm{e}}, k_\gamma\rangle\\
=& \frac{1}{N_{p-p}}\int_{K_x\times K_x} dk_{1, x}dk_{2, x} \frac{\hbar}{L_\perp} \tilde{f}(k_{1, x}, k_{2, x}) \delta(p_{\mathrm{e}}+\hbar k_{1, x})\delta(k_\gamma -k_{1, x})\delta(p_{\mathrm{e}}+\hbar k_{2, x})\delta(k_\gamma -k_{2, x})\\
=& \frac{1}{N_{p-p}}\lim_{L_\perp\rightarrow\infty} \frac{\hbar}{L_\perp} \tilde{f}(k_\gamma, k_\gamma) \delta^2(\hbar k_\gamma+p_{\mathrm{e}}),\\
=& \frac{1}{N_{p-p}}\tilde{f}(k_\gamma, k_\gamma) \delta(\hbar k_\gamma+p_{\mathrm{e}}).
    \end{aligned}
\end{equation}
whose normalization constant is given by
\begin{equation}\label{eq:Normalization_Np}
    N_{p-p}=\int_{K_x} dk_\gamma\tilde{f}(k_\gamma, k_\gamma).
\end{equation}
The expected momentum correlation is demonstrated in this expression, which arises from momentum conservation and corresponds to the wave number distribution represented by $\tilde{f}(k_\gamma, k_\gamma)$.

Regarding the simultaneous measurement of both particles' positions, we obtain the matrix elements of our reduced density operator in Eq.~\eqref{eq:red_Density_x} in the position representation as follows:  
\begin{equation}\label{eq:rhoX_constraint_pos_general}
    \begin{aligned}
\rho_{\mathrm{e}\mhyphen\gamma, x}(x_{\mathrm{e}, 1}, x_{\gamma, 1}; x_{\mathrm{e}, 2}, x_{\gamma, 2})=&\langle x_{\mathrm{e}, 1}, x_{\gamma, 1}| \hat{\rho}_{\mathrm{e}\mhyphen\gamma, x}|x_{\mathrm{e}, 2}, x_{\gamma, 2}\rangle \\
\propto& \int_{K_x\times K_x} dk_{1, x} dk_{2, x} \tilde{f}(k_{1, x}, k_{2, x})\langle x_{\mathrm{e}, 1}, x_{\gamma, 1}| -\hbar k_{1,x}, k_{1,x}\rangle \langle -\hbar k_{2, x}, k_{2, x}|x_{\mathrm{e}, 2}, x_{\gamma, 2}\rangle\\
\propto& \int_{K_x\times K_x} dk_{1, x} dk_{2, x} \tilde{f}(k_{1, x}, k_{2, x}) e^{-i(x_{1, \mathrm{e}}-x_{1, \gamma})k_{1,x}} e^{i(x_{2, \mathrm{e}}-x_{2, \gamma})k_{2,x}}.
    \end{aligned}
\end{equation}  
{These elements depend on the relative position coordinates $ x_{j, \mathrm{e}} - x_{j, \gamma}$ for $ j=\{1, 2\}$. Therefore, the density matrix is uniform with respect to the midpoint coordinates $\frac{1}{2}(x_{j, \mathrm{e}}+x_{j, \gamma})$, meaning that the quantum state is characterised by the same distribution throughout. }
{We set $x_{\mathrm{e}}, x_\gamma \in [-x_{\text{max}}, x_{\text{max}}] $ allowing for multiple periods of the periodic basis, ensuring that the joint probabilities converge with respect to $x_{\text{max}}$. 
This procedure provides a reliable estimation of position correlations while effectively capturing the essential features of an infinite spatial distribution in both position variables.}

Considering only the diagonal elements of the density matrix in the position representation in Eq.~\eqref{eq:rhoX_constraint_pos_general}, we obtain the normalised probability density over the possible position values:  
\begin{equation}\label{eq:Pi_xx_general}
    \begin{aligned}
\tilde{\Pi}(x_{\mathrm{e}}, x_\gamma)=& \frac{1}{N_{x-x}}\bra{x_{\mathrm{e}}, x_\gamma} \int_{K_x\times K_x} dk_{1,x}dk_{2, x} \tilde{f}(k_{1, x}, k_{2, x}) \ket{-\hbar k_{1, x}, k_{1, x}} \bra{-\hbar k_{2, x}, k_{2, x}} x_{\mathrm{e}}, x_\gamma\rangle\\
= & \frac{1}{N_{x-x}}\int_{K_x\times K_x} dk_{1, x}dk_{2, x} \frac{1}{(2\pi)^2\hbar}\tilde{f}(k_{1, x}, k_{2, x}) e^{i(k_{1, x}-k_{2, x})(x_{\mathrm{e}}-x_\gamma)}\,,
    \end{aligned}
\end{equation}  
where the normalization constant is given by  
\begin{equation}
    N_{x-x}=\int_{-x_{\text{max}}}^{x_{\text{max}}}dx_{\mathrm{e}}\int_{-x_{\text{max}}}^{x_{\text{max}}}dx_\gamma\int_{K_x\times K_x} dk_{1, x}dk_{2, x} \frac{1}{(2\pi)^2\hbar}\tilde{f}(k_{1, x}, k_{2, x}) e^{i(k_{1, x}-k_{2, x})(x_{\mathrm{e}}-x_\gamma)}.
\end{equation}  
In fact, the probability density in Eq.~\eqref{eq:Pi_xx_general} explicitly depends on the distance between the two particles. Unlike the strong one-to-one anti-correlations observed in the momentum, the position correlations are generally weaker and do not map a specific position of the electron directly onto a corresponding photon position. Indeed, they are affected by the mixing with the traced degrees of freedom, represented by the dependence of the EPR amplitudes on the other momentum components, electron spin, and polarization.

The probability density for the combined measurement of the electron position with the photon momentum is given by:
\begin{equation}\label{eq:Pi_px_general}
\begin{aligned}
\Pi(x_{\mathrm{e}}, k_\gamma)=&\bra{x_{\mathrm{e}}, k_\gamma} \int_{K_x\times K_x} dk_{1,x}dk_{2,x} \frac{\hbar}{L_\perp} \tilde{f}(k_{1, x}, k_{2, x}) \ket{-\hbar k_{1, x}, k_{1, x}} \bra{-\hbar k_{2, x}, k_{2, x}} x_{\mathrm{e}}, k_\gamma \rangle\\
=&\int_{K_x\times K_x} dk_{1,x}dk_{2,x} \frac{\hbar}{L_\perp} \tilde{f}(k_{1, x}, k_{2, x}) \frac{1}{2\pi \hbar}e^{-i(k_{1,x}-k_{2,x})x_{\mathrm{e}}}\delta(k_\gamma-k_{1,x})\delta(k_{2,x}-k_\gamma)\\
=& \frac{1}{2\pi L_\perp}\tilde{f}(k_\gamma, k_\gamma). 
\end{aligned}
\end{equation}
This probability density is normalised taking into account the positions $x_{\mathrm{e}}, x_\gamma\in [-x_{\text{max}}, x_{\text{max}}]$. Therefore,
\begin{equation}\label{eq:Pi_xp_general}
    \begin{aligned}
\tilde{\Pi}(x_e,k_\gamma)=&\frac{1}{N_{x-p}}\frac{1}{2\pi L_\perp}\tilde{f}(k_\gamma, k_\gamma)=\frac{1}{2x_{\text{max}} N_{p-p}}\tilde{f}(k_\gamma, k_\gamma), \quad \text{where}\\
N_{x-p}=&\int_{K_x}dk_\gamma \int_{-x_{\text{max}}}^{x_{\text{max}}}dx_e \frac{1}{2\pi L_\perp}\tilde{f}(k_\gamma, k_\gamma)\\
=&\frac{x_{\text{max}}}{\pi L_\perp} N_{p-p}.
    \end{aligned}
\end{equation}
Similarly the probability density for the simultaneous measurement of electron momentum with the photon position corresponds to
\begin{equation}
    \Pi(p_{\mathrm{e}}, x_\gamma)=\frac{1}{2\pi L_\perp \hbar} \tilde{f}(p_{\mathrm{e}}/\hbar, p_{\mathrm{e}}/\hbar), 
\end{equation}
and the normalised probability density considering the detection ranges of $x_\gamma$ is
\begin{equation}
    \tilde{\Pi}(p_{\mathrm{e}}, x_\gamma)=\frac{1}{2x_{\text{max}}\hbar N_{p-p}}\tilde{f}(p_{\mathrm{e}}/\hbar, p_{\mathrm{e}}/\hbar).
\end{equation}
These probability densities for measurements of position and momentum for each particle exhibit a dependence on the momentum of the respective particle, but not on the position of the other. Consequently, determining one variable for one particle results in complete uncertainty for the conjugate variable in the other particle, which is necessary for constructing a mutually unbiased bases. Furthermore, the probability densities obtained from these combined measurements require an additional normalization based on the range of position values considered by the detector.

\newpage
\section{Cherenkov Electron-Photon Pair State}

In this Appendix, we aim to explain an idealised model of the quantum state of the electron-Cherenkov-photon pair based on~\cite{2016Kaminer, ChaikovskaiaKarlovetsSerbo2024}. This model uses the framework of quantum electrodynamics (QED) to describe how the Dirac electron interacts with the quantised electromagnetic field within the dielectric material. 
Specifically, one starts from the minimal coupling term in the Dirac Hamiltonian, expressed as $-\hat{\psi}^\dagger \gamma^0 \gamma^\mu {q}\hat{A}_\mu \hat{\psi}$, which describes the interaction between the electron and the photon. In this expression, $\hat{\psi}$ and $\hat{\psi}^\dagger$ represent the electron field operators, respectively, while $\gamma^0$ and $\gamma^\mu$ are the Dirac gamma matrices, $q$ is the charge of the particle and $\hat{A}_\mu$ corresponds to the operator for the electromagnetic field. 

 The evolution resulting from this interaction is described through the scattering operator, which acts directly on an incident electron with momentum $\vec{p}_\text{i} = p_{\text{i},z} \vec{e}_z$, energy $E_\text{i}=\sqrt{(cp_{\text{i},z})^2+(m_{\mathrm{e}}c^2)^2}$ and spin $s_\text{i}=\{\uparrow,\downarrow\}$, along with an initial vacuum photon state. 
 The leading contribution to single-photon production arises from the first-order term in the electron-photon interaction within the scattering matrix, whereas a second-order contribution does not result in the production of a single photon in the final state.
 Therefore, elements of the scattering matrix up to first-order give us information about the probability amplitudes for having a scattered electron with momentum $\vec{p}$, energy $E_\text{f}=\sqrt{c^2|\vec{p}|^2+(m_{\mathrm{e}}c^2)^2}$, and spin $s_\text{f}=\{\uparrow,\downarrow\}$, along with a photon having a wave vector $\vec{k}$, frequency $\omega_k$, and polarization label $\epsilon=\{\text{azi}, \text{rad}\}$ (denoting azimuthal and radial polarization), from the interaction of the incident electron with the electromagnetic field of the dielectric medium. In fact the resultant (unnormalised) electron-photon pair state corresponds to:
\begin{equation}
    \begin{split}
        \ket{\psi_{\mathrm{e}\mhyphen\gamma}} = \sum_{s_\text{f}, \epsilon} \int d^3 p \int d^3 k \,  \bar{S}_{\text{f\,i}}^{\epsilon, s_\text{i}, s_\text{f}}(\vec{p}_\text{i}, \vec{p}, \vec{k}) \ket{\vec{p}, s_\text{f}, \vec{k}, \epsilon}\,.
    \end{split} \label{eq:final_state_cherenkov}
\end{equation}
where $\bar{S}_{\text{f}\,\text{i}}^{\epsilon, s_\text{i}, s_\text{f}}(\vec{p}_\text{i}, \vec{p}, \vec{k})$ refers to the scattering matrix elements. 
These are calculated considering a finite quantization volume $V$ and a finite time evolution $T$
in an intermediate step to simplify calculations. $V$ and $T$ are eventually sent to infinity.
As a consequence, the matrix elements are given by~\cite{2016Kaminer}:
\begin{equation}\label{eq:amplitudes_cherenkov}
    \begin{split}
       {\bar{S}}_{\text{f}\,\text{i}}^{\epsilon, s_\text{i}, s_\text{f}}({\vec{p}_\text{i},\vec{p},\vec{k}})=&{\lim_{V,T\to \infty}} \frac{{i}}{\hbar}\int_V d^3x \int_T dt\bra{\vec{p}, s_\text{f}, \vec{k}, \epsilon}\hat{\psi}^\dagger \gamma^0 \gamma^\mu {q}\hat{A}_\mu \hat{\psi} \ket{\vec{p}_\text{i}, s_\text{i}}_{\text{e}}\otimes\ket{\text{vac}} \\
       =&{\lim_{V,T\to \infty}} {i}\bar{N}(\vec{p}_\text{i}, \vec{p}, \vec{k}) \text{Sp}_{\epsilon, s_\text{i}, s_\text{f}}({\vec{p}_\text{i},\vec{p},\vec{k}}) \int_V d^3x \int_T dt  \,e^{i\left(\frac{E_\text{i}}{\hbar}-\frac{E_\text{f}}{\hbar}-\omega\right)t} e^{i\left(\frac{\vec{p}_\text{i}}{\hbar}-\frac{\vec{p}_\text{f}}{\hbar}-\vec{k}\right)\cdot \vec{r}}\\
       =&{\lim_{V,T\to \infty}} {i} \bar{N}(\vec{p}_\text{i}, \vec{p}, \vec{k}) \text{Sp}_{\epsilon, s_\text{i}, s_\text{f}}(\vec{p}_\text{i}, \vec{p}, \vec{k})  \bar{\Delta}(\vec{p}_\text{i}, \vec{p}, \vec{k}),
    \end{split}
\end{equation}
where the normalization factor
\begin{equation}
   \bar{N}(\vec{p}_\text{i}, \vec{p}, \vec{k}) = \frac{c^{\frac{3}{2}}}{\sqrt{\hbar \omega} \tilde n} \sqrt{\frac{\alpha}{2 \pi^2 \hbar^2 V}} \frac{1}{\sqrt{4 E_\text{i} E_\text{f}}}, \label{eq:norm_finite_length_plane} 
\end{equation}
where $\alpha$ is the fine-structure constant, and the amplitudes
\begin{equation}
    \text{Sp}_{\epsilon, s_\text{i}, s_\text{f}}({\vec{p}_\text{i},\vec{p},\vec{k}})=u^\dagger_{s_\text{f}}(\Vec{p})\gamma^0\gamma^\mu{\left(\vec{\varepsilon}_k\right)_{\mu}} u_{s_\text{i}}(\vec{p_\text{i}})
\end{equation}
are further developed in~\cite{2016Kaminer}. Additionally, $u_{s_\text{i}}(\vec{p}_\text{i}),\;u_{s_\text{f}}(\vec{p})$ represents the Dirac bispinor of an electron with a certain spin $s_\text{i},\;s_\text{f}$ and momentum $\vec{p},\;\vec{p}_\text{f}$ respectively and $\vec{\varepsilon}_k$ is the polarization vector of a photon with polarization label $\epsilon$, and wave vector $\vec{k}$.

Since the position and time dependencies are exclusively introduced through the plane waves used in the expansion of the electromagnetic and electron fields,
\begin{equation}
    \begin{split}
  \bar{\Delta}(\vec{p}_\text{i}, \vec{p}, \vec{k}) &= (2\pi)^4\delta_T\left(\omega - \frac{1}{\hbar} \left( E_\text{i} - E_\text{f} \right)\right) \delta_{L}\left( \frac{p_x}{\hbar} + k_x \right) \delta_{L}\left( \frac{p_y}{\hbar} + k_y \right)\delta_{L}\left( \frac{1}{\hbar} \left( p_{\text{i},z} - p_z \right) - k_z \right), 
\end{split}
\end{equation}
which explicitly defines the energy and momentum conservation relations for a finite volume $V=L^3$ and a finite time $T$ with 
\begin{subequations}
    \begin{align}
\begin{split}
    \delta_T\left(\omega-\frac{1}{\hbar}(E_\text{i} - E_\text{f})\right) 
&=  \frac{T}{2\pi} \, \text{sinc}\left( \left( \omega - \frac{1}{\hbar}(E_\text{i} - E_\text{f})\right) \frac{T}{2}\right) \\
&\stackrel{T \rightarrow \infty}{=}  \delta\left( \omega - \frac{1}{\hbar}(E_\text{i} - E_\text{f})\right)\,,
\end{split}\\
\begin{split}
    \delta_{L}\left( \frac{p_x}{\hbar}+ k_x\right) &= \frac{L}{2\pi} \, \text{sinc}\left( \left(\frac{p_x}{\hbar}+ k_x \right) \frac{L}{2} \right) \\
&\stackrel{L \rightarrow  \infty}{=}  \delta\left( \frac{p_x}{\hbar}+ k_x \right)\quad\text{and analogously for $y$,}
\end{split}\\
\begin{split}
    \delta_{L}\left( \frac{1}{\hbar} (p_{\text{i},z} - p_z) - k_z\right) &=  \frac{L}{2\pi} \, \text{sinc}\left( \left( \frac{1}{\hbar}(p_{\text{i},z} - p_z) - k_z \right) \frac{L}{2} \right) \\
&\stackrel{L \rightarrow \infty}{=}  \delta\left( \frac{1}{\hbar}(p_{\text{i},z} - p_z) - k_z \right)\,.
\end{split}
    \end{align}
\end{subequations}
In the following, we adjust the above model to our purposes by considering a finite interaction region of size $L_z$ in the $z$-direction which defines the boundary condition on the slab for the finite sample size. 
This leads to imperfect momentum conservation along the $z$-direction which implies that the dielectric slab itself becomes a source of momentum. 
We study the case where the slab is also infinite in the propagation direction as a limiting case to understand the consequences of having perfect energy and momentum conservation.

\subsection{Assumptions and Implications for a Realistic Model} \label{ap:assumptions}

Our model for the electron-photon pair final state corresponds exclusively to the production of electron-photon pairs via the Cherenkov phenomenon. Therefore, certain assumptions about a realistic scenario were made, and their implications are briefly discussed in this section. Firstly, the slab was considered as a poor resonator, allowing a wide spectrum of modes to cover many (thousands) of free spectral ranges of the slab. In principle, the slab acts as a Fabry-P\'erot cavity, modulating the Cherenkov spectrum to select certain resonant modes~\cite{Chang2010}. This situation potentially affects the phase matching conditions between the scattered electron and the emitted photon, as certain frequencies are enhanced due to reflectivity inside the slab~\cite{deAbajo2004}. 
On the other hand, multiple scattering of the electron is possible due to the thickness of the samples, which can contain thousands of atomic layers. Inelastic scattering of the electron beam in the material may also lead to charge build-up through, for example, secondary electron emission, affecting the electron exit trajectory and effectively blurring the resolution. Charging effects can be mitigated via an appropriate choice of material or by coating an insulating sample with a conductive layer. These processes inside the dielectric are not considered here, but we highlight that they can cause incoherences between the particle pairs. However, these incoherences can be largely filtered out by choosing the detection scheme appropriately.

Additionally, the production of electron-photon pairs at the boundary due to transition radiation, the Smith-Purcell effect, and surface plasmon polaritons is not included in this model~\cite{2010deAbajo}. As a result, a matching relation determined by a post-selection process based on the energy and arrival time of the particles and the forward production of radiation should be used to approximately obtain the particle pairs described here by the Cherenkov phenomenon. Furthermore, the effects of refraction and diffraction at the boundary should be considered when relating the photon produced in the slab to the photon outside the dielectric material, depending on the specific experimental conditions.
In our model, we account for refraction at the boundary by applying a mapping to the photon states based on the law of refraction. However, for the electron, we assume the interface to be transparent, as additional scattering processes at the boundary are not included in our treatment.
Ultimately, the present model is idealised, providing a proof of principle for witnessing the entanglement between the deflected electron and the emitted photon, as will be clarified in the following sections.

\newpage
\section{Reduced Density Operator in Momentum Representation }\label{ap:reduce_matrix}

The final state presented in Eq.~\eqref{eq:final_state_cherenkov} should describe the physical situation under the assumptions previously stated, while ensuring normalizability. This condition is often not fulfilled because we initially consider a non-normalizable plane wave as the incident electron beam wave function. 
Divergences in the probability occur when electron-photon pair production is unbounded, as is the case when a continuous plane wave enters the dielectric slab. 
 
To model the finite duration of the Cherenkov process, we describe the initial electron state as a Gaussian wave packet in the $z$-dimension which we identify with the propagation direction. 
The main part of this wave packet (aside from the exponential tails) remains in the interaction region (the dielectric) for only a limited time, resulting in a finite emission probability. 
Normalizable states can also be obtained by considering Gaussian wave packets in the other spatial dimensions, but since this is not required for the present calculations, we omit it for simplicity. 
The incident electron state is then given by a Gaussian wavepacket with mean momentum $\bar{p}_{\text{i}, z}$ and standard deviation $\Delta p_z/\bar{p}_{\text{i}, z} =  10^{-6} $ (using the convention of small angles), producing a narrow Gaussian wavepacket in momentum or a quasi-plane wave profile in position. Consequently, the  unnormalised final state conditioned on the production of an electron photon pair (post-selection through coincidence measurements) is:
\begin{equation}
    \begin{split}
        \ket{\psi_{\mathrm{e}\mhyphen\gamma}}= &\sum_{s_\text{f}, \epsilon}\int{d^3p}\int{d^3 k}\int dp_{\text{i}, z} \psi_G(p_{\text{i},z})S_{\text{f}\,\text{i}}^{\epsilon, s_\text{i}, s_\text{f}}({\vec{p}_\text{i},\vec{p},\vec{k}}) \ket{\vec{p}, s_\text{f}, \vec{k},\epsilon}\,,\\
        \psi_{G}(p_{\text{i}, z})=&\frac{1}{(2\pi \Delta p^2_z)^{1/4}}\exp\left\{-\frac{(p_{\text{i}, z}-\bar{p}_{\text{i}, z})^2}{4 \Delta p_z}\right\},
    \end{split} \label{eq:final_state_cherenkov_gaussian}
\end{equation}
where the new amplitudes are 
\begin{equation}\label{eq:amplitudes_cherenkov_new}
    \begin{split}
       S_{\text{f}\,\text{i}}^{\epsilon, s_\text{i}, s_\text{f}}({\vec{p}_\text{i},\vec{p},\vec{k}})=&\lim_{L_\perp,T\to \infty} i N(\vec{p}_\text{i}, \vec{p}, \vec{k}) \text{Sp}_{\epsilon, s_\text{i}, s_\text{f}}(\vec{p}_\text{i}, \vec{p}, \vec{k})  \Delta(\vec{p}_\text{i}, \vec{p}, \vec{k}),
    \end{split}
\end{equation}
the new normalization constant is
\begin{equation}
    {N}(\vec{p}_\text{i}, \vec{p}, \vec{k})=\frac{c^{\frac{3}{2}}}{\sqrt{\hbar \omega} \tilde n} \sqrt{\frac{\alpha}{2 \pi^2 \hbar^3 L_\perp^2}} \frac{1}{\sqrt{4 E_\text{i} E_\text{f}}}\,,
\end{equation}
and
\begin{equation}
    \begin{split}
  \Delta(\vec{p}_\text{i}, \vec{p}, \vec{k}) &= (2\pi)^4\delta_T\left(\omega - \frac{1}{\hbar} \left( E_\text{i} - E_\text{f} \right)\right) \delta_{L_\perp}\left( \frac{p_x}{\hbar} + k_x \right) \delta_{L_\perp}\left( \frac{p_y}{\hbar} + k_y \right)\delta_{L_z}\left( \frac{1}{\hbar} \left( p_{\text{i},z} - p_z \right) - k_z \right)\,.
\end{split}
\end{equation}
In the following, we will refrain from writing the limit of $L_\perp$ and $T$ to infinity explicitly in the equations and just note that this limit is taken in the end.

The model for certifying entanglement which we describe in this article requires measuring a certain momentum variable and its conjugate position in both particles (the Cherenkov photon and the scattered electron). Therefore, it becomes necessary to trace the final density matrix over the remaining degrees of freedom. Considering an initial plane wave incident electron, we deal with a statistical mixture of spin `up' and `down', which includes the initial spin in the traced degrees of freedom. Upon tracing over the spin and polarization degrees of freedom and utilising the orthogonality relations of the bispinors and polarization vectors, the resulting reduced unnormalised density operator in terms of the momentum variables has the form
\begin{equation}
    \begin{aligned}
\hat{\rho}_{\mathrm{red}}=&\int {d^3p_1 d^3p_2}\int {d^3 k_1 d^3 k_2} \int dp_{\text{i}, z} dp'_{\text{i}, z}  \psi_G(p_{\text{i},z}) \psi^*_G(p'_{\text{i},z})\\
& \times \sum_{s_\text{i}, s_\text{f}, \epsilon}\left(S_{\text{f}\,\text{i}}^{\epsilon, s_\text{i}, s_\text{f}}({\vec{p}_\text{i},\vec{p}_1,\vec{k}_1})\right)\left(S_{\text{f}\,\text{i}}^{\epsilon, s_\text{i}, s_\text{f}}({\vec{p}'_\text{i},\vec{p}_2,\vec{k}_2})\right)^*\ket{\vec{p}_1, \vec{k}_1}\bra{\vec{p}_2, \vec{k}_2}\\
=& \int{d^3p_1 d^3p_2}\int{d^3 k_1 d^3 k_2} \int dp_{\text{i}, z} dp'_{\text{i}, z}  {N}(\vec{p}_\text{i}, \vec{p}_1, \vec{k}_1) {N}(\vec{p}'_\text{i}, \vec{p}_2, \vec{k}_2) \psi_G(p_{\text{i},z}) \psi^*_G(p'_{\text{i},z}) \\
&\times \Delta(\vec{p}_\text{i}, \vec{p}_1, \vec{k}_1) \Delta(\vec{p}'_\text{i}, \vec{p}_2, \vec{k}_2)\frac{1}{2} \sum_{s_\text{i}, s_\text{f}, \epsilon}\left(\text{Sp}_{\epsilon, s_\text{i}, s_\text{f}}({\vec{p}_\text{i},\vec{p}_1,\vec{k}_1})\right)\left(\text{Sp}_{\epsilon, s_\text{i}, s_\text{f}}({\vec{p}'_\text{i},\vec{p}_2,\vec{k}_2})\right)^*\ket{\vec{p}_1, \vec{k}_1}\bra{\vec{p}_2, \vec{k}_2}\,.
    \end{aligned}
\end{equation}

Experimentally, the photons emitted due to the Cherenkov effect have energies four orders of magnitude lower than the incident electron. This condition defines the classical limit of this process and establishes the following relations in the momentum variables: $p_{x}, p_{y}, \hbar k_x, \hbar k_y, \hbar k_z \ll p_{\text{i}, z}, p_{z}$. Those inequalities allow us to define a factor of smallness $\tilde{\epsilon}\sim p_{x}/p_{\text{i}, z}\sim p_{y}/p_{\text{i}, z}\sim \hbar k_x/p_{\text{i}, z}\sim \hbar k_y/p_{\text{i}, z}\sim \hbar k_z/p_{\text{i}, z}\sim (p_z-p_{\text{i}, z})/p_{\text{i}, z}$. As predicted in experimental realizations and indicated in~\cite{2016Kaminer}, the spin-polarization amplitude that is significant in this limit is the one which corresponds to having no spin flip and radial polarization of the created photon. This becomes clear by establishing the following approximations and scale relations:
\begin{subequations}
    \begin{align}
\begin{split}
   E_\text{f} =& \sqrt{E_\text{i}^2+(p^2_z-p^2_{\text{i}, z}+p_x^2+p_y^2)c^2}\approx E_\text{i} \left[1+\frac{E_\text{i}+mc^2}{ 2E_\text{i}}\chi\right]\,,
\end{split} \\  
\begin{split}
  ( E_\text{f} + mc^2)^\ell\approx & (E_\text{i} +mc^2)^\ell\left[1+\frac{\ell}{2}\chi \right]\,,
\end{split}\\  
\begin{split}
  (E_\text{i}+E_\text{f} + 2mc^2)\approx & 2(E_\text{i} +mc^2)\left[1+\frac{1}{4}\chi\right]\,,
\end{split}
    \end{align}
\end{subequations}
where $\ell\in\mathbb{R}$, and we define the function with small values
\begin{equation}
    \begin{split}
\chi :=&  \frac{(p^2_z-p^2_{\text{i}, z}+p_x^2+p_y^2)c^2}{E_\text{i}(E_\text{i}+mc^2)}\sim \tilde{\epsilon}\,.
    \end{split}
\end{equation}
Then, applying the above approximations to Equations S6 and S7 in the Supplementary Material of~\cite{2016Kaminer}, which summarise the amplitudes $\text{Sp}_{\epsilon, s_\text{i}, s_\text{f}}$, results in the following scaling
 relations: 
\begin{subequations}
\allowdisplaybreaks
    \begin{align}
\begin{split}
\Re(\text{Sp}_{\text{azi},\uparrow,\uparrow})\approx & \frac{2c(p_yk_{x}-p_{x}k_y)}{\sqrt{k_{ x}^2+k_y^2}}\left(1+\mathcal{O}(\chi^2)\right)   \sim \tilde{\epsilon} E_\text{i}\,,   
\end{split} \\ 
\begin{split}
\Im(\text{Sp}_{\text{azi},\uparrow,\uparrow})\approx & -c\hbar\sqrt{k_{x}^2+k_y^2}\left(1+\frac{\chi}{4}+\mathcal{O}(\chi^2)\right)-\frac{c(p_{x}k_{x}+p_yk_y)}{\sqrt{k_{x}^2+k_y^2}}\left(\frac{\chi}{2}+\mathcal{O}(\chi^2)\right) \sim \tilde{\epsilon} E_\text{i}\,,
\end{split} \\ 
\begin{split}
\Re(\text{Sp}_{\text{azi},\uparrow,\downarrow})\approx & \frac{ck_y}{\sqrt{k_{x}^2+k_y^2}}\left[(p_z-p_{\text{i},z})-\frac{\chi}{4}(p_z+p_{\text{i},z})\right] \sim \tilde{\epsilon} E_\text{i}\,,
\end{split} \\ 
\begin{split}
\Im(\text{Sp}_{\text{azi},\uparrow,\downarrow})\approx & -\frac{ck_x}{\sqrt{k_{x}^2+k_y^2}}\left[(p_z-p_{\text{i},z})-\frac{\chi}{4}(p_z+p_{\text{i},z})\right] \sim \tilde{\epsilon}   E_\text{i}\,,
\end{split} \\ 
\begin{split}\label{eq:Re_Sp_rad_upup}
\Re(\text{Sp}_{\text{rad},\uparrow,\uparrow})\approx & -\frac{ck_z}{k}\left[\left(1+\mathcal{O}(\chi^2)\right)\frac{2(p_{x }k_{x}+p_yk_y)}{\sqrt{k_{x}^2+k_y^2}}+\left(1+\frac{\chi}{4} +\mathcal{O}(\chi^2)\right)\hbar \sqrt{k_{x}^2+k_y^2}\right]\\
&+\frac{c\sqrt{k_{x}^2+k_y^2}}{k}\left[(p_z+p_{\text{i},z})-\frac{\chi}{4}(p_z-p_{\text{i},z})\right] \sim c p_z\,, 
\end{split} \\ 
\begin{split}
\Im(\text{Sp}_{\text{rad},\uparrow,\uparrow})\approx & \frac{c k_z}{k}\frac{(p_{x}k_y-p_yk_{x})}{\sqrt{k_{x}^2+k_y^2}}\left(\frac{\chi}{2}+\mathcal{O}(\chi^2)\right)\sim \tilde{\epsilon}^2 E_\text{i}\,,
\end{split} \\ 
\begin{split}
\Re(\text{Sp}_{\text{rad},\uparrow,\downarrow})\approx & \frac{c \sqrt{k_{x}^2+k_y^2}}{k}\left[-\hbar k_x-\frac{\chi}{4}(2p_x+\hbar k_x)\right]+\frac{c k_z}{k}\frac{k_{x}}{\sqrt{k_{x}^2+k_y^2}}\left[(p_z-p_{\text{i},z})-\frac{\chi}{4}(p_z+p_{\text{i},z})\right] \sim \tilde{\epsilon} E_\text{i}\,,
\end{split} \\ 
\begin{split}
\Im(\text{Sp}_{\text{rad},\uparrow,\downarrow})\approx & \frac{c \sqrt{k_{x}^2+k_y^2}}{k}\left[-\hbar k_y-\frac{\chi}{4}(2p_y+\hbar k_y)\right]+\frac{c k_z}{k}\frac{k_{y}}{\sqrt{k_{x}^2+k_y^2}}\left[(p_z-p_{\text{i},z})-\frac{\chi}{4}(p_z+p_{\text{i},z})\right] \sim \tilde{\epsilon} E_\text{i}\,.
\end{split} 
    \end{align}
\end{subequations}
We find that only the real part of $\text{Sp}_{\text{rad},\uparrow,\uparrow}$ (Eq.~\eqref{eq:Re_Sp_rad_upup}), which corresponds to the amplitude for generating a radially polarised photon without flipping the electron spin, scales with the electron momentum in the $z$ direction, while the other terms scale with the small constant $\tilde{\epsilon}$.

In fact, a zeroth order approximation of the trace over spin and polarization degrees of freedom leads to
\begin{equation}
\begin{aligned}
    \mathcal{S}(\vec{p}_\text{i},\vec{p}, \vec{k})=&\frac{1}{2}\sum_{s_\text{i}, s_\text{f}, \epsilon}\left(\text{Sp}_{\epsilon, s_\text{i}, s_\text{f}}({\vec{p}_\text{i},\vec{p}_1,\vec{k}_1})\right)\left(\text{Sp}_{\epsilon, s_\text{i}, s_\text{f}}({\vec{p}_\text{i},\vec{p}_2,\vec{k}_2})\right)^*\\
    \approx& \Re({\text{Sp}_{\text{rad}, \uparrow, \uparrow}({\vec{p}_\text{i},\vec{p}_1,\vec{k}_1})})\Re({\text{Sp}_{\text{rad}, \uparrow, \uparrow}({\vec{p}_\text{i},\vec{p}_2,\vec{k}_2})})\,\\
    \approx & 4c^2p^2_{\text{i},z} \frac{k_{1, \perp}k_{2, \perp}}{k_1k_2}\,,
\end{aligned}
\end{equation}
where $k_{j, \perp}=\sqrt{k_{j,x}^2+k_{j,y}^2}$ and $k_j=\sqrt{k_{j, \perp}^2+k_{j,z}^2}$ with $j=1,2$.
This term is sufficient to recover the classical emission rate, because if we use Equation S24 from the Supplementary Material of~\cite{2016Kaminer}, we find that the emission rate is given by
\begin{equation}
\begin{split}
    \Gamma_\omega \approx &\frac{\alpha}{ \pi \beta_\text{i} }\int \frac{\sin(\theta_\gamma)d\theta_\gamma}{4 E_\text{i}^2}\frac{4c^2p^2_{\text{i},z}\sin^2(\theta_\gamma)}{\sqrt{(\sin(\theta_\text{i})\sin(\theta_{\text{CR}}))^2-(\cos(\theta_\gamma)-\cos(\theta_\text{i})\cos(\theta_{\text{CR}}))^2}}\\
    \approx& \alpha \beta_\text{i}\left[1-\cos^2\theta_\text{i}\cos^2\theta_{\text{CR}}-\frac{1}{2}\sin^2\theta_\text{i}\sin^2\theta_{\text{CR}}\right]\,,
\end{split}
\end{equation}
where we introduce $\cos \theta_\text{i} = p_{\text{i},z}/|\vec{p}_\text{i}|$, and $\beta_\text{i}=c p_{\text{i},z}/E_\text{i}=v/c$ (electron's velocity divided by the speed of light),  $\cos \theta_\gamma=k_z/|\vec{k}_j|$ and the conventional Cherenkov emission angle $\cos\theta_{\text{CR}}=({\beta_\text{i} \tilde n})^{-1}$; we also used the relations S25 in~\cite{2016Kaminer}. When the incident beam is only along the $z$ axis, i.e. $\theta_\text{i}=0$, the emission rate is given by
\begin{equation}\label{eq:emission_rate_low_photon}
    \begin{split}
\Gamma_\omega 
\approx &  \alpha \beta_\text{i}  \sin^2(\theta_{\text{CR}})\,,
    \end{split}
\end{equation}
which is exactly the classical result. 

Since this test of the approximations to leading order satisfactorily recovers the classical limit, we can apply them, and notice that our reduced density matrix corresponds to that of a pure state. The final state after the Cherenkov process under the limit of infinite transversal length and interaction time is given by:
\begin{equation}
    \begin{aligned}
\ket{\psi_{\mathrm{e}\mhyphen\gamma}}
=& \int d^3k dp_{z}dp_{\text{i},z} \psi_G(p_{\text{i}, z})\sqrt{\frac{\alpha c^3}{2\pi^2 L_\perp^2 \tilde{n}^2}}\frac{2c p_{\text{i}, z}}{\sqrt{4 \omega E_\text{i}(E_\text{i}-\hbar \omega)}}\frac{k_{\perp}}{k}\delta \left(\omega-\frac{1}{\hbar}(E_\text{i}-E_\text{f})\right) \\
& \times L_z \text{sinc}\left(\left(\frac{1}{\hbar}(p_{\text{i},z}-p_{z})-k_z\right)\frac{L_z}{2}\right)\ket{(-\hbar k_x, -\hbar k_y, p_{z}), \vec{k}}.
    \end{aligned}
\end{equation}
Then, employing the energy conservation condition, it is possible to solve for $p_z$, as
\begin{equation}\label{eq:pzsol_leading}
    \begin{aligned}
0=&~ \omega-\frac{E_\text{i}}{\hbar}+\frac{1}{\hbar}\sqrt{c^2\hbar^2 k_x^2+c^2\hbar^2 k_y^2+c^2p_z^2+m_{\mathrm{e}}^2c^4}\,,\\
p_z^{\text{con}}=&~ \frac{1}{c}\sqrt{c^2p_{\text{i},z}^2-2\hbar \omega E_\text{i} +(\hbar \omega)^2-c^2\hbar^2 k_x^2-c^2\hbar^2 k_y^2}\\
\approx &~ p_{\text{i}, z}-\frac{\hbar \omega E_\text{i}}{c^2 p_{\text{i}, z}}    { = p_{\text{i}, z}-\frac{\hbar k E_\text{i}}{c p_{\text{i}, z}\tilde{n}} }\,,
    \end{aligned}
\end{equation}
leading to 
\begin{equation}
  \delta\left(\omega-\frac{1}{\hbar}(E_\text{i}-E_\text{f})\right)  = \frac{\hbar E_\text{f}}{c^2p_z^{\text{con}}}\delta(p_z-p_z^{\text{con}}).  
\end{equation}
This expression allows us to rewrite the final state as:
\begin{equation}\label{eq:final_state_finiteLz}
    \begin{aligned}
\ket{\psi_{\mathrm{e}\mhyphen\gamma}}= &\int d^3k dp_{\text{i},z} \frac{1}{L_\perp} \mathcal{C}(\vec{k}, p_{\text{i}, z}) \ket{(-\hbar k_x, -\hbar k_y, p_{z}^{\text{con}}), \vec{k}}.
    \end{aligned}
\end{equation}
where
\begin{equation}
    \begin{aligned}
\mathcal{C}(\vec{k}, p_{\text{i}, z})=& \psi_G(p_{\text{i}, z})\sqrt{\frac{\alpha c^3}{2\pi^2 \tilde{n}^2}}\frac{2c p_{\text{i}, z}}{\sqrt{4 \omega E_\text{i}(E_\text{i}-\hbar \omega)}}\frac{k_{\perp}}{k}  L_z \text{sinc}\left(\left(\frac{1}{\hbar}(p_{\text{i},z}-p^{\text{con}}_{z})-k_z\right)\frac{L_z}{2}\right)\frac{\hbar (E_\text{i}-\hbar \omega)}{c^2p_z^{\text{con}}}\\
\approx & \psi_G(p_{\text{i}, z}) \sqrt{\frac{\alpha \hbar ^2 }{2\pi^2 \tilde{n}}}\frac{k_\perp}{k^{3/2}}  L_z \text{sinc}\left(\left(\frac{1}{\hbar}(p_{\text{i},z}-p^{\text{con}}_{z})-k_z\right)\frac{L_z}{2}\right).
    \end{aligned}
\end{equation}
 where we use in the second line a leading order approximation such that
\begin{equation}
    \begin{aligned}
\frac{p_{\text{i},z}}{p_z^{\text{con}}} \frac{E_\text{i}-\hbar \omega}{\sqrt{\omega E_\text{i} (E_\text{i}-\hbar \omega)}}&\approx\frac{1}{\sqrt{\omega}}=\sqrt{\frac{\tilde{n}}{c k}}.
    \end{aligned}
\end{equation}
In the following section, the normalizability of this state is studied when the emission rate is calculated.

\subsection{Emission Angle Profile and Normalizability of the Electron-Photon Pair State}\label{ap:emmsion_angle}

 Based on our model for the Cherenkov pair wave function, we analyze the emission angle profile by realising that the total probability is proportional to the integral over the absolute square of the scattering matrix 
\begin{equation} \label{eq:proba_emission_original}
    \begin{aligned}
\mathcal{P} & { = \sum_{s_\text{f}, \epsilon} \int d^3 p \int d^3 k \,  S_{\text{f}\,\text{i}}^{\epsilon, s_\text{i}, s_\text{f}}(\vec{p}_\text{i}, \vec{p}, \vec{k}) S_{\text{f}\,\text{i}}^{\epsilon, s_\text{i}, s_\text{f}}(\vec{p}_\text{i}, \vec{p}, \vec{k})^* }{\propto}  \langle \psi_{\mathrm{e}\mhyphen\gamma} \ket{\psi_{\mathrm{e}\mhyphen\gamma}}\\
&\approx \int d^3k_1 dp_{\text{i},z}\int d^3k_2 dp'_{\text{i},z} \frac{1}{L^2_\perp}  \mathcal{C}(\vec{k}_1, p_{\text{i}, z}) \mathcal{C}(\vec{k}_2, p'_{\text{i}, z})^* \langle (-\hbar k_{2, x}, -\hbar k_{2, y}, p_{2, z}^{\text{con}}), \vec{k}_2\ket{(-\hbar k_{1, x}, -\hbar k_{1, y}, p_{1, z}^{\text{con}}), \vec{k}_1} \\
& \approx \int d^3k_1 dp_{\text{i},z}\int d^3k_2 dp'_{\text{i},z} \mathcal{C}(\vec{k}_1, p_{\text{i}, z}) \mathcal{C}(\vec{k}_2, p'_{\text{i}, z})^*\frac{1}{\hbar^2}\delta (\vec{k}_1-\vec{k}_2)\delta(p_{1, z}^{\text{con}}-p_{2, z}^{\text{con}}),
    \end{aligned}
\end{equation}
where we use a similar identity  as in Eq.~\eqref{eq:doubledeltaremoval}
for the double delta functions in $x$ and $y$ momenta.
Notice that the last delta function in Eq.~\eqref{eq:proba_emission_original} under the condition $\vec{k}_1=\vec{k}_2$ and approximated to leading order in its expansion over $\hbar\omega/E_\text{i}$ is given by
\begin{equation}
  \begin{aligned}
\delta(p_{1, z}^{\text{con}}-p_{2, z}^{\text{con}})\approx & \delta \left(p_{\text{i}, z}-\frac{\hbar \omega E_\text{i}}{c^2 p_{\text{i}, z}}- p'_{\text{i}, z}+\frac{\hbar \omega E'_\text{i}}{c^2 p'_{\text{i}, z}}\right)\\
\approx & \left(1+\frac{\hbar \omega }{E_\text{i}\gamma_\text{i}^2\beta_\text{i}^2}\right)^{-1}\delta(p_{\text{i}, z}-p'_{\text{i}, z}) \approx \delta(p_{\text{i}, z}-p'_{\text{i}, z}),
  \end{aligned}  
\end{equation}
where $\gamma_\text{i}=(1-\beta_\text{i}^2)^{-1/2}$.
Therefore, we obtain
\begin{equation} \label{eq:proba_emission_original2}
    \begin{aligned}
\mathcal{P}{\approx} & \int d^3k d p_{\text{i}, z} \frac{1}{\hbar^2} \mathcal{C}(\vec{k}, p_{\text{i}, z}) \mathcal{C}(\vec{k}, p_{\text{i}, z})^*\\
\approx & \int d^3k  \frac{\alpha}{2\pi^2\tilde n}\frac{k_{\perp}^2}{k^3} \left[\int d p_{\text{i}, z} |\psi_G(p_{\text{i}, z})|^2L_z^2 \text{sinc}^2\left(\left(\frac{k E_\text{i}}{c p_{\text{i}, z} \tilde n}-k_{z}\right)\frac{L_z}{2}\right) \right].
    \end{aligned}
\end{equation}
If we apply an additional approximation regarding the narrow distribution in momentum of the incident electron beam, we find
\begin{equation}
    \frac{E_\text{i}}{cp_{\text{i}, z}}\approx\beta^{-1}-{ \frac{p_{\text{i}, z}-\bar{p}_{\text{i}, z}}{\beta \gamma_L^2 \bar{p}_{\text{i}, z}}}\,,
\end{equation}
in which the following variables are given by the mean momentum in $z$ ($\bar{p}_{\text{i}, z}$), such that the mean energy $E_0=\sqrt{c^2\bar{p}_{\text{i}, z}^2+m_{\mathrm{e}}^2c^4}$, $\beta=\frac{c \bar{p}_{\text{i}, z}}{E_0}$ and the Lorentz factor $\gamma_L = \frac{E_0}{m_{\mathrm{e}}c^2}$. In addition,
\begin{equation}\label{eq:approx_sinc}
    \text{sinc}\left(\left(\frac{k E_\text{i}}{c p_{\text{i}, z} \tilde n}-k_{z}\right)\frac{L_z}{2}\right)\approx  \text{sinc}\left(\left(\frac{k }{\beta \tilde n}-k_{z}\right)\frac{L_z}{2}\right)+ {\mathcal{O}\left(\frac{k L_z(p_{\text{i},z}-\bar{p}_{\text{i}, z})}{\beta \tilde{n}\gamma_L^2\bar{p}_{\text{i}, z}}\right)} \,,
\end{equation}
but with the integral over $p_{\text{i},z}$, only the second order terms remain  which can be considered to be small, as for plane waves and the parameters we have set, the width of the Gaussian typically satisfies $\Delta p / \bar{p}_{\text{i}, z} \ll \frac{\beta \tilde{n} \gamma_L^2}{ k L_z}$.
Consequently, the resulting correction is significantly smaller than the previously neglected terms. Given the chosen parameters, these correction terms are squared, leading to an expected order of magnitude of approximately $10^{-10}$. In contrast, the previously neglected terms scale as the transverse photon momentum relative to $\bar{p}_{\text{i}, z}$, which is on the order of $10^{-6}$.
It is important to note that the argument of the sinc function vanishes when $\theta_\gamma = \arcsin(k_{\perp}/k) = \theta_{\text{CR}}$, which precisely corresponds to the non-vanishing condition in the limit $L_z \to \infty$. This Cherenkov angle arises naturally from the constraints of energy and momentum conservation.

As a consequence the emission probability corresponds to 
\begin{equation} \label{eq:proba_emission}
    \begin{aligned}
\mathcal{P}{\approx} &\frac{\alpha}{2\pi^2{\tilde n}} \int d^3k    \frac{k_{\perp}^2}{k^{3}}L_z^2\text{sinc}^2\left(\left(\frac{k}{\beta \tilde n}-k_{z}\right)\frac{L_z}{2}\right) \\
\approx &\frac{\alpha }{\pi \tilde n} \int  k 
 \sin \theta_\gamma  dk d\theta_\gamma \sin^2\theta_\gamma L_z^2\text{sinc}^2\left(\left(\frac{k}{\beta \tilde n}-k\cos\theta_\gamma\right)\frac{L_z}{2}\right)\,.
    \end{aligned}
\end{equation}
In our study, we consider a photon energy filter that limits the wave vector magnitude of the photon by defining a $k_{\text{min}}=\frac{\tilde n}{c}\omega_{\text{min}}$ and $k_{\text{max}}=\frac{\tilde n}{c}\omega_{\text{max}}$ as the photon energy is limited to $E_\gamma\in[\hbar \omega_{\text{min}}, \hbar \omega_{\text{max}}]$. In that case, the integral over $k$ is given by:
\begin{equation}
    \int_{k_{\text{min}}}^{k_{\text{max}}}~ kd k~ \text{sinc}^2(ky)=\frac{1}{2y^2}\left(\log\left(\frac{k_{\text{max}}}{k_{\text{min}}}\right)-\text{Ci}(2 k_{\text{max}} y)+\text{Ci}(2 k_{\text{min}} y)\right),
\end{equation}
where Ci is the cosine-integral function and $y=\frac{L_z}{2}\left(\cos\theta_\gamma -\frac{1}{\beta \tilde n}\right)$.
Hence, the probability is given by:
\begin{equation}
    \begin{aligned}
\mathcal{P}{\approx}\frac{2\alpha}{\pi \tilde n}\int d(\cos(\theta_\gamma)) \sin^2\theta_\gamma\frac{\log(\frac{k_{\text{max}}}{k_{\text{min}}})-\text{Ci}(k_{\text{max}}L_z\left(\cos\theta_\gamma-\frac{1}{\beta \tilde n}\right))+\text{Ci}(k_{\text{min}}L_z\left(\cos\theta_\gamma-\frac{1}{\beta \tilde n}\right))}{\left(\cos\theta_\gamma -\frac{1}{\beta \tilde n}\right)^2}\,,
    \end{aligned}
\end{equation}
and angular emission profile 
\begin{equation}\label{eq:prob_cherenkov_emmission}
    \frac{d\mathcal{P}}{d\Omega}{\approx}\frac{\alpha}{\pi^2 \tilde n}\sin^2\theta_\gamma \frac{\log(\frac{k_{\text{max}}}{k_{\text{min}}})-\text{Ci}(k_{\text{max}}L_z\left(\cos\theta_\gamma-\frac{1}{\beta \tilde n}\right))+\text{Ci}(k_{\text{min}}L_z\left(\cos\theta_\gamma-\frac{1}{\beta \tilde n}\right))}{\left(\cos\theta_\gamma -\frac{1}{\beta \tilde n}\right)^2},
\end{equation}
where $\Omega$ is the solid angle.
This angular emission profile can be also calculated outside the sample under the assumption that the transmission at the lower boundary is independent of the Cherenkov emission processes, such that:
\begin{equation}\label{eq:prob_cherenkov_emmission_out}
    \frac{d\mathcal{P}_{\text{out}}}{d\Omega}=\frac{d\mathcal{P}}{d\Omega}\mathcal{T}_\parallel(\theta_\gamma)\,,
\end{equation} 
where the transmission probability of p-polarised light is
\begin{equation}
    \mathcal{T}_\parallel(\theta_\gamma) = \frac{4 \tilde{n} \cos\theta_\gamma\cos\theta_t}{(\cos \theta_\gamma + \tilde{n} \cos \theta_t)^2}\,,
\end{equation}
with the transmitted angle $\theta_t$ given by $\sin \theta_t = \tilde{n} \sin \theta_\gamma$ (see Chapter 4.6 in \cite{hecht2002optics}). Since the radiation emitted in the Cherenkov 
effect is radially polarised in the low-energy photon regime, as previously shown, it corresponds to p-polarised light when viewed from the perspective of incidence on the dielectric material~\cite{yamamoto1996}. 

\begin{figure}[h]
    \centering
    \includegraphics[width=0.7\linewidth]{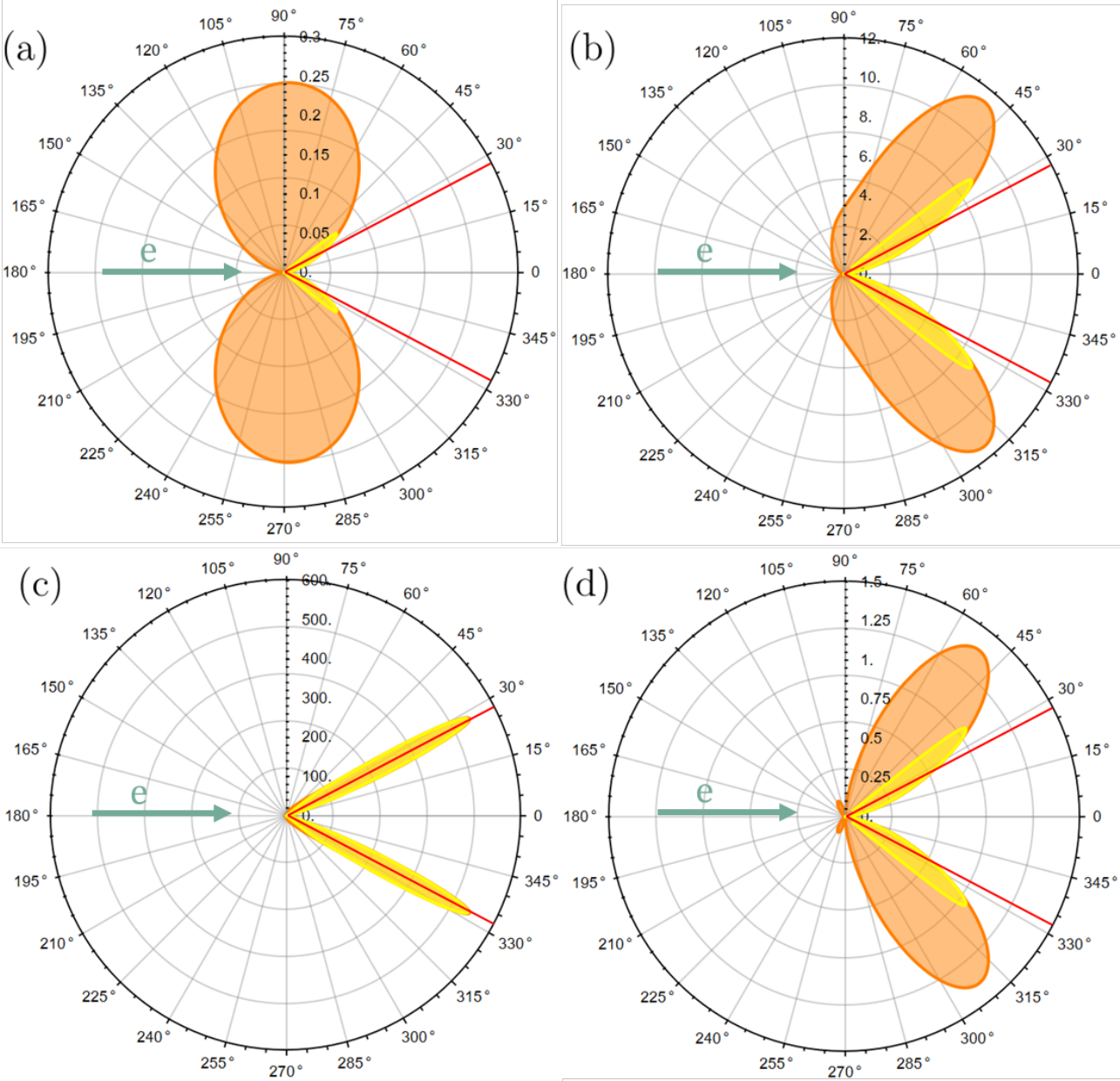}
    \caption{Angular probability density profile of Cherenkov radiation emission as function of $\theta_\gamma$ inside the dielectric material (in orange) according to Eq.~\eqref{eq:prob_cherenkov_emmission} and the angular profile of the portion of radiation that is transmitted outside the sample (in yellow) according to Eq.~\eqref{eq:prob_cherenkov_emmission_out}. Both are in units of $\frac{\alpha}{\pi^2 \tilde n}$ with a sample thickness $L_z=200$ nm for a different energy windows set by (a) $k_{\text{min}}L_z=0$ and $k_{\text{max}}L_z=1$ (for $\hbar \omega= 0.28$ eV), (b) $k_{\text{min}}L_z=0$ and $k_{\text{max}}L_z=10$ (for $\hbar \omega= 2.8$ eV), (c) $k_{\text{min}}L_z=0$ and $k_{\text{max}}L_z=100$ (for $\hbar \omega= 28$ eV), and (d) $k_{\text{min}}L_z=6.2$ (correspondent to $\hbar \omega= 3.5$ eV) and $k_{\text{max}}L_z=7.1$ (for $\hbar \omega= 4.0$ eV). The emission probabilities have a different radial scale, as a larger energy window results in increased radiation emission, which tends to be more concentrated near the Cherenkov angle $\theta_{\text{CR}}$, which is represented by the red lines. In graph (c), the emission profile inside the dielectric material overlaps with the transmitted emission. This indicates that a significant portion of the generated light is successfully transmitted into the vacuum. We set the values $\beta = 0.7$, and $\tilde{n} = 1.6$, resulting in $\theta_{\text{CR}} = 26.77^\circ$ and critical angle $\theta_{\text{crit}}=38.7^\circ>\theta_{\text{CR}}$.
}
    \label{fig:ang-profile}
\end{figure}

Considering the angular probability density profile of the radiation transmitted outside the sample, represented by the yellow regions in Fig.~\ref{fig:ang-profile}, we observe that the critical angle for total internal reflection restricts the amount of Cherenkov radiation that can escape.  However, within the range $\theta_{\text{CR}}\leq\theta_\gamma\leq \theta_{\text{crit}}$, the angular probability density profile of the transmitted radiation closely matches that of the Cherenkov radiation emission inside the dielectric. This is reflected in the full overlap of the orange and yellow regions, indicating that the probability is not significantly reduced in this angular range due to transmission across the interface.
This constraint reduces the amount of detectable radiation. Furthermore, a distribution peaked at the Cherenkov angle transmits the majority of the radiation, as shown in Fig.~\ref{fig:ang-profile}~c, while also restricting the range of possible wave numbers along $z$. This implies that increasing the energy range or sample thickness could enhance the detected photon counts from Cherenkov processes. Indeed, without an energy filter, allowing a broad energy range results in a pronounced peak at the Cherenkov angle, as inferred from the trend in Fig.~\ref{fig:ang-profile}.

Cherenkov radiation is frequently detected at optical or near-UV energies, and TEM sample thicknesses are typically up to hundreds of nanometers, consistent with the inelastic mean free path of electrons travelling through the samples~\cite{yamamoto1996}. To explore a more realistic scenario, we consider photon energies ranging from 3.5 eV to 4.0 eV. For these energies, we find $k_{\text{min}} L_z = 6.2$ and $k_{\text{max}} L_z = 7.1$, using a dielectric slab with a thickness of $L_z = 200$ nm. The resulting angular profile, shown in Fig.~\ref{fig:ang-profile}~d, is as narrow as the profile in Fig.~\ref{fig:ang-profile}~b and follows the trend of a distribution tilted toward $\theta_\gamma = \theta_{\text{CR}}$, albeit with a reduced magnitude due to the narrower energy range considered.

Although a significant portion of the radiation produced is transmitted, this does not guarantee an identical $k_z$ value for all emitted radiation. This condition is necessary to relate deflection angles to momentum, as will be discussed in Appendix~\ref{ap:deflection_vs_momenta}. 

Note that the probabilities are related to the normalization constant of our electron-photon state. For the entire angular range, the experimental parameters and energy filter values considered in Fig.~\ref{fig:ang-profile}~d (identical to those used in the main text) yield a finite normalization constant outside the dielectric of $\mathcal{P}_{\text{out}}=5.65 \times 10^{-5}$. This result demonstrates that a finite slab thickness and a Gaussian wave packet in the $z$-direction are sufficient to ensure a normalizable state suitable for further analysis.

\subsection{Reduced Density Operator in $x$-Momentum}\label{ap:red_density_x}

Based on our results above, we work with a relatively narrow energy window so that the emitted radiation is closer to the Cherenkov angle and is transmitted outside the dielectric in a significant proportion. Therefore, our model for the unnormalised final state of the Cherenkov pair, including an energy filter in the range of $\hbar \omega_{\text{min}}$ to $\hbar \omega_{\text{max}}$ and the approximation of the final electron $z$-momentum in Eq.~\eqref{eq:pzsol_leading}, is given by 
\begin{equation}\label{eq:psi_egamma_in}
    \begin{split}
        \hat{\tilde{\rho}}_{\mathrm{e}\mhyphen\gamma}=&\ket{\tilde{\psi}_{\mathrm{e}\mhyphen\gamma}}\bra{\tilde{\psi}_{\mathrm{e}\mhyphen\gamma}},\\
        \ket{\tilde{\psi}_{\mathrm{e}\mhyphen\gamma}}=& \int d^3k dp_{\text{i},z} \frac{1}{L_\perp} \mathcal{C}(\vec{k}, p_{\text{i}, z}) \eta({k}) \ket{(-\hbar k_x, -\hbar k_y, p_{\text{i}, z}-\frac{\hbar kE_\text{i}}{c p_{\text{i}, z} \tilde n}), \vec{k}},
    \end{split}
\end{equation}
in which the energy filter function
\begin{equation}
    \eta({k})=\left\{\begin{matrix}
        1, \quad& \omega_{\text{min}}\leq \frac{c}{\tilde n}{k}\leq \omega_{\text{max}} \\
        0, \quad&\text{otherwise}
    \end{matrix}\right..
\end{equation}
However, the state in Eq.~\eqref{eq:psi_egamma_in} describes the Cherenkov pair inside the dielectric slab. Due to the slab's thickness, we assume that, for the electron, multiple scattering processes are negligible, the lower-layer interface is transparent, and momentum is conserved. For the photon, we can define a map to model the refraction of light as it exits the dielectric material. 
This transformation is based on the law of refraction and the Fresnel coefficients, assuming a planar lower interface, which results in a probability that depends on the incidence angle $\theta_\gamma$ for $\theta_\gamma=\arccos{\left(\frac{\vec{k}\cdot \vec{n}}{|\vec{k}|}\right)} \leq \theta_{\text{crit}}$ (where $\theta_{\text{crit}}$ is the critical angle for total internal reflection and $\vec{n}$ is the surface normal vector), and zero otherwise. Mathematically, this unitary map in the subset of transmitted radiation is expressed as:
\begin{equation}
    \mathcal{\hat{M}} = \int_{\theta_\gamma \leq \theta_{\text{crit}}} d^3k \, {\tau_\parallel(\theta_\gamma)}\ket{k_x, k_y, k'_z}\bra{\vec{k}},
\end{equation}
where the transformed longitudinal wave vector is given by $k'_z = \frac{1}{\tilde{n}} \sqrt{(1 - \tilde{n}^2)(k_x^2 + k_y^2) + k_z^2}$, since the longitudinal component is reduced when the light enters the vacuum while the transverse components are conserved. Additionally, the transmission Fresnel coefficient  
\begin{equation}
    \tau_\parallel(\theta_\gamma) = \frac{2\tilde{n}\cos\theta_\gamma}{\left(\cos\theta_\gamma + \tilde{n} \sqrt{1 - \tilde{n}^2 \sin^2 \theta_\gamma}\right)}
\end{equation}
defines the probability for detecting light with a given momentum outside the dielectric slab (see Chapter 4.6 in~\cite{hecht2002optics}). This coefficient corresponds to p-polarised light (parallel to the plane of incidence) since Cherenkov radiation, to leading order in scattering theory, is emitted with radial polarization, as explicitly shown in~\cite{yamamoto1996}. The radial polarization vector lies in  the plane formed by the normal vector of the surface and the wave vector $\vec{k}$.

The resulting state outside the dielectric slab is:
\begin{equation}
    \begin{aligned}
        \ket{\tilde{\psi}_{\mathrm{e}\mhyphen\gamma, \text{out}}} &= \mathcal{\hat{M}} \ket{\tilde{\psi}_{\mathrm{e}\mhyphen\gamma}} \\
        &= \int_{\theta_\theta \leq \theta_{\text{crit}}} d^3k \, dp_{\text{i},z} \, \frac{1}{L_\perp} \mathcal{C}(\vec{k}, p_{\text{i}, z}) \eta(k) \tau_\parallel(\theta_\gamma)\big|-\hbar k_x, -\hbar k_y, p_{\text{i}, z} - \frac{\hbar k E_\text{i}}{c p_{\text{i}, z} \tilde n}\big\rangle \otimes \ket{k_x, k_y, k'_z},
    \end{aligned}
\end{equation}
where the transmission outside the dielectric includes a dependence on the incidence angle $\theta_\gamma$ and projects the momentum state onto the subspace of incident angles that are not totally internally reflected. 
Hence, after applying this energy filtering in momentum representation, the density matrix corresponds to
\begin{equation}
    \begin{aligned}
\hat{\tilde{\rho}}_{\mathrm{e}\mhyphen\gamma, \text{out}}\approx &\int d^3k_1 d^3k_2 dp_{\text{i},1, z} dp_{\text{i},2, z} g(\vec{k}_1, p_{\text{i}, z}; \vec{k}_2,  p'_{\text{i}, z}) \\
&\big|-\hbar k_{1, x}, -\hbar k_{1, y}, p_{\text{i},1, z}-\frac{\hbar k_1E_{\text{i},1}}{c p_{\text{i},1,z} \tilde n}\big\rangle\otimes \ket{k_{1,x}, k_{1,{y}}, k'_{1, z}}\big\langle-\hbar k_{2, x}, -\hbar k_{2, y}, p_{\text{i},2, z}-\frac{\hbar k_2E_{\text{i},2}}{c p_{\text{i},2, z} \tilde n}\big|\otimes \bra{k_{2,x}, k_{2,{y}}, k'_{2, z}},
    \end{aligned}
\end{equation}
where 
\begin{align}\label{eq:kprimej}
    k'_{j, z}=&\frac{1}{\tilde{n}} \sqrt{(1 - \tilde{n}^2)(k_{j, x}^2 + k_{j, y}^2) + k_{j, z}^2}\quad\text{and}\\
    E_{\text{i},j} =& \sqrt{m^2_e c^4 + p^2_{\text{i},j,z}c^2}
\end{align}
with $j=\{1,2\}$. In addition, the amplitudes correspond to
\begin{equation}
    \begin{aligned}
g(\vec{k}_1, p_{\text{i},1, z}; \vec{k}_2,  p'_{\text{i},2, z})= &\frac{1}{L^2_\perp} \mathcal{C}(\vec{k}_1, p_{\text{i},1, z}) \eta({k_1}) \tau_\parallel(\theta_{1, \gamma}) \mathcal{C}(\vec{k}_2, p_{\text{i},2, z}) \eta({k_2}) \tau_\parallel(\theta_{2, \gamma})\\
    =&\tilde{g}(\vec{k}_1 ; \vec{k}_2)\psi_G(p_{\text{i},1,z})\psi_G(p_{\text{i},2, z}),
    \end{aligned}
\end{equation}
where $\cos({\theta_{j, \gamma}})=\frac{k_{j,z}}{k_j}$. 
In the second line, we make explicit the presence of the Gaussian wavepackets as contributors to the incident momentum by defining
\begin{align}
    \tilde{g}(\vec{k}_1 ; \vec{k}_2):=\frac{\hbar^2}{L_\perp^2} \mathcal{A}(\vec{k}_1)\mathcal{A}(\vec{k}_2)\tau_\parallel(\theta_{1, \gamma}) \tau_\parallel(\theta_{2, \gamma})\eta({k_1})\eta({k_2}),     
\end{align}
with the Cherenkov amplitudes
\begin{equation}
    \mathcal{A}(\vec{k})=\sqrt{\frac{\alpha}{2\pi^2\tilde n}}\frac{k_{\perp}}{k^{3/2}}L_z\text{sinc}\left(\frac{L_z}{2}(k_{z}-\frac{k}{\beta \tilde{n}})\right),
\end{equation}
in which $k=\sqrt{k_{x}^2+k_{y}+k^{2}_z}$, and $k_{\perp}=\sqrt{k_{x}^2+k_{y}^2}$. The sinc function argument is approximated using Eq.~\eqref{eq:approx_sinc}.

Given that the entanglement certification procedure is applied solely to the arbitrarily chosen variable $x$ (transverse to the propagation of the electron beam) and its conjugate momentum, we trace the $y$ and $z$ momenta of both particles. This trace yields the following unnormalised reduced density matrix 
\begin{equation}\label{eq:rhoX_constraint_pre_first}
    \begin{aligned}
\hat{\tilde \rho}_{\mathrm{e}\mhyphen\gamma, x}\approx&\text{Tr}_{p_y, k_y, p_z, k_z}\left(\hat{\tilde{\rho}}_{\mathrm{e}\mhyphen\gamma, \text{out}}\right)
=&{ \int dk_{1, x} \int dk_{2, x} \frac{\hbar}{L_\perp} {f}(k_{1, x}, k_{2, x}) \ket{-\hbar k_{1, x}, k_{1,x}}\bra{-\hbar k_{2, x}, k_{2, x}}} \,,
    \end{aligned}
\end{equation}
where the amplitudes are
\begin{equation}
    \begin{aligned}
    f(k_{1, x}, k_{2, x})={\int dk_{1, y} dk'_{1, z}\int dk_{2, y} dk'_{2, z} \left|\frac{\tilde{n}^2 k'_{1,z}}{k_{1,z}}\right| \left|\frac{\tilde{n}^2 k'_{2,z}}{k_{2,z}}\right|   \frac{L_\perp}{\hbar}} & R_0((k_{1,x}, k_{1,y}, k_{1,z});(k_{2,x}, k_{2,y}, k_{2,z}))\\
    &\times R_G((k_{1,x}, k_{1,y}, k_{1,z});(k_{2,x}, k_{2,y}, k_{2,z}) \,,
\end{aligned}
\end{equation}
in which $d^3k'_j=dk_{j,x}dk_{j,y}dk'_{j,z}$ and the prefactors arise from the Jacobian of the transformation of wavevector variables from inside to outside the dielectric slab, where $k_{j,z}=\sqrt{(\tilde{n}^2-1)(k_{j,x}^2+k_{j,y}^2)+\tilde{n}^2k^{'2}_{j,z}}$ is a function of $k_{j,x}$, $k_{j,y}$ and $k'_{j,z}$.
In addition, the trace over $y$ and $z$ momentum of both particles lead to the following functions
\begin{subequations}
    \begin{align}
    \begin{split}
R_0(\vec{k}_1; \vec{k}_2)=&\int dp_y dk_y dk_z \tilde{g}(\vec{k}_1; \vec{k}_2)\langle p_y, k_y |-\hbar k_{1, y}, k_{1, y} \rangle \langle -\hbar k_{2, y}, k_{2, y}  | p_y, k_y\rangle \langle k_z | k'_{1, z} \rangle \langle  k'_{2, z}  | , k_z\rangle \\
=& \frac{L_\perp}{\hbar} \tilde{g}(\vec{k}_1; \vec{k}_2) \delta(k_{1, y} - k_{2, y}) \delta(k'_{1, z} - k'_{2, z})\,,     \end{split}\label{eq:R0_integral}\\ 
\begin{split}
   R_G (\vec{k}_1; \vec{k}_2)=& \int dp_{\text{i},1, z} dp_{\text{i},2,z} dp_z  \psi_G(p_{\text{i},1,z})\psi_G(p_{\text{i},2,z}) \langle p_z \big|p_{\text{i},1, z}-\frac{\hbar k_1E_{\text{i},1}}{c p_{\text{i},1, z} \tilde n} \big\rangle \big\langle p_{\text{i},2, z}-\frac{\hbar k_2E_{\text{i},2}}{c p'_{\text{i}, z} \tilde n} \big| p_z\rangle \\
=& \int dp_{\text{i},1, z} dp_{\text{i},2,z}   \psi_G(p_{\text{i},1,z})\psi_G(p_{\text{i},2, z}) \delta\left(p_{\text{i},1, z}-\frac{\hbar k_1E_{\text{i},1} }{c p_{\text{i},1, z} \tilde n} - p_{\text{i},2, z}+\frac{\hbar k_2E_{\text{i},2} }{c p_{\text{i},2, z} \tilde n}\right).
\end{split}  \label{eq:RG_integral} 
    \end{align}
\end{subequations}
Note that we have reduced the square of a delta function in Eq.~\eqref{eq:R0_integral} to a single delta function through the analogue of the identity in Eq.~\eqref{eq:doubledeltaremoval} for the $y$-momentum.
The overlap of the $z$ wave vector states in Eq.~\eqref{eq:R0_integral} imposes a condition of equality for the wave vectors along the $z$ direction outside the dielectric material.

The argument of the delta function in Eq.~\eqref{eq:RG_integral} can be rewritten using the fact that $p_{\text{i},j,z}$ is distributed according to a narrow Gaussian wavepacket which implies that $q_{\text{i},j} = p_{\text{i},j,z} - \bar{p}_{\text{i}, z}$ is much smaller than $\bar{p}_{\text{i}, z}$ since $\Delta p_z \ll \bar{p}_{\text{i}, z}$. As a consequence,
\begin{equation}\label{eq:delta_pi1zpi2z}
    \begin{aligned}
\delta\left(p_{\text{i},1, z}-\frac{\hbar k_1E_{\text{i},1}}{c p_{\text{i},1, z} \tilde n}- p_{\text{i},2, z}+\frac{\hbar k_2E_{\text{i},2}}{c p_{\text{i},2, z} \tilde n}\right)\approx& \delta \left(q_{\text{i},1}-q_{\text{i},2}+\frac{\hbar}{\gamma_L^2\beta \bar{p}_{\text{i}, z} \tilde n}(k_1q_{\text{i},1}-k_2q_{\text{i},2}) - \frac{\hbar(k_1-k_2)}{\beta \tilde n}\right)\\
\approx & \left|1+\frac{\hbar k_2}{\gamma_L^2\beta \bar{p}_{\text{i}, z}  \tilde n}\right|^{-1}\delta\left(q_{\text{i},2}-q_{\text{i},2}^{\text{con}}\right)\approx\delta\left(q_{\text{i},2}-q_{\text{i},2}^{\text{con}}\right),
    \end{aligned}
\end{equation}
where
\begin{equation}
q_{\text{i},2}^{\text{con}}\approx q_{\text{i},1}\left(1+\frac{\hbar(k_1-k_2)}{\gamma_L^2\beta \bar{p}_{\text{i}, z} \tilde n}\right)- \frac{\hbar(k_1-k_2)}{\beta \tilde n}.
\end{equation}
Applying Eq.~\eqref{eq:delta_pi1zpi2z} to Eq.~\ref{eq:RG_integral}, the Gaussian integrals from the incident electron wave function take the form:
\begin{equation}
    \begin{aligned}
R_G (\vec{k}_1; \vec{k}_2) 
\approx& \int dp_{\text{i},1, z}   \psi_G(p_{\text{i},1,z})\psi_G(\bar{p}_{\text{i}, z}+q_{\text{i},2}^{\text{con}})\\
\approx& \int d q_{\text{i}}\frac{\exp\left\{-\frac{q_{\text{i}}^2}{4\Delta p_z^2}\right\}}{\sqrt{2\pi \Delta p_z^2}}\exp\left\{-\frac{1}{4\Delta p_z^2}\left(q_{\text{i}}\left(1+\frac{\hbar(k_1-k_2)}{\gamma_L^2\beta \bar{p}_{\text{i}, z} \tilde n}\right)-\frac{\hbar(k_1-k_2)}{\beta \tilde n}\right)^2\right\}\\
=& \sqrt{\frac{2}{\sigma(k_1,k_2)}}\exp\left\{-\frac{\hbar^2(k_1-k_2)^2}{4\beta^2 \tilde n^2\Delta p_z^2\sigma(k_1,k_2)}\right\} \approx \exp\left\{-\frac{\hbar^2(k_1-k_2)^2}{8\beta^2 \tilde n^2\Delta p_z^2}\right\},
    \end{aligned}
\end{equation}
where the modified variance $\sigma(k_1,k_2)=1+\left(1+\frac{\hbar (k_1-k_2)}{\gamma_L^2 \beta \bar{p}_{\text{i}, z} \tilde n }\right)^2$ is approximated to leading order in the last step.
This factor refers to the overlap between two Gaussian distributions from the initial state and its dual with a certain deviation of each momentum value due to the energy transferred to the photon.

 With the above approximations, we can re-write the matrix elements $f(k_{1, x},k_{2, x})$ as  
\begin{equation}
    \begin{aligned}
    f(k_{1, x}, k_{2, x})
    =&{\int dk_{y} dk'_{z}\left|\frac{\tilde{n}^2 k'_{z}}{k_{1,z}}\right| \left|\frac{\tilde{n}^2 k'_{z}}{k_{2,z}}\right|   \frac{L^2_\perp}{\hbar^2} \tilde{g}((k_{1,x}, k_y, k_{1,z});(k_{2,x}, k_y, k_{2,z}))R_G((k_{1,x}, k_y, k_{1,z});(k_{2,x}, k_y, k_{2,z}))}\,,
\end{aligned}
\end{equation}
where $k_{j,z}=\sqrt{(\tilde{n}^2-1)(k_{j,x}^2+k_y^2)+\tilde{n}^2k^{'2}_{z}}$ is now a function of $k_{j,x}$, $k_{y}$ and $k'_{z}$.
As discussed before, totally reflected photons are not considered and do not appear in the final state outside the medium. Therefore, the $k'_z$-integral is performed over the positive part of the real axis. Furthermore, detected photon momenta are limited by an energy filter and the transverse momentum range $[-k_{x,\text{max}}, k_{x,\text{max}}] \times [k_{y,\text{min}}, k_{y,\text{max}}]$ that is covered by the detector. In the experiment, the wave function is not expected to extend beyond these limits. These conditions define boundaries for the integrals in the density matrix and $f(k_{1, x}, k_{2, x})$ that we discuss further below. We write
\begin{equation}\label{eq:rhoX_constraint_pre}
    \begin{aligned}
\hat{\tilde \rho}_{\mathrm{e}\mhyphen\gamma, x}\approx&\int_0^{k_{x, \text{max}}} dk_{1, x} \int_0^{k_{x, \text{max}}} dk_{2, x}{ \frac{\hbar}{L_\perp}} {f}(k_{1, x}, k_{2, x}) \left(\ket{-\hbar k_{1, x}, k_{1, x}}+\ket{\hbar k_{1, x}, -k_{1, x}} \right)\left(\bra{-\hbar k_{2, x}, k_{2, x}}+\bra{\hbar k_{2, x}, -k_{2, x}} \right),
    \end{aligned}
\end{equation}
{where we use the fact that the matrix elements $f(k_{1, x},k_{2, x})$ are symmetric under both, $k_{1, x}\rightarrow-k _{1, x}$ and $k_{2, x}\rightarrow-k _{2, x}$ separately. Furthermore,\begin{equation}\label{eq:felements_finitetiem}
    \begin{aligned}
f(k_{1, x}, k_{2, x})   
=&\int_{K_{y\mhyphen z}} dk_{y} dk'_{z}\mathcal{A}((k_{1,x}, k_y, k_{1,z}))\mathcal{A}((k_{2,x}, k_y, k_{2,z}))\mathcal{G}((k_{1,x}, k_y, k_{1,z});(k_{2,x}, k_y, k_{2,z}))\,\tau_\parallel(\theta_{\gamma, 1})\tau_\parallel(\theta_{\gamma, 2}),
    \end{aligned}
\end{equation}
where $\cos\theta_{\gamma, j}=\frac{k_{j,z}}{k_j}$, $k_j=\tilde{n}\sqrt{k_{j,x}^2+k_{y}+k^{'2}_{z}}$, and $K_{y\mhyphen z}=[k_{y,\text{min}},k_{y,\text{max}}]\times[0,\infty]$.
Meanwhile, the Jacobian from the change of variables outside the dielectric slab, the energy filter functions and the condition of a longitudinal Gaussian wavepacket leads to the contribution
\begin{equation}\label{eq:gaussian_spread_term}
 \mathcal{G}(\vec{k}_1, \vec{k}_2)=\eta(k_1)\eta(k_2)\frac{\tilde{n}^4 k^{'2}_z}{k_{1, z}k_{2,z}} \exp\left\{-\frac{\hbar^2(k_1 - k_2)^2}{8\beta^2 \tilde n^2 \Delta p_z^2}\right\}.   
\end{equation}
$\mathcal{G}(\vec{k}_1, \vec{k}_2)$ is a function centered at $k_1=k_2$, and in the limit of a narrow Gaussian wavepacket ($\Delta p_z\rightarrow 0$) is nonzero when $|k_{1,x}|=|k_{2,x}|$.

We find the correspondent normalised reduced density operator 
\begin{equation}\label{eq:rhoX_constraint}
    \begin{aligned}
\hat{\rho}_{\mathrm{e}\mhyphen\gamma, x}=& \frac{1}{\tilde N} \hat{\tilde{\rho}}_{\mathrm{e}\mhyphen\gamma, x}\\
\approx&\int_0^{k_{x, \text{max}}} dk_{1, x} \int_0^{k_{x, \text{max}}} dk_{2, x} {\frac{\hbar}{L_\perp}}\tilde{f}(k_{1, x}, k_{2, x}) \left(\ket{-\hbar k_{1, x}, k_{1, x}}+\ket{\hbar k_{1, x}, -k_{1, x}} \right)\left(\bra{-\hbar k_{2, x}, k_{2, x}}+\bra{\hbar k_{2, x}, -k_{2, x}} \right),
    \end{aligned}
\end{equation}
in which $\tilde{f}(k_{1, x}, k_{2, x})=\frac{1}{\tilde N} {f}(k_{1, x}, k_{2, x})$ with a normalization constant $\tilde{N}$. The normalization constant corresponds to:
\begin{equation}\label{eq:norm_Ntilde}
    \begin{aligned}
\tilde{N}=& \text{Tr}_{p_x, k_x}\left(\hat{\tilde{\rho}}_{\text{e}\mhyphen \gamma,x}\right)\\
\approx&\int dk_x dp_x \int_{-k_{x, \text{max}}}^{k_{x, \text{max}}} d k_{1, x} \int_{-k_{x, \text{max}}}^{k_{x, \text{max}}} d k_{2, x} \frac{\hbar}{L_\perp}f(k_{1,x}, k_{2,x})\bra{p_x, k_x}-\hbar k_{1, x}, k_{1,x}\rangle \bra{-\hbar k_{2, x}, k_{2, x}}p_x,k_x\rangle\\
=&\int_{-k_{x, \text{max}}}^{k_{x, \text{max}}} d k_{x} f(k_{x}, k_{x})\\
=&{\int_{K} d^3k' \frac{\alpha L_z^2}{2\pi^2 }\frac{k_{\perp}^2}{k^{3} }\frac{\tilde{n}^3 k^{'2}_z}{k^2_z}\text{sinc}^2\left(\frac{L_z}{2}(k_{z}-\frac{k}{\beta n})\right)\tau^2_\parallel(\theta_\gamma)\eta^2(k),}
    \end{aligned}
\end{equation}}
 in which clearly $k_{z}=\sqrt{(\tilde{n}^2-1)(k_{x}^2+k_y^2)+\tilde{n}^2k^{'2}_z}$ and $\cos\theta_{\gamma}=\frac{k_{z}}{k}.$

\subsection{Numerical Evaluation of Matrix Elements}

\begin{figure}[h]
    \centering
    \includegraphics[width=0.85\linewidth]{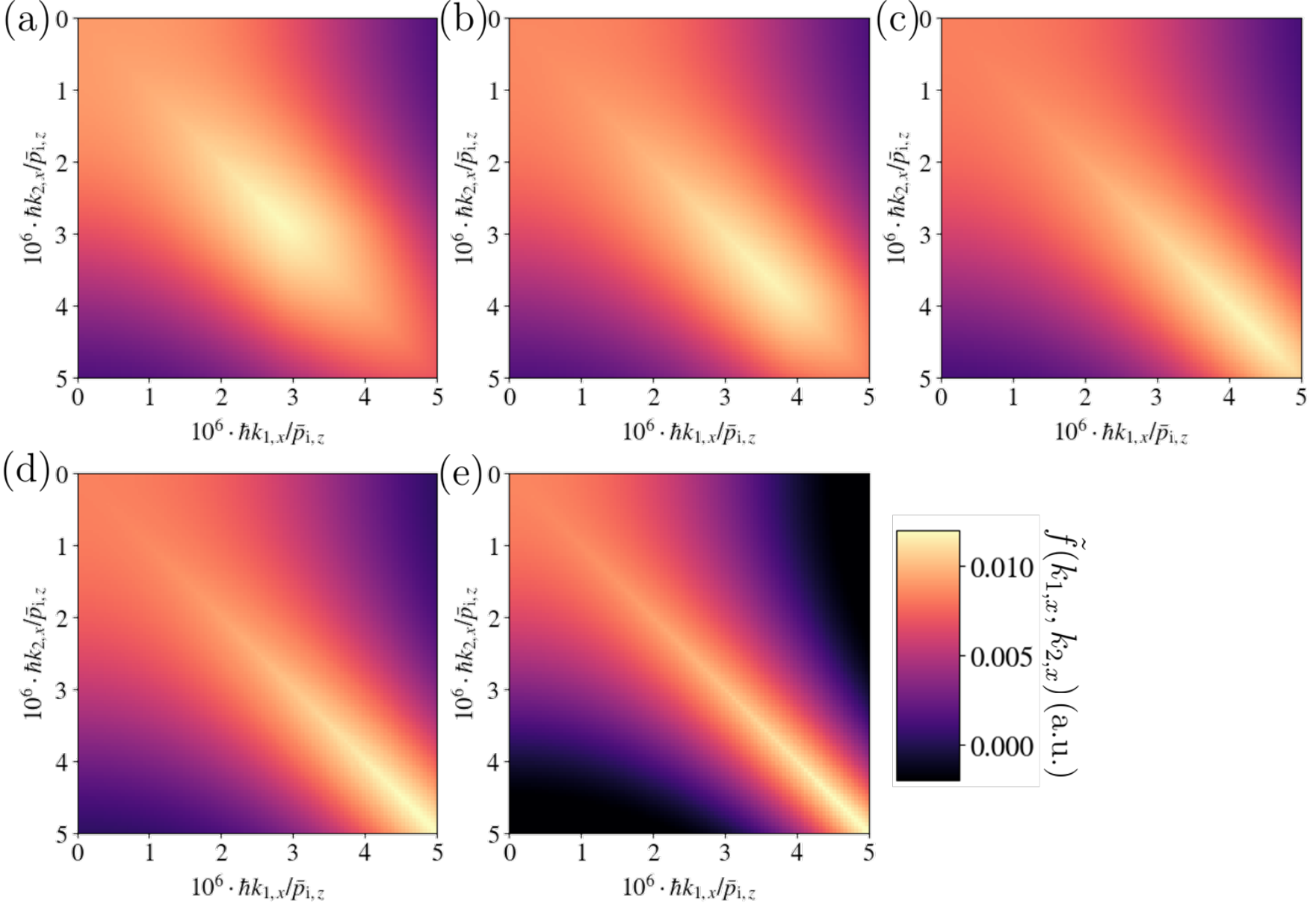}
    \caption{Reduced density matrix elements $\tilde{f}(k_{1, x}, k_{2, x})$ in $x$ momenta representation for different energy ranges (a) $E_\gamma\in [1.5, 4.0]$ eV, (b) $E_\gamma\in [2.0, 4.0]$ eV, (c) $E_\gamma\in [2.5, 4.0]$ eV, (d) $E_\gamma\in [3.0, 4.0]$ eV, (e) $E_\gamma\in [3.5, 4.0]$ eV.}
    \label{fig:rho_x_with_filters}
\end{figure}

The density matrix elements in Eq.~\eqref{eq:felements_finitetiem} are calculated numerically and depicted in Fig.~\ref{fig:rho_x_with_filters} for different energy filters. To this end, we take into account the momentum filter imposed by the measurement limits. Further limits in $k_{x}$ come from the fact that with a minimum momentum in $k_{y}$, $k_{x}$ is limited by the maximum energy possible. $k_y$ is constrained by the maximum possible $k$-vector amplitude in vacuum once we choose a certain $k_x$ and by the limits $k_{y, \text{min(max)}}$ set by the data analysis constraints. $k_z$ is also bounded in its upper and lower limit by the maximum and minimum $k$, given the energy filtering and the previous selection of $k_x$ and $k_y$.
Consequently, the limits are defined as follows:
\begin{equation}\label{eq:momentum_limits}
    \begin{aligned}
        |k_{1, x}|\leq &\text{min}\left( \sqrt{\left(\frac{\omega_{\text{max}}}{c} \right)^2-k_{y, \text{min}}^2}, k_{x, \text{max}}\right) \quad \text{and} \quad |k_{2, x}|\leq \text{min}\left( \sqrt{\left(\frac{\omega_{\text{max}}}{c} \right)^2-k_{y, \text{min}}^2}, k_{x, \text{max}}\right),\\
         &k_{y,\text{min}}\leq k_{y} \leq \text{min}\left( \sqrt{\left(\frac{\omega_{\text{max}}}{c}\right)^2-\max(|k_{1, x}|, |k_{2, x}|)^2}, k_{y, \text{max}}\right),\\
        &\sqrt{\max\left(\left(\frac{ \omega_{\text{min}}}{c}\right)^2 - \min(|k_{1, x}|, |k_{2, x}|)^2 - k_{y}^2, 0\right)} \leq k'_{z} \leq \sqrt{\left(\frac{\omega_{\text{max}}}{c}\right)^2 - \max(|k_{1, x}|, |k_{2, x}|)^2 - k_{y}^2},
    \end{aligned}
\end{equation}
where $k_y = k_{1,y} = k_{2,y}$ and $k'_z = k'_{1,z} = k'_{2,z}$ due to the delta functions in Eq.~\eqref{eq:R0_integral}. 
Additionally, the use of the max and min functions applied to $k_{1,x}$ and $k_{2,x}$ ensures that valid wavenumbers for $k_y$ and $k'_z$ exist within the specified energy range and imposed bounds.
The combination of these limits configures the region $K$ shown in Fig.~\ref{fig:simulations_model} of the allowed momenta for $k_{y, \text{min}}\geq 0$. For simplicity, we consider $\forall k_{1, x}, k_{2, x}$, $k_{x, \text{max}}< \sqrt{\left(\frac{\omega_{\text{max}}}{c} \right)^2-k_{y, \text{min}}^2}$, $k_{y,\text{min}}=0$, and $\sqrt{\left(\frac{\omega_{\text{max}}}{c}\right)^2-\max(|k_{1, x}|, |k_{2, x}|)^2}<k_{y, \text{max}}$. Taking into account those limits over $k_y$ and $k_z$ in Eq.~\eqref{eq:momentum_limits} for a given $k_{1 x}$ and $k_{2 x}$, we define an integration region $K_{y\mhyphen z}$.

Notably, the elements of the density matrix $\hat{\rho}_{\mathrm{e}\mhyphen\gamma, x}$ are primarily distributed along the diagonal $|k_{1, x}| = |k_{2, x}|$. The specific pattern of this distribution depends on the energy and momentum constraints, as well as the set of allowed values, which are graphically represented in Fig.~\ref{fig:simulations_model}. A narrow energy filter, as illustrated in Fig.~\ref{fig:rho_x_with_filters}~e, results in a more uniform probability distribution extending to higher values of momentum in the $x$-direction. In contrast, a broader energy range leads to a concentration of probabilities around values close to zero.
This behaviour can be attributed to the emission angle distribution depicted in Fig.~\ref{fig:ang-profile}, where angles between the Cherenkov angle and the critical angle are preferred. Additionally, the broader energy window allows for a larger range of permissible $\vec{k}$ values, which collectively increase the probability for $k_x$ values near zero compared to higher $k_x$. 

\begin{figure}[h]
    \centering
    \includegraphics[width=0.5\linewidth]{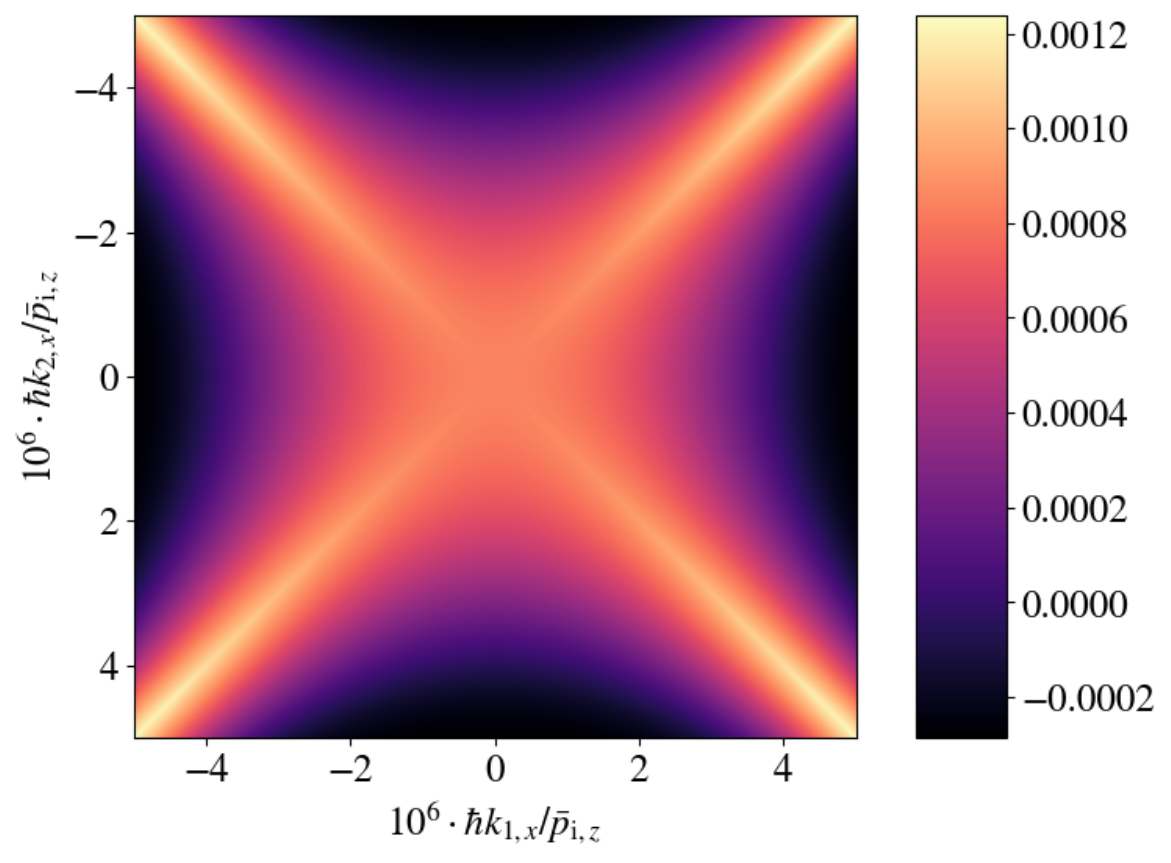}
    \caption{Matrix elements $\tilde{f}(k_{1, x}, k_{2, x})$ for the reduced electron-photon pair density matrix in the $x$ momenta of both particles for a photon energy window $E_\gamma=[3.5, 4.0]$ eV, and the momentum limits described in Eq.~\eqref{eq:momentum_limits}. The Gaussian wavepacket width of the incident electron is $\Delta p_z/\bar{p}_{\text{i}, z}=10^{-6}$. }
    \label{fig:density_E16}
\end{figure}

Due to the necessity of a narrow energy filter, which we will discuss in detail in Appendix~\ref{ap:deflection_vs_momenta}, we focus on the case chosen for the main text, with density matrix elements shown in Fig.~\ref{fig:density_E16}. The finite sample thickness, the Gaussian wave packet of the incident electron in the longitudinal direction, and refraction at the boundary all contribute to the presence of nonzero elements beyond the diagonal and anti-diagonal, as seen in Fig.~\ref{fig:density_E16}.
Specifically, in Eq.~\eqref{eq:felements_finitetiem}, the finite length in the $z$ direction is reflected by the sinc functions, while the incident Gaussian wave function introduces a Gaussian factor. The anti-diagonal elements specifically represent the coherences expected for an entangled state. Moreover, this structure resembles the maximally entangled state described in Sec.~\ref{sec:what_is_entanglement}, where the only nonzero entries are located along the diagonal and anti-diagonal.
In the limits $\Delta p_z \rightarrow 0$ and $L_z \rightarrow \infty$, these functions approach delta functions, ensuring perfect conservation of energy and momentum. Under such ideal conditions, the matrix elements are confined to the diagonal and anti-diagonal, with $f(k_x, k_x) = f(k_x, -k_x)$. This results in a density operator that represents a statistical mixture of maximally entangled momentum states within disjoint subspaces spanned by $\ket{k_x}$ and $\ket{-k_x}$. However, this idealised state would not be normalizable. The inclusion of a finite sample thickness and a Gaussian wave packet resolves this issue, as demonstrated in Appendix~\ref{ap:emmsion_angle}.

\subsection{Peres-Horodecki Criterion over the Electron-Photon Pair} \label{app:criterion}

In the context of bipartite systems, a sufficient condition for entanglement is the Peres-Horodecki Criterion~\cite{1998Horodecki}. It signals entanglement when the state's partial transpose leads to negative eigenvalues. In this case we take the partial transpose $\intercal_\gamma$ over the photon momentum space in  Eq.~\eqref{eq:rhoX_constraint} as follows:
\begin{equation}
\begin{split}
    \hat{\rho}_{\mathrm{e}\mhyphen\gamma, x}^{\intercal_\gamma}=&\int_{K_x} dk_{1, x}dk_{2, x} \frac{1}{L_\perp} \tilde{f}(k_{1, x}, k_{2, x}) \ket{-k_{1, x}}_{\mathrm{e}} \otimes \ket{k_{2, x}}_\gamma  \bra{-k_{2, x}}_{\mathrm{e}} \otimes \bra{k_{1, x}}_\gamma\,,
\end{split}
\end{equation}
where $k_x, k'_x\in K_x=[-k_{x, \text{max}}, k_{x, \text{max}}]$, and we add the sub labels $\mathrm{e}$ and $\gamma$ to specify the subspace of each state. The density matrix can be decomposed into orthogonal subspaces due to the orthogonality of the momentum eigenbasis. Specifically,
\begin{equation}\label{eq:transposed_rho_red}\allowdisplaybreaks
    \begin{aligned}
\hat{\rho}_{\mathrm{e}\mhyphen\gamma, x}^{\intercal_\gamma}=&2 k_{x, \text{max}}\int_{K_x} dk_{x} \frac{1}{L_\perp}\tilde{f}(k_{x}, k_{x}) \ket{-k_{x}}_{\mathrm{e}} \otimes \ket{k_{x}}_\gamma  \bra{-k_{x}}_{\mathrm{e}} \otimes \bra{k_{x}}_\gamma \\
&+ \int_{k_{1, x}, k_{2, x}\in {K_x} \& k_{1, x}\neq k_{2, x}} dk_{1, x}dk_{2, x} \frac{1}{L_\perp} \tilde{f}(k_{1, x}, k_{2, x}) \ket{-k_{1, x}}_{\mathrm{e}} \otimes \ket{k_{2, x}}_\gamma  \bra{-k_{2, x}}_{\mathrm{e}} \otimes \bra{k_{1, x}}_\gamma\\
=&2 k_{x, \text{max}}\int_{K_x} dk_{x}\frac{1}{L_\perp}\tilde{f}(k_{x}, k_{x}) \ket{-k_{x}}_{\mathrm{e}} \otimes \ket{k_{x}}_\gamma  \bra{-k_{x}}_{\mathrm{e}} \otimes \bra{k_{x}}_\gamma \\
&+ \int_{k_{1, x}, k_{2, x}\in {K_x} \& k_{1, x}< k_{2, x}} dk_{1, x}dk_{2, x} \frac{1}{L_\perp}\tilde{f}(k_{1, x}, k_{2, x}) \left( \ket{-k_{1, x}}_{\mathrm{e}} \otimes \ket{k_{2, x}}_\gamma  \bra{-k_{2, x}}_{\mathrm{e}} \otimes \bra{k_{1, x}}_\gamma\right.\\
&\left.\hspace{8cm}+\ket{-k_{2, x}}_{\mathrm{e}} \otimes \ket{k_{1, x}}_\gamma  \bra{-k_{1, x}}_{\mathrm{e}} \otimes \bra{k_{2, x}}_\gamma\right),\\
=&2 k_{x, \text{max}}\int_{K_x} dk_{x}\frac{1}{L_\perp}\tilde{f}(k_{x}, k_{x}) \ket{-k_{x}}_{\mathrm{e}} \otimes \ket{k_{x}}_\gamma  \bra{-k_{x}}_{\mathrm{e}} \otimes \bra{k_{x}}_\gamma \\
&+ \int_{k_{1, x}, k_{2, x}\in {K_x} \& k_{1, x}< k_{2, x}} dk_{1, x}dk_{2, x} \frac{1}{L_\perp}\tilde{f}(k_{1, x}, k_{2, x}) \left( \ket{\psi_+(k_{1, x}, k_{2, x})} \bra{\psi_+(k_{1, x}, k_{2, x})}\right.\\
&\left.\hspace{8cm}-\ket{\psi_-(k_{1, x}, k_{2, x})} \bra{\psi_-(k_{1, x}, k_{2, x})}\right),
    \end{aligned}
\end{equation}
where in the diagonalization, we define the unnormalizable eigenstates
\begin{equation}
    \ket{\psi_\pm(k_{1,x }, k_{2, x})}:=\frac{1}{\sqrt{2}}(\ket{-k_{1, x}}_{\mathrm{e}}\ket{k_{2, x}}_\gamma\pm \ket{-k_{2, x}}_{\mathrm{e}}\ket{k_{1, x}}_\gamma)
\end{equation}
for a given $k_{1, x}, k_{2, x}$, revealing a structure similar to a pair of Bell states. The first integral in Eq.~\eqref{eq:transposed_rho_red} sums over the projector of $\ket{-k_{x}}_{\mathrm{e}} \otimes \ket{k_{x}}_\gamma$ indexed by $k_x$, when $k_{1, x} = k_{2, x}$. In contrast, the second term represents the $2 \times 2$ matrices when $k_{1, x} \neq k_{2, x}$ in the basis ${\ket{-k_{1, x}}_\text{e} \otimes \ket{k_{2, x}}_\gamma, \ket{-k_{2, x}}_\text{e} \otimes \ket{k_{1, x}}_\gamma }$. Additionally, $\tilde{f}(k_{1, x}, k_{2, x}) = \tilde{f}(k_{2, x}, k_{1, x})$.

These block matrices are orthogonal to each other. Consequently, the eigenvalues from the first type of matrix are $\tilde{f}(k_{x}, k_{x})$, which are positive since they arise from the diagonal terms of the reduced density matrix $\rho_{\mathrm{e}\mhyphen\gamma, x}$. However, diagonalization of the $2 \times 2$ matrices yields the eigenvalues $\frac{1}{L_\perp}|\tilde{f}(k_{1, x}, k_{2, x})|$ and $-\frac{1}{L_\perp}|\tilde{f}(k_{1, x}, k_{2, x})|$, where the latter contributes to the negativity of the Cherenkov pair density operator. In effect, the negativity is given by:
\begin{equation}
    \mathcal{N}(\hat{\rho}_{\mathrm{e}\mhyphen\gamma, x}^{\intercal_\gamma})\sim \int_{k_{1, x}, k_{2, x}\in {K_x} \& k_{1, x}< k_{2, x}} dk_{1, x}dk_{2, x} \frac{1}{L_\perp}|\tilde{f}(k_{1, x}, k_{2, x})|,\,
\end{equation}
which is not an exact expression because it depends on the normalization of plane waves. Up to a positive prefactor arising from the eigenstates normalization, the states shown in Fig.~\ref{fig:rho_x_with_filters} satisfy $\mathcal{N}(\hat{\rho}_{\mathrm{e}\mhyphen\gamma, x}^{\intercal_\gamma})>0$, verifying the presence of entanglement. 
Additionally, increasing the energy range results in a more mixed reduced density matrix. This occurs because a broader energy range expands the set of unmeasured degrees of freedom, increasing the overall mixture in the system.

\subsection{Position and Momentum Measurement over the Cherenkov Pair} \label{ap:pos_mom_measurement}

\begin{figure}[h]
    \centering
    \includegraphics[width=0.55\linewidth]{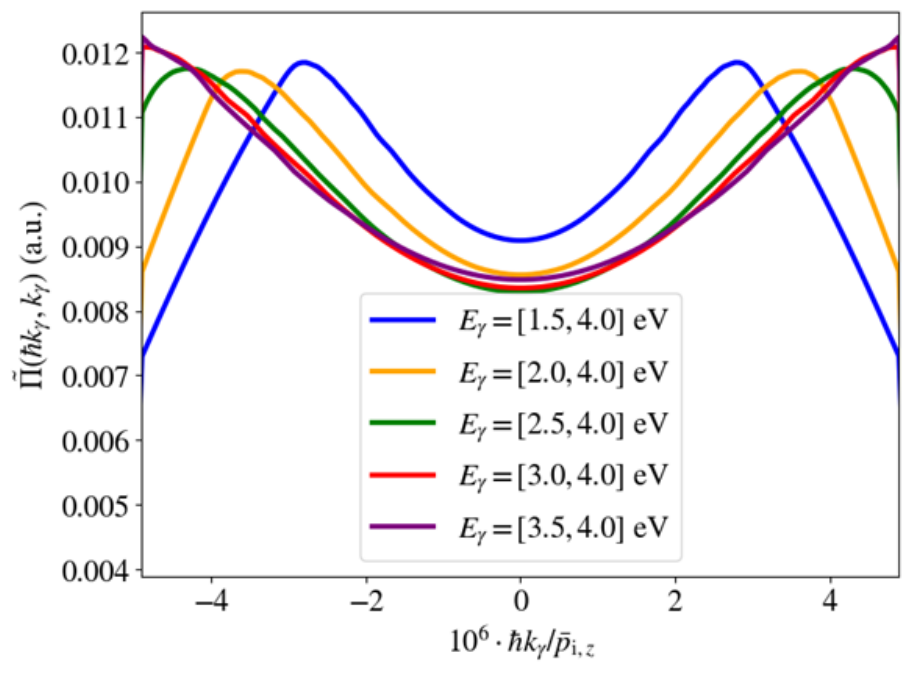}
    \caption{Electron momentum and photon momentum joint distribution, which has non-zero values when $\hbar k_\gamma =-p_{\mathrm{e}}$, as a function of the photon wave number $k_\gamma$. This distribution also corresponds to the diagonal terms of the reduced density matrix in Eq.~\eqref{eq:rhoX_constraint}.}
    \label{fig:momentum_dist}
\end{figure}

For the Cherenkov pair described in Appendix~\ref{ap:red_density_x}, the joint probability densities derived in Appendix~\ref{ap:position_momentum_measurements_electron_photon} reveal clear signatures of momentum and position correlations for this type of CL. Using the concrete analytical expression~\eqref{eq:felements_finitetiem} for the reduced density matrix elements, we find that the simultaneous measurement of momentum variables given by Equation~\eqref{eq:Pi_pp_general} leads to the probability density:
\begin{equation}\label{eq:Pi_pp}
    \begin{aligned}
\tilde{\Pi}(p_{\mathrm{e}}, k_\gamma)
=& \tilde{f}(k_\gamma, k_\gamma) \delta(\hbar k_\gamma+p_{\mathrm{e}}).
    \end{aligned}
\end{equation}
 where $N_{p-p}=1$ since the state normalization already accounts for the momentum detection range $K$.
 As  discussed in Appendix~\ref{ap:position_momentum_measurements_electron_photon}, the momentum anti-correlation is represented by the delta function dependence from momentum conservation and along the nonzero values the distribution is given by $\tilde{f}(k_\gamma, k_\gamma)$. In Fig.~\ref{fig:momentum_dist}, this distribution is shown for different energy windows, where the momentum constraints modify its shape along the momentum conservation condition $\hbar k_\gamma + p_{\mathrm{e}}=0$. As mentioned above, $k_x$ defines the allowed momentum values within the momentum boundaries, where each wave vector has a probability that scales with its magnitude and how closely $\theta_\gamma$ approaches the Cherenkov angle $\theta_{\text{\text{CR}}}$ and its probability to be transmitted outside the dielectric sample.

\begin{figure}[h]
    \centering
    \includegraphics[width=0.55\linewidth]{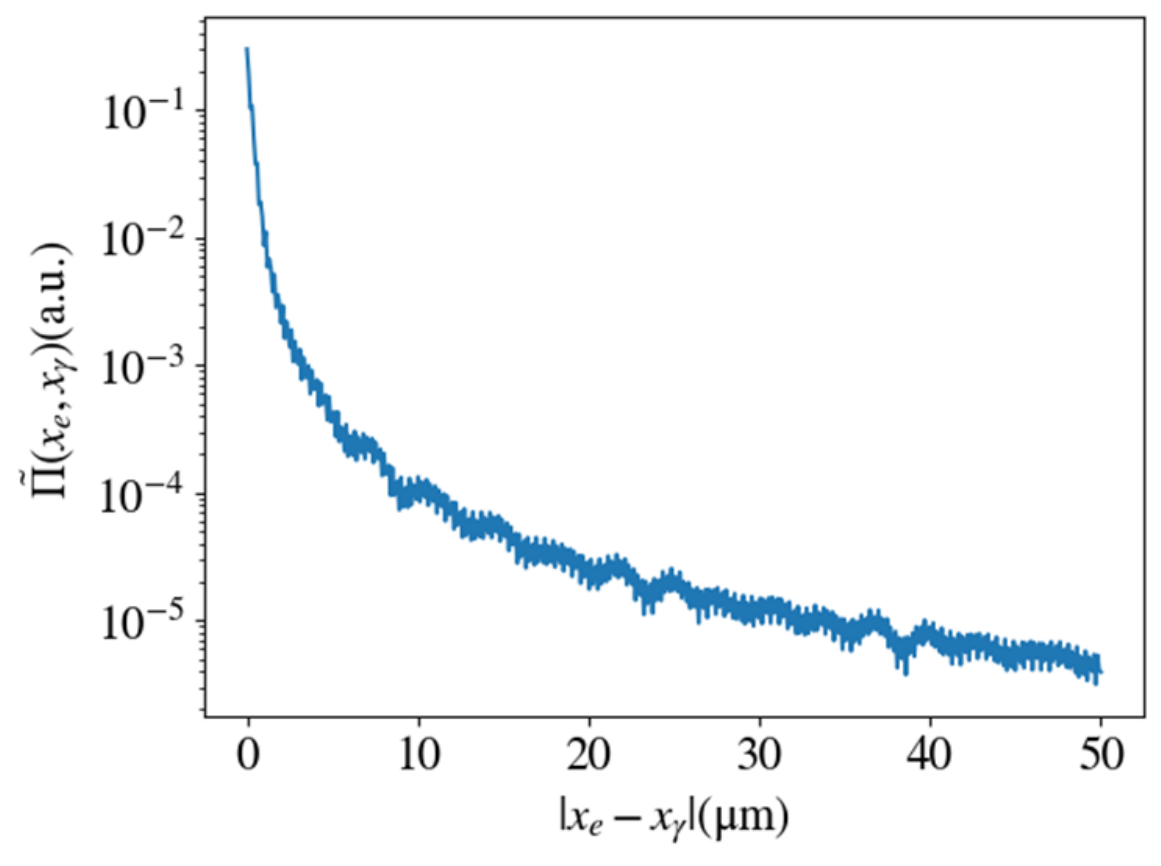}
    \caption{Electron position and photon position joint distribution as a function of the distance between each other for the electron-photon pair state in $x$ position after considering a energy window $E_\gamma \in [3.5, 4.0] \text{eV}$.}
    \label{fig:pos_dist}
\end{figure}

Considering the matrix elements in Eq.~\eqref{eq:felements_finitetiem} into Eq.~\eqref{eq:Pi_xx_general}, we find the probability density in that depends on the distance between the two particles, as shown also in Fig.~\ref{fig:pos_dist}. Numerical integration reveals a peaked distribution at zero distance $(|x_e - x_\gamma| = 0)$ along the $x$-axis, indicating the expected position correlation. The probability density decays monotonically for larger distances with minor oscillations. Notably, these oscillations occur systematically at certain larger periodic distances depending on the precision of the coarse-graining in reciprocal space. This numerical artifact can be overcome by considering either a shorter spatial range or finer divisions in momentum. Similarly, small-length-scale oscillations in position arise from the maximum momentum $\hbar k_{x, \text{max}}$, thereby depending on the momentum filtering applied to our state.

Probability densities involving the position of one particle and the momentum of the other, as given in Eqs.\eqref{eq:Pi_px_general} and\eqref{eq:Pi_xp_general}, depend solely on the momentum variable. The resulting uniformity in position reflects complete uncertainty in the conjugate observable, consistent with the properties of mutually unbiased bases.

Together, these results demonstrate that the Cherenkov emission process generates electron-photon pairs exhibiting quantum correlations in both momentum and position, with the structure of these correlations determined by the emission angle, spectral window, and the kinematic constraints of the process.

\clearpage
\newpage
\section{Discussion of Deflection Angles and Transverse Momentum}\label{ap:deflection_vs_momenta}

Practical momentum measurements are performed via measurements of the deflection angle relative to the incident beam. This angle depends not only on the transverse components of the wave vector, but also on its longitudinal component, $p_z$. Consequently, a one-to-one correspondence between angles and transverse momenta can only be established when the photon's energy or its longitudinal momentum $p_z$ is fixed.

 As demonstrated in Appendix~\ref{ap:emmsion_angle}, the radiation transmitted outside the dielectric does not have a unique momentum along $z$. Thus, achieving this uniqueness requires the use of a narrow energy filter.
The relationships between deflection angles and momentum variables are expressed as 
\begin{equation}\label{eq:deflections_pxkx}
    \tan \varphi_{\mathrm{e}, x} :=\frac{p_{x}}{p_{z}}, \quad \tan \varphi_{\gamma, x} :=\frac{k_{x}}{k_{z}}
\end{equation}
and in a similar way for the $y$-direction. 
Then, using momentum conservation (which is implied when assuming a large thickness dielectric slab in comparison to the photon wavelengths for the sake of simplicity) and energy conservation, we have:
\begin{equation}\label{eq:pxkx_deflections_conservation}
    \begin{aligned}
p_{x}=& (p_{\text{i},z}-\hbar k \cos\theta_\gamma)\tan\varphi_{\mathrm{e}, x}\\
=&\frac{1}{\beta c} (\beta^2E_\text{i}-E_\gamma)\tan \varphi_{\mathrm{e}, x},\\
k_{x}=& k_{z}\tan\varphi_{\gamma, x}\\
=&\frac{E_\gamma}{c\hbar \beta}\tan \varphi_{\gamma, x},
    \end{aligned}
\end{equation}
where $E_\gamma=\hbar \omega$ is the photon energy.

\begin{figure}
    \centering
    \includegraphics[width=0.85\linewidth]{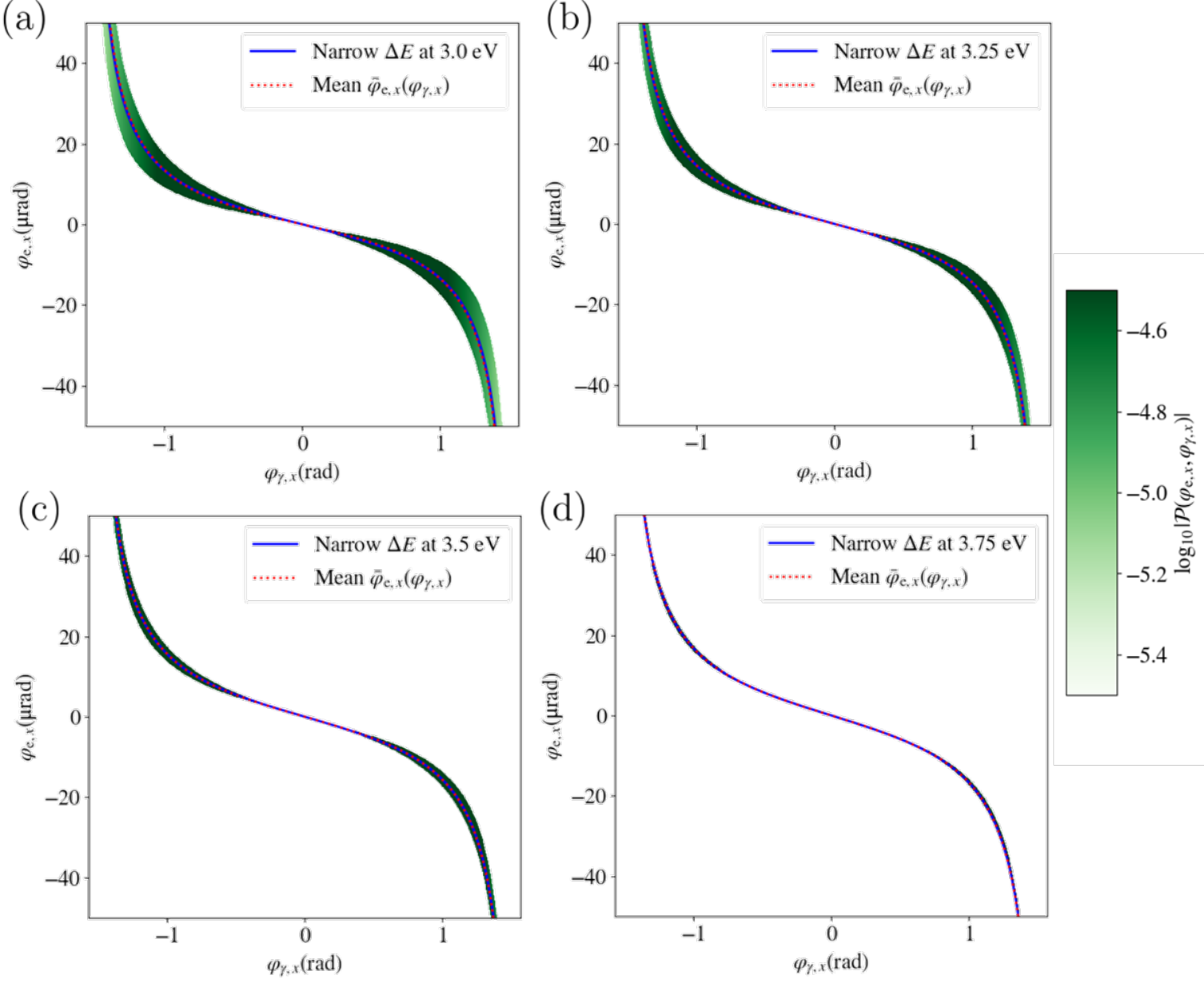}
    \caption{$\cP(\varphi_{\mathrm{e}, x}|\varphi_{\gamma, x})$, in logarithmic scale, is plotted according to Eq.~\eqref{eq:angular_prob_window} for different energy windows: (a) $E_\gamma \in [2.0, 4.0]\;$eV, (b) $E_\gamma \in [2.5, 4.0]\;$eV, (c) $E_\gamma \in [3.0, 4.0]\;$eV, and (d) $E_\gamma \in [3.5, 4.0]\;$eV. The narrow distribution given by Eq.~\eqref{eq:angular_prob_narrow} is represented by the blue curve at the median energy of each respective range. The mean value of $\varphi_{\mathrm{e}, x}$ under the distribution in Eq.~\eqref{eq:angular_prob_window} is shown as a dashed red curve, demonstrating good agreement between the two.
}
    \label{fig:deflection_Ewindow}
\end{figure}

 Based on Equations~\eqref{eq:deflections_pxkx} and \eqref{eq:pxkx_deflections_conservation}  and taking into account that momentum conservation implies $k_x=-p_x/\hbar$, we can relate the deflection angle corresponding to the electron's $x$ momentum and the respective angle for the photon as
\begin{equation}
    \begin{aligned}
\varphi_{\mathrm{e}, x}=&\arctan\left(\frac{\beta c \hbar k_x}{\beta^2E_\text{i}-E_\gamma}\right)\\
=& \arctan\left(\frac{E_\gamma \tan\varphi_{\gamma,x}}{\beta^2E_\text{i}-E_\gamma}\right).
    \end{aligned}
\end{equation}
Note that the deflection angles can be related one-to-one if we have complete certainty about the incident electron energy and produced photon energy. In fact, we can define the following probability relation:
\begin{equation}
    \cP(\varphi_{\mathrm{e}, x}, \varphi_{\gamma,x}|E_\gamma)\propto \delta\left(\varphi_{\mathrm{e}, x}+\arctan\left(\frac{E_\gamma \tan \varphi_{\gamma, x}}{\beta ^2E_\text{i}-E_\gamma}\right)\right)
\end{equation}
A probabilistic analysis of the joint probability between angular variables reveals how the probability $\cP(\varphi_{\mathrm{e}, x},\varphi_{\gamma, x})$ depends on energy filtering. Therefore,  
\begin{equation}
    \begin{aligned}
    \cP(\varphi_{\mathrm{e}, x},\varphi_{\gamma, x}) =& \int dE_\gamma \, \cP(\varphi_{\mathrm{e}, x}, \varphi_{\gamma,x} | E_\gamma) \cP(E_\gamma)\\
    \propto & \int dE_\gamma \, \delta\left(\varphi_{\mathrm{e}, x}+\arctan\left(\frac{E_\gamma \tan \varphi_{\gamma, x}}{\beta ^2E_\text{i}-E_\gamma}\right)\right) \frac{1}{\Delta E}  \Theta(E_\gamma - E_{\gamma,\text{min}}) \Theta(\Delta E + E_{\gamma,\text{min}} - E_\gamma)\\
    = & \int_{E_{\gamma,\text{min}}}^{E_{\gamma,\text{min}}+\Delta E} \frac{dE_\gamma}{\Delta E}  \delta\left(\varphi_{\mathrm{e}, x}+\arctan\left(\frac{E_\gamma \tan \varphi_{\gamma, x}}{\beta ^2E_\text{i}-E_\gamma}\right)\right).
    \end{aligned}
\end{equation}
The probability density for a Cherenkov photon with energy $E_\gamma$ is given by $\cP(E_\gamma) \propto \Gamma_\gamma$. In the low-energy photon regime, the emission rate is independent of the photon frequency, as recovered in Eq.~\eqref{eq:emission_rate_low_photon}.  When this probability is normalised over the observed energy range, we obtain  
\begin{equation}
    \cP(E_\gamma) = \frac{1}{\Delta E} \Theta(E_\gamma - E_{\gamma,\text{min}}) \Theta(\Delta E + E_{\gamma,\text{min}} - E_\gamma),
\end{equation}
where $\Delta E$ is the width of the energy range and $\Theta(x)$ is the Heaviside function.
In the limit where $\Delta E\rightarrow 0$ or if we work with a narrow energy filter at $E_\gamma$, we conclude that:
\begin{equation}\label{eq:angular_prob_narrow}
    \begin{aligned}
\cP(\varphi_{\mathrm{e}, x},\varphi_{\gamma, x})
\propto\delta\left(\varphi_{\mathrm{e}, x}+\arctan\left(\frac{E_\gamma \tan \varphi_{\gamma, x}}{\beta ^2E_\text{i}-E_\gamma}\right)\right),
    \end{aligned}
\end{equation}
in which the perfect correlation in momentum is transferred to the deflection angles. 
Otherwise, the condition imposed by the delta function can be solved for $E_\gamma$ giving
\begin{equation}
    E_\gamma^{\text{con}}=\beta^2E_\text{i} \frac{\tan \varphi_{\mathrm{e}, x}}{\tan \varphi_{\mathrm{e}, x}-\tan \varphi_{\gamma, x}},
\end{equation}
which is achievable under the conditions $\tan\varphi_{\mathrm{e}, x}\tan \varphi_{\gamma, x}<0$ and $E_\gamma^{\text{con}}\in [E_{\gamma,\text{min}}, E_{\gamma,\text{min}}+\Delta E]$. Then
\begin{equation}\label{eq:angular_prob_window}
    \begin{aligned}
\cP(\varphi_{\mathrm{e}, x},\varphi_{\gamma, x})
\propto&\left\{\begin{matrix}
\left|\frac{\beta^4E_\text{i}^2-2\beta^2E_\text{i}E_\gamma^{\text{con}}+\sec^2\varphi_{\gamma, x}E_\gamma^{\text{con} 2}}{\tan \varphi_{\gamma, x}\beta^2 E_\text{i}\Delta E}\right|, &\quad \tan\varphi_{\mathrm{e}, x}\tan \varphi_{\gamma, x}<0 ~~~~\text{and} ~~~~ E_\gamma^{\text{con}}\in [E_{\gamma,\text{min}}, E_{\gamma,\text{min}}+\Delta E]\\
0, &\quad  \text{otherwise}
\end{matrix}\right.
    \end{aligned}
\end{equation}
which leads to the plots in Fig.~\ref{fig:deflection_Ewindow} for different energy windows.

As a result, an exact correlation between the deflection angles due to momentum conservation requires a narrow energy window. The narrower the range of allowed photon energies, the more peaked the distribution becomes, leading to a closer 
correspondence between the deflection angles. This reinforces the necessity of using a narrow energy window in the results. Therefore, we use the energy range shown in Fig.~\ref{fig:ang-profile}~d and Fig~\ref{fig:deflection_Ewindow}~e, specifically $E_\gamma \in [3.5, 4.0]$ eV, but this value, in experimental realizations, can be narrower with a resolution of $0.1$ eV. 
Additionally, due to the bi-valued correspondence between momentum and deflection angles, it is necessary to restrict the analysis to $|\varphi_{\gamma, x}| \leq \frac{\pi}{2}$. Alternatively, working with the average value of $\varphi_{\mathrm{e}, x}$ for a given energy window could provide a good fit with the narrow distribution case at the middle of the energy range.

As discussed above, momentum measurements pose challenges in establishing a reliable connection between deflection angle counts and transverse momentum counts. This effect is even more pronounced for photons than for electrons, as small deflections in the electron trajectory can result in large photon deflection angles due to momentum and energy conservation constraints, which are influenced by energy uncertainty.  
However, a perfectly narrow energy filter that fixes the photon energy $E_\gamma$ allows for a direct correspondence between deflection angles and transverse momentum. In particular, deflection counts (for photons, but equivalently for electrons) can be translated into transverse momentum counts using the following relationships:
\begin{subequations}
    \begin{align}
        k_x &= k_\gamma \sin\varphi_{\gamma, x} \sqrt{\frac{\cos^2\varphi_{\gamma, y}}{1 - \sin^2\varphi_{\gamma, x} \sin^2\varphi_{\gamma, y}}}, \\
        k_y &= k_\gamma \sin\varphi_{\gamma, y} \sqrt{\frac{\cos^2\varphi_{\gamma, x}}{1 - \sin^2\varphi_{\gamma, x} \sin^2\varphi_{\gamma, y}}},
    \end{align}
\end{subequations}
where $k_\gamma = E_\gamma / \hbar c$. These relations allow for two approaches: either designing a tailored mask in terms of deflection angles based on the desired coarse-graining for momentum projective maps, or performing post-processing on deflection data to reconstruct the expected results in transverse momentum.

\newpage
\section{Experimental Values for the FWHM} \label{ap:experimenta_resolution_values}

As explained in the main text, we model the limited precision of the position and momentum measurements via a Gaussian PSF. 
It is characterised by the FWHM, $\delta\suptiny{0}{0}{(\kappa)}_\zeta$, where, as before, $\kappa\in\{x,p\}$ and $\zeta\in\{\mathrm{e},\gamma\}$. 
In order to gauge the experimental feasibility of this entanglement witness we estimate what values for $\delta\suptiny{0}{0}{(\kappa)}_\zeta$ are realistically attainable. 

We have extracted estimates from literature on comparable experiments \cite{MatsukataOguraGarciadeAbajo2022,KisielowskietAl2008,Yamamoto2016_Highres_CL_STEM}, which investigate position- and momentum-resolved CL or conduct TEM imaging. 
While this provides an overview of the state of the art, the measurements originate from distinct experiments which were not designed for studying electron-photon entanglement. 

The first steps in building a similar experiment to the one depicted in Fig.~\mbox{\ref{fig:setup}} have recently been published~\mbox{\cite{preimesbergerExperimentalVerificationElectronPhoton2025}}.
In the following, we will provide estimates of $\delta\suptiny{0}{0}{(\kappa)}_\zeta$ for this setup. We omit a separate discussion of $\delta\suptiny{0}{0}{(x)}_\mathrm{e}$ as the required position resolution is easily attainable in any TEM~\cite{ReimerKohl2008}.

\subsection{Electron Momentum - $\delta\suptiny{0}{0}{(p)}_\mathrm{e}$}
Measurements were taken on a Tecnai G2 F20 transmission electron microscope at a primary beam energy of $E_\mathrm{kin} = 200\;\mathrm{keV} \pm 0.45\;\mathrm{eV}$ (FWHM). 
We illuminated a 30$\,$nm thick, mono-crystalline Si membrane with a collimated electron beam. 
We operated the microscope in low angle diffraction mode and used a Timepix3-based detector (Advascope ePix) that is retrofitted to the TV-port of a Gatan GIF 2001 post-column imaging filter.

\begin{figure}[H]
\includegraphics[width = \linewidth]{"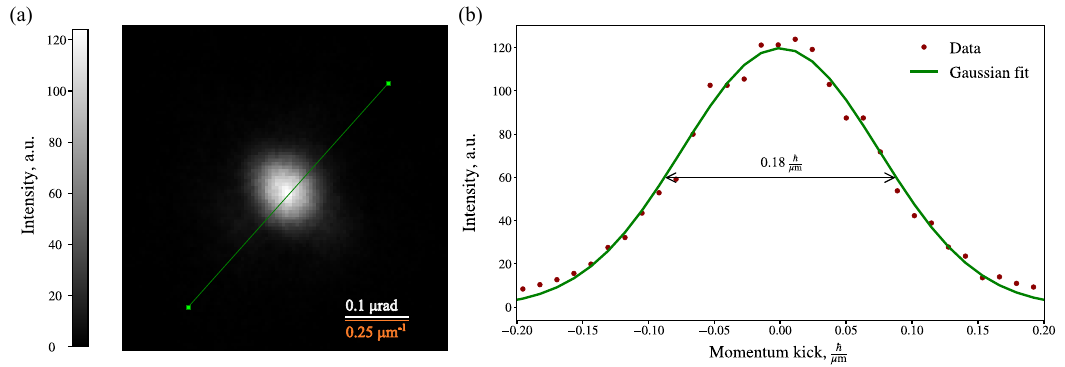"}
\caption{Measuring the attainable momentum resolution for the electron: (a) shows a low angle diffraction image of the undeflected electron beam after passing through a crystalline sample. (b) shows a line profile of the measurement in (a), fitted with a Gaussian distribution in order to gauge the momentum resolution.}
\label{fig:e_mom_res}
\end{figure}

The momentum space images are calibrated by extrapolating the calibration curve from lower nominal camera lengths to the extremely high camera lengths needed for characterising such low angle scattering. 
To calibrate the lower camera lengths, we use the diffraction pattern from a diffraction grating replica (Agar Scientific, S106, $2160\;$lines/mm). Fig.~\ref{fig:e_mom_res} shows the null beam in diffraction mode.
By taking a line profile and fitting a Gaussian to the intensity distribution we obtain a value of $\delta\suptiny{0}{0}{(p)}_\mathrm{e} = 0.18\,\frac{\hbar}{\text{\textmu} \text{m}}$ for the best-case resolution.

\subsection{Photon Position - $\delta\suptiny{0}{0}{(x)}_\mathrm{\gamma}$}
We measured the spatial resolution of our experimental optical setup by collecting the CL emission generated by an electron beam. 
Using the same TEM in scanning mode, we illuminated a $\text{100\, nm}$ thick monocrystalline silicon membrane to produce CL. Then, we collected the CL emission using a custom-made parabolic mirror, which directed it through an optical window out of the TEM column into the optical system. Subsequently, a series of lenses focused the CL beam onto a CMOS camera.
Despite the scanning electron beam having a diameter of approximately 5$\,$nm, the spatial resolution of the emitted CL beam is limited by the diffraction limit of the photons. 
Additionally, imperfections of optical components such as the parabolic mirror and lenses may further reduce the quality of the recorded image. 

After acquiring a picture of the beam on the CMOS, we shifted the electron beam by $1-10\,$\textmu m in the transverse direction of the parabolic mirror and recorded a second image of the CL on the CMOS. 

\begin{figure}[H]
\centering
\includegraphics[width= 0.9\linewidth]{"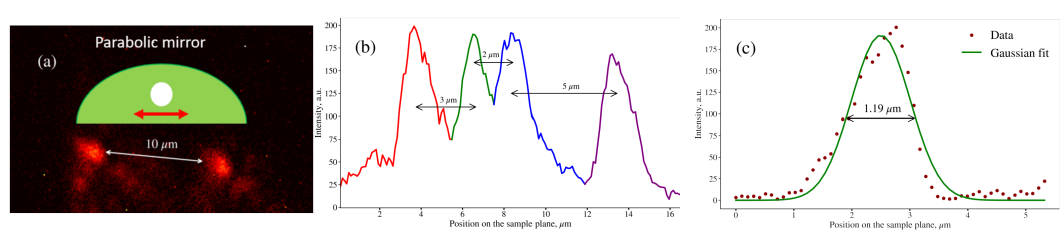"}
\caption{Imaging the CL beam in position space: (a) Two overlaid CL images with $10\,$\textmu m shift of the electron beam. The sketch of the parabolic mirror shows the direction of electron beam shifts (not to scale).
(b) Line profile of overlaid CL images with electron beam shifts by $3$, $2$ and $5\,$\textmu m. 
(c) Line profile of the CL beam from one of the acquisition frames shown in (a).}
\label{fig:CL_position}
\end{figure}

We use these images to calibrate the magnification of our CL imaging system. To quantify its spatial resolution, we fit the shape of the CL signal with a Gaussian distribution. The experimental results are shown in Fig.~\ref{fig:CL_position}. 
In Fig.~\ref{fig:CL_position}~a, one can see the original and shifted beam in position space separated by a distance of $10\,$\textmu m. 
The observed aberrations in the CL beam, mainly resulting from the roughness of the parabolic mirror surface, degrade the spatial resolution of the imaging system. 
Fig.~\ref{fig:CL_position}~b depicts the sum of multiple line profiles, showing the CL beam in different positions. The electron beam is shifted by $3$, $2$ and $5\,$\textmu m. 
Two CL peaks separated by a distance of $2\,$\textmu m are clearly spatially resolved according to the Rayleigh criterion \cite{Rayleigh01101879,Rayleigh01081896}. 
Fig.~\ref{fig:CL_position}~c shows a line profile from the CL beam in a single acquisition frame. 
The fit reveals a spatial resolution of $\sim1.19\,$\textmu m.

\subsection{Photon Momentum - $\delta\suptiny{0}{0}{(p)}_\mathrm{\gamma}$}
\begin{figure}[H]
\centering
\includegraphics[width= 0.9\linewidth]{"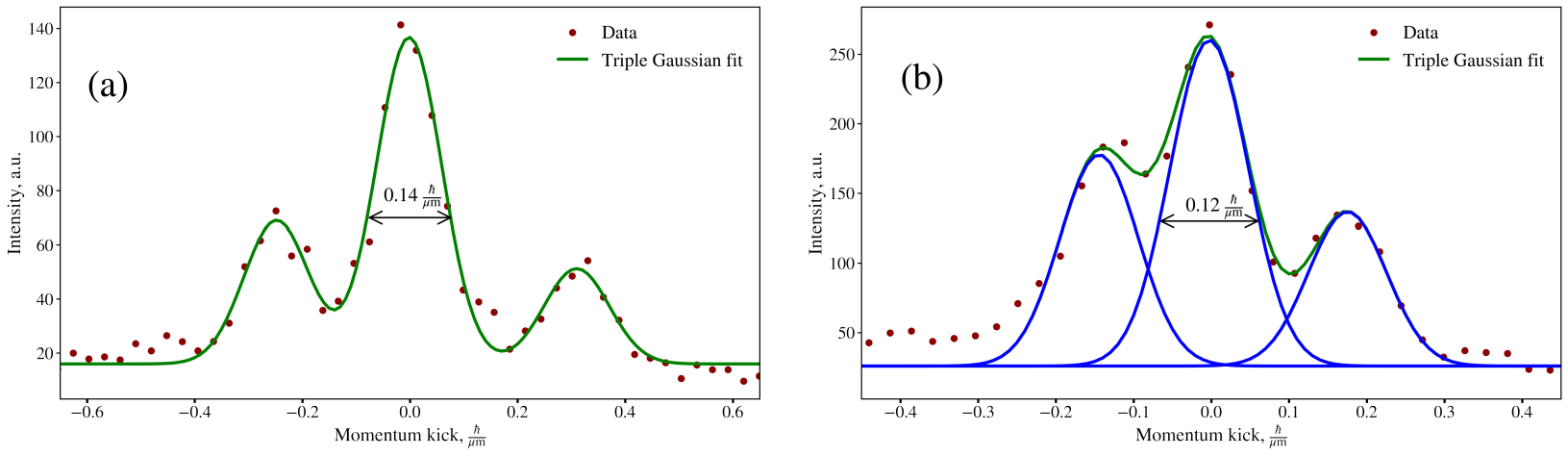"}
\caption{
Interference patterns from different gratings:
(a) Line profile of the interference pattern of a grating with a periodicity of $20\,$\textmu m. 
(b) Line profile of the interference pattern of a grating with a periodicity of $40\,$\textmu m.
}
\label{fig:CL_momentum}
\end{figure}

We measured the momentum resolution of the experimental optical setup and the custom-made parabolic mirror by creating interference patterns on an optical bench outside of the TEM. 
A green point light source was used to illuminate different grating lines.
The light was reflected by the parabolic mirror before it propagated through the optical setup. 
The grating lines were approximately parallel to the optical axis of the parabolic mirror. 
Thus, we determine the momentum resolution of our optical system only in the transverse direction of parabolic mirror. 
We focussed the interference pattern onto the CMOS camera and captured a single frame. 
Due to the dynamic range of the CMOS, one can capture only the 0th and 1st interference maxima in a single image.
Since the period of the grating is known, we can determine the momentum transfer associated with the first maximum and thus calibrate the momentum scale. 
The obtained experimental interference patterns are shown in Fig.~\ref{fig:CL_momentum}.

Fig.~\ref{fig:CL_momentum}~a shows the spatially resolved 0th and 1st interference orders of a grating with a periodicity of $20\,$\textmu m.  
To find the momentum resolution we fit the data with a triple Gaussian function with identical FWHMs for each Gaussian. 
In contrast, Fig.~\ref{fig:CL_momentum}~b shows the line profile of an interference pattern resulting from a $40\,$\textmu m grating. 
As a result, we obtain a momentum resolution of the optical system of $\sim0.14\,\frac{\hbar}{\text{\textmu} \text{m}}$.
\newpage

\section{Robustness}\label{app:robustness}
Measurement errors can either be considered during the data analysis or by adapting the provided entanglement bounds~\cite{morelli_entanglement_2022,LiHuberFriis2024}. As such, any experimental implementation of the presented protocol would ideally be accompanied by an in-depth characterisation of the underlying errors and noise. In this appendix, we discuss two types of expected deviations from the ideal measurements.
Firstly, statistical errors will mostly result in a coarser and worse resolution as described in Sec.~\ref{sec:electron_cherenkov_photon_pair}. Below, we estimate how far this blurring can go before our witness stops detecting entanglement.
Secondly, misalignment errors which affect the unbiasedness of the projectors, like periodicity errors, can be characterised in order to adapt the bounds according to the closeness of the experimental and theoretical measurements. Here, we discuss the protocol's robustness to them.

\subsection{Minimal Variances for Detection}

To provide a rough estimate of the resolution limits compatible with entanglement certification in CV, one may consider a simplified distribution. We model the normalised joint probability distributions as infinitely wide in position and momentum with Gaussian correlation/anti-correlation profiles, characterised by the standard deviations $\Sigma_x$ and $\Sigma_p$:
\begin{subequations}
    \begin{align}
        \tilde{\Pi}(p_{\rm e}, \hbar k_\gamma) &\propto
        \exp\left\{-\frac{1}{2}\frac{(\hbar k_\gamma + p_{\rm e})^2}{\Sigma_p^2}\right\}, \\
        \tilde{\Pi}(x_{\rm e}, x_\gamma) &\propto
        \exp\left\{-\frac{1}{2}\frac{(x_\gamma - x_{\rm e})^2}{\Sigma_x^2}\right\}.
    \end{align}
\end{subequations}
These distributions are not physical, as they have infinite energy, but incorporate generic blurring effects without attributing them to specific mechanisms such as detector response or decoherence.

\begin{figure}[H]
    \centering
    \includegraphics[width= 0.5\linewidth]{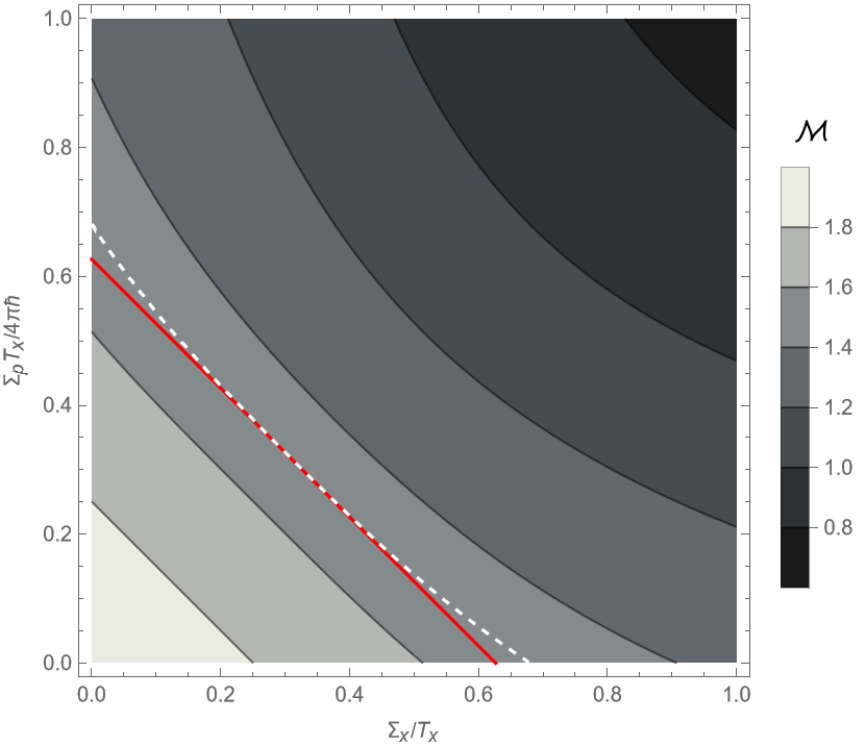}
    \caption{
    Entanglement bound $\mathcal{M}$ as a function of the standard deviation of the position-position and momentum-momentum probability distribution. 
    The dashed white curve defines the values of this standard deviation where the bound is exactly 1.5, and the red curve defines a linear bound below which it is always possible to have a correlation measure larger than 1.5 and certify entanglement. 
    }
    \label{fig:bound_sxsp}
\end{figure}

To relate the distribution widths to measurable correlations, we consider periodic coarse-graining with spatial- and momentum-bin sizes $T_x$ and $T_p$.
\begin{equation}
    \begin{aligned}
        \mathcal{P}(0^{(p)}, 0^{(p)}) = \mathcal{P}(1^{(p)}, 1^{(p)}) 
        &= \mathcal{N} \int_{-T_p/2}^{T_p/2} dp_{\rm e} \int_{-T_p/2}^{T_p/2} d(\hbar k_\gamma)~\tilde{\Pi}(p_{\rm e}, \hbar k_\gamma) \\
        &= \frac{1}{\sqrt{2\pi}} \frac{\Sigma_p}{T_p} \left( \exp\left\{-\frac{T_p^2}{2\Sigma_p^2}\right\} - 1 \right) + \frac{1}{2} \mathrm{Erf}\left[\frac{T_p}{\sqrt{2} \Sigma_p}\right],
    \end{aligned}
\end{equation}
with the normalisation constant $\mathcal{N}$.
Analogously, in position space:
\begin{equation}
    \mathcal{P}(0^{(x)}, 0^{(x)}) = \mathcal{P}(1^{(x)}, 1^{(x)}) = \frac{1}{\sqrt{2\pi}} \frac{\Sigma_x}{T_x} \left( \exp\left\{-\frac{T_x^2}{2\Sigma_x^2}\right\} - 1 \right) + \frac{1}{2} \mathrm{Erf}\left[\frac{T_x}{\sqrt{2} \Sigma_x}\right].
\end{equation}
The total correlation measure, $\mathcal{M}$, which defines the witness is given by:
\begin{equation}
    \begin{aligned}
        \mathcal{M} = &~\mathcal{P}(0^{(x)}, 0^{(x)}) + \mathcal{P}(1^{(x)}, 1^{(x)}) + \mathcal{P}(0^{(p)}, 0^{(p)}) + \mathcal{P}(1^{(p)}, 1^{(p)}) \\
        = &~\frac{2}{\sqrt{2\pi}} \frac{\Sigma_p}{T_p} \left( \exp\left\{-\frac{T_p^2}{2\Sigma_p^2}\right\} - 1 \right) + \mathrm{Erf}\left[\frac{T_p}{\sqrt{2} \Sigma_p} \right] 
        + \frac{2}{\sqrt{2\pi}} \frac{\Sigma_x}{T_x} \left( \exp\left\{-\frac{T_x^2}{2\Sigma_x^2}\right\} - 1 \right) + \mathrm{Erf}\left[\frac{T_x}{\sqrt{2} \Sigma_x} \right].
    \end{aligned}
\end{equation}
As illustrated in Fig.~\ref{fig:bound_sxsp}, large blurring (i.e., large $\Sigma_x$ or $\Sigma_p$ compared to $T_x$ or $T_p$) reduces the observed correlation $\mathcal{M}$ below the threshold of $1.5$, preventing entanglement certification.

Imposing $T_x T_p = 4\pi\hbar$ as in the proposal, we can define a tight linear lower bound for observing entanglement.
\begin{equation}
     \frac{\Sigma_p }{4\pi \hbar/T_x}+ \frac{\Sigma_x}{T_x}\leq\sqrt{\frac{a}{\pi}},
\end{equation}
where $a=1.235$.
Rewriting the equality as a quadratic and solving for $T_x$ shows that it must lie within the interval $T_x \in \left[T_{x}^{(-)}, T_{x}^{(+)}\right]$, where
\begin{equation}
    T_{x}^{(\pm)} = \frac{2 \hbar\sqrt{\pi a}}{\Sigma_p} \left( 1 \pm \sqrt{1 - \frac{\Sigma_x \Sigma_p}{a\hbar}} \right).
\end{equation}
If the discriminant becomes negative, i.e., $\Sigma_x \Sigma_p > a\hbar$, then $T_x$ does not have real solutions. However, if $\Sigma_x \Sigma_p \lesssim a\hbar$, for $T_x \in \left[T_{x}^{(-)}, T_{x}^{(+)}\right]$, it is guaranteed that $\mathcal{M}\geq 1.5$. In this regime, periodic binning can certify entanglement, providing a rough upper bound on the tolerable level of blurring in the joint distribution.

\subsection{Periodicity}

Generally, the mutual unbiasedness of two bases made up of rank-one projectors 
can be verified by sequential measurements in different elements of each basis. 
The absence of correlations proves their unbiasedness.  

For CV projectors, a similar argument as above has been used to experimentally validate the required conditions on the product of the periodicities~\cite{TascaSanchezPieroWalbornRudnicki2018}. 
Here, the initial state passes two filters corresponding to projections onto periodic basis elements. 
The unbiased bases then produce uniform outcome statistics. 
For higher-rank projectors this condition is necessary but not sufficient to prove unbiasedness because they have multiple,
even infinitely many, eigenstates. 
A sequential measurement, however, only shows the absence of correlations for one of these eigenstates. 
To give further evidence for unbiasedness, one can add a third measurement in the initial basis, which determines whether the encoded information was retained. 
This can be tested using numerical models, e.g., considering Gaussian initial wave functions, one can show that the uniformity of the outcome statistics is particularly robust for two-dimensional measurements, which can be explained by the symmetry of the bases. 
Here, as a worst case, we have considered an initial Gaussian distribution in position with a resolution width of $0.1\,$\textmu m and an optimal periodicity of $10\,$\textmu m. 
As a result, differences of $20\%$ in the periodicity only lead to a reduction in the distribution entropy of $0.8\%$ (see Ref.~\cite{TascaSanchezPieroWalbornRudnicki2018} for comparison), meaning that for small variations in periodicity, the information cannot be retained and the bound holds.
Finally, considering any experimental misalignments as combinations of phase errors, which do not affect the unbiasedness, and periodicity errors, the bounds are robust to such experimental factors.

If the correlations are not compatible with mutual unbiasedness but the measurement bases are nevertheless known, 
the criterion given in Eq.~\eqref{eq:lower bound on I of rho 2} can be modified as shown in Ref.~\cite{LiHuberFriis2024}. Deviations from mutual unbiasedness generally lead to increased thresholds for entanglement detection, which can be problematic even for small deviations when aiming to quantify high-dimensional entanglement~\cite{LiHuberFriis2024}, but small deviations are typically not prohibitive for detecting bipartite entanglement.

When the imperfections are not known exactly, but can be bounded, methods from Ref.~\cite{MorelliYamasakiHuberTavakoli2022} can be used to determine increased entanglement-detection thresholds for typical witnesses (including those we describe here). As in the previous case, the penalties can be significant for detecting high-dimensional entanglement and can prevent the detection of weakly entangled states even in the bipartite case (specifically, the separability thresholds can increase by adding terms proportional to $\sqrt{\epsilon}$ for small measurement inaccuracies $\epsilon$), but they do not prevent the detection of entanglement for moderately entangled states.

The strategies described above still fall into the category of device-dependent scenarios, which means that some trust in or knowledge of the correct functioning of the measurement devices is assumed. In cases where these assumptions are not reasonable, one may switch entirely into the device-independent setting, where one considers the violation of Bell-type inequalities for entanglement detection (see, e.g., Ref.~\cite{BrunnerCavalcantiPironioScaraniWehner2014} for a review). In such scenarios, the thresholds for entanglement detection (violation of a Bell inequality) are generally higher (although not prohibitively so), but have the advantage that they can be used to detect entanglement irrespective of any detector imperfections. 

\end{document}